%% file: CaCoTin-paper.tex
\documentclass[fleqn,twoside]{article}
\usepackage{epsfig}
\usepackage{xspace}
\usepackage{longtable}
\usepackage{graphicx}
\usepackage[figuresright]{rotating}

\newcommand{\MeV}{\ensuremath{\mbox{MeV}}\xspace}
\newcommand{\GeVc}{\ensuremath{\mbox{GeV}/c}\xspace}
\newcommand{\MeVc}{\ensuremath{\mbox{MeV}/c}\xspace}
\newcommand{\T}{\ensuremath{\mbox{T}}\xspace}

\newcommand{\mm}{\ensuremath{\mbox{mm}}\xspace}

\newcommand{\mrad}{\ensuremath{\mbox{mrad}}\xspace}
\newcommand{\rad}{\ensuremath{\mbox{rad}}\xspace}

\newcommand{\ps}{\ensuremath{\mbox{ps}}\xspace}

\newcommand{\dedx}{\ensuremath{\mbox{d}E/\mbox{d}x}\xspace}
\newcommand{\pip}{\ensuremath{\pi^+}\xspace}
\newcommand{\pim}{\ensuremath{\pi^-}\xspace}
\newcommand{\piz}{\ensuremath{\pi^0}\xspace}
\newcommand{\bfpip}{\ensuremath{\mathbf {\pi^+}}\xspace}
\newcommand{\bfpim}{\ensuremath{\mathbf {\pi^-}}\xspace}

\newcommand{\dzeroprime}{\ensuremath{d'_0}\xspace}
\newcommand{\zzeroprime}{\ensuremath{z'_0}\xspace}
\newcommand{\evtspill}{\ensuremath{N_{\mathrm{evt}}}\xspace}

\def\be{\begin{equation}}
\def\ee{\end{equation}}
\def\bea{\begin{eqnarray}}
\def\eea{\end{eqnarray}}

\newcommand{\bfGeVc}{\ensuremath{\mathbf {\mbox{\bf GeV}/c}}\xspace}
\begin{document}
\title{\bf Large-angle production of charged pions by 3~\bfGeVc--12~\bfGeVc protons on 
 carbon, copper and tin  targets}

\author{HARP Collaboration}

\maketitle

\begin{abstract}
  A measurement of the  double-differential $\pi^{\pm}$ production
  cross-section  in proton--carbon, proton--copper and proton--tin
  collisions  in the range of pion
  momentum $100~\MeVc \leq p < 800~\MeVc$ 
  and angle $0.35~\rad \le \theta <2.15~\rad$
  is presented. 
  The data were taken  with the HARP detector in the T9 beam
  line of the CERN PS.
  The pions were produced by proton beams in a momentum range from
  3~\GeVc to  12~\GeVc hitting a target with a thickness of
  5\% of a nuclear interaction length.  
  The tracking and identification of the
  produced particles was done using a small-radius
  cylindrical time projection chamber (TPC) placed in a solenoidal
  magnet. 
  An elaborate system of detectors in the beam line ensured the
  identification of the incident particles.
  Results are shown for the double-differential cross-sections 
  $
  %{{d^2 \sigma^{\pi}}}/{{dpd\Omega }}
  %%{{d^2 \sigma}}/{{dpd\Omega }}
  {{\mathrm{d}^2 \sigma}}/{{\mathrm{d}p\mathrm{d}\theta }}
  $
  at four incident proton beam momenta (3~\GeVc, 5~\GeVc, 8~\GeVc 
  and 12~\GeVc).
\end{abstract}
\begin{center}
(Submitted to The European Physical Journal C)
\end{center}

\clearpage
\input{harpauthors}
\clearpage
%\tableofcontents
%\clearpage
%\listoffigures
%\clearpage
%\listoftables
%\clearpage

\section{Introduction}

 %+++ short intro HARP goals +++
The HARP experiment~\cite{harp-prop} 
makes use of a large-acceptance spectrometer for
a systematic study of hadron
production on a large range of target nuclei for beam momenta from 1.5~\GeVc to 15~\GeVc. 
The main motivations are the measurement of pion yields for a quantitative
design of the proton driver of a future neutrino factory~\cite{ref:nufact}, 
a substantial improvement of the calculation of the atmospheric neutrino
flux~\cite{Battistoni,Stanev,Gaisser,Engel,Honda} 
and the measurement of particle yields as input for the flux
calculation of accelerator neutrino experiments, 
such as K2K~\cite{ref:k2k,ref:k2kfinal},
MiniBooNE~\cite{ref:miniboone} and SciBooNE~\cite{ref:sciboone}. 

 %+++++++++++++++++++++++++++physics motivation intro++++++++++++++

The measurement of the double-differential cross-section, 
$
%{{d^2 \sigma^{\pi}}}/{{dpd\Omega }}
{{\mathrm{d}^2 \sigma^{\pi}}}/{{\mathrm{d}p\mathrm{d}\theta }}
$
%of positive and negative pion production for 
for $\pi^{\pm}$ production by
protons of 3~\GeVc, 5~\GeVc, 8~\GeVc and 12~\GeVc momentum impinging
on a thin carbon, copper or tin target of 5\% nuclear interaction length
are  presented. 

Especially for carbon targets it is interesting to measure pion
production cross-sections in the framework of the HARP measurement programme
for neutrino flux calculations.
Carbon targets are frequently used as hadron production targets in
neutrino beam lines.
In addition, measurements on carbon can be used to predict pion
production off nitrogen and oxygen nuclei without a large extrapolation
in the production models.
The knowledge of the latter production cross-sections are needed to
model atmospheric muon and neutrino fluxes.
Owing to the relatively low incoming beam momenta the data are
especially interesting for the calculation of hadron production in
secondary interactions in extended production targets and in atmospheric
flux calculations.
The comparison of the 
measurements on copper and tin targets with the carbon data in this
paper and with the tantalum data obtained with the same apparatus
described in Ref.~\cite{ref:harp:tantalum} can be used to check the
dependence on the atomic number $A$ in hadron production models.
Copper and tin are interesting target materials as their atomic 
numbers are midway between light target materials, such as Be, Al and C
(used in targets for conventional neutrino beams) and heavy targets 
such as Ta (relevant for the optimization of neutrino factory designes). 
%+++++++++++intro part of detector description++++++++++++++++++
%

Data were taken in the T9 beam of the CERN PS.
For this analysis,  about 1,159,000 (1,066,000 and 1,284,000)  incoming protons 
were selected which gave an interaction trigger in the Large Angle spectrometer,
resulting in 235,000 (209,500 and 243,400) well-reconstructed secondary pion tracks
for the carbon (copper and tin) target.
The different settings have been taken within a short running period so that
in their comparison detector variations are minimized.

%
%+++++++++++short description of analysis++++++++++++++++++
%

The analysis proceeds by selecting tracks in the Time Projection
Chamber (TPC) in events with incident beam protons.  
Momentum and polar angle measurements and particle identification are
based on the measurements of track position and energy deposition in
the TPC.
An unfolding method is used to correct for experimental resolution,
efficiency and acceptance and to obtain the double-differential pion
production cross-sections, with a full error 
%discussion.
evaluation.
A comparison with available data is 
presented. 
The analysis follows the same methods as the ones used for the determination of
$\pi^{\pm}$ production cross-sections by
protons on a tantalum target described in Ref.~\cite{ref:harp:tantalum}.
We refer to Ref.~\cite{ref:harp:tantalum} for a detailed description of
the analysis, only the main points and differences with respect to the
latter are described here.

\section{Experimental apparatus}
\label{sec:apparatus}
\begin{figure}[tbp]
  \begin{center}
    \hspace{0mm} \epsfig{file=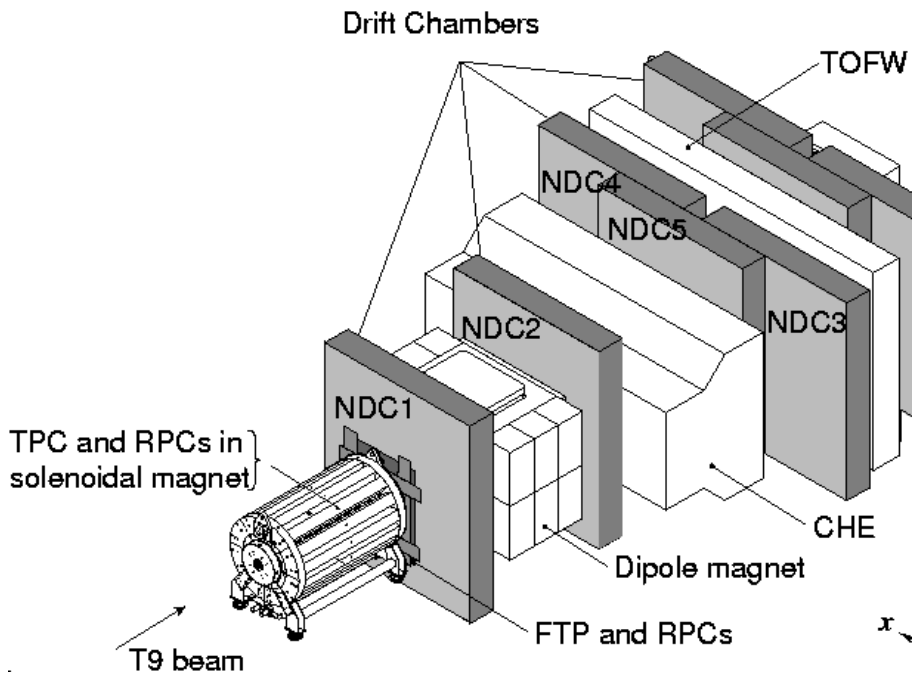,width=12cm}
  \end{center}
\caption{Schematic layout of the HARP detector. 
The convention for the coordinate system is shown in the lower-right
corner. 
The three most downstream (unlabelled) drift chamber modules are only partly
equipped with electronics and are not used for tracking.
}
\label{fig:harp}
\end{figure}

 The HARP detector is shown in Fig.~\ref{fig:harp}.
% together with 
% the convention used for the coordinate system.
%
 The forward spectrometer is built around a dipole magnet 
 for momentum analysis, with large planar drift chambers
 (NDC)~\cite{NOMAD_NIM_DC} for particle tracking and  a time-of-flight wall
 (TOFW)~\cite{ref:tofPaper}, a threshold Cherenkov detector (CHE) 
 and an electromagnetic calorimeter (ECAL)
 used for particle identification.
 The forward spectrometer covers an acceptance for tracks originating
 from the target with polar angles up to 250~\mrad.
 This is well matched to the angular range of interest for the 
 measurement of hadron production to calculate the properties of
 conventional accelerator neutrino beams~\cite{ref:alPaper,ref:bePaper}.
 In the large-angle region a cylindrical TPC with a radius of 408~\mm
 is  positioned in a solenoidal magnet with a field of 0.7~\T. 
 The target is inserted into the inner field cage of the TPC.
 The TPC is used
 for tracking, momentum determination and the measurement of the
 energy deposition \dedx for particle identification~\cite{ref:tpc:ieee}.
 A set of resistive plate chambers (RPC) form a barrel inside the solenoid 
 around the TPC to measure the arrival time of the secondary
 particles~\cite{ref:rpc}. 
 Beam instrumentation provides identification of the incoming
 particle, the determination of the time when it hits the target, 
 and the impact point and direction of the beam particle
 on the target.  
 Several trigger detectors are installed to select events with an
 interaction and to define the normalization.
 
In addition to the data taken with the thin carbon, copper and tin targets of
5\% nuclear interaction length ($\lambda_{\mathrm{I}}$),
runs were also taken with an empty target holder, a 
thin 2\%~$\lambda_{\mathrm{I}}$ target and a
thick 100\%~$\lambda_{\mathrm{I}}$ target.
Data taken with a liquid hydrogen target at 3~\GeVc, 5~\GeVc and
8~\GeVc incident beam momentum together with cosmic-ray data were used 
to provide an absolute calibration of the efficiency, momentum scale and
resolution of the detector. 
Moreover, tracks produced in runs with Pb and Ta targets in
the same period and with the same beam settings were used for
the calibration of the detector, verification of the event reconstruction and analysis
procedures (see Ref.~\cite{ref:harp:tantalum} for further details). 

The momentum of the T9 beam is known with a precision of
the order of 1\%~\cite{ref:t9}. 
The absolute normalization of the number of incident protons was
performed using 300~000, 240~000 and 280~000 `incident-proton' triggers
for the carbon, copper and tin data, respectively. 
These are triggers where the same selection on the beam particle was
applied but no selection on the interaction was performed.
The rate of this trigger was down-scaled by a factor 64.
A cross-check of the absolute normalization was provided by counting
tracks in the forward spectrometer.

A detailed description of the HARP apparatus is
given in Ref.~\cite{ref:harpTech}. 
In this analysis the detector components
of the large-angle spectrometer and the beam instrumentation are employed
and are briefly summarized in the following.

A set of four multi-wire
proportional chambers (MWPCs) measures the position and direction of
the incoming beam particles with an accuracy of $\approx$1~\mm in
position and $\approx$0.2~\mrad in angle per projection.
At low momenta the precision of the prediction at the target is
limited by multiple scattering.
A beam time-of-flight system (BTOF)
measures time difference of particles over a $21.4$~m path-length. 
It is made of two
identical scintillation hodoscopes, TOFA and TOFB (originally built
for the NA52 experiment~\cite{ref:NA52}),
which, together with a small target-defining 
%trigger counter 
scintillator
(TDS),
also used for the trigger, provide particle
identification at low energies. This provides separation of pions, kaons
and protons up to 5~\GeVc and determines the initial time at the
interaction vertex ($t_0$). 
The timing resolution of the combined BTOF system is about 70~\ps.
%
%\item 
A system of two N$_2$-filled Cherenkov detectors (BCA and BCB) is
used to tag electrons at low energies and pions at higher energies. 
The electron and pion tagging efficiency is found to be close to
100\%.
The proton fraction in the incoming beam varies from 35\% at
3 GeV/c to 92\% at 12 GeV/c. 
The length of the accelerator spill is 400~ms with a typical intensity
of 15~000 beam particles per spill.
The average number of events recorded by the data acquisition ranges
from 300 to 350 per spill for the four different beam momenta.

The target is placed inside the inner field cage (IFC) of the TPC such that,
in addition to particles produced in the forward direction, 
backward-going tracks can be measured.
All three targets have a nominal thickness of
5\%~$\lambda_{\mathrm{I}}$ and a cylindrical shape with a nominal
diameter of 30~\mm. 
The 99.95\% pure carbon target used for the 
measurement described here has a thickness of 18.94~mm with a variation
of $\pm 0.02$~mm.
Its density was measured to be 1.88~g/cm$^3$.
The copper target has a purity of 99.99\% with a thickness of 7.52~mm
with a variation of $\pm 0.01$~mm and a density of 8.92~g/cm$^3$.
The tin target has a purity of 99.99\% with a thickness of 11.04~mm
with a variation of $\pm 0.04$~mm and a density of 7.29~g/cm$^3$.
A set of trigger detectors completes the beam instrumentation: a
thin scintillator slab covering the full aperture of the last
quadrupole magnet in the beam line to start the trigger logic
decision (BS); a small scintillator disk, TDS, positioned
upstream of the target to ensure that only 
particles hitting the target cause a trigger; and `halo' counters
(scintillators with a hole to let the beam particles pass) to veto
particles too far away from the beam axis. 
A cylindrical detector (inner trigger cylinder, ITC) made of six
layers of 1~\mm thick scintillating fibres is positioned inside the
inner field cage of the TPC and surrounds the target.
It provides full coverage of the acceptance of the TPC.
The large-angle spectrometer consists of a TPC and a set of RPC 
detectors inside the solenoidal magnet. 
The TPC detector was designed to measure and identify tracks in the
angular region from 0.25~\rad to 2.5~\rad from the beam axis.
%The momentum of produced particles is obtained from the curvature of
%their trajectories in the magnetic field.
%The emission angle is given by the direction of the trajectory near
%the interaction point.
Charged particle identification (PID) can be achieved by measuring the 
ionization per unit length in the gas (\dedx) as a function of the total
momentum of the particle. 
%This  measurement is obtained with the TPC~\cite{ref:tpc:ieee}. 
Additional PID can be performed through a time-of-flight 
measurement with the RPCs.
%~\cite{ref:rpc,ref:barr:rpc,ref:ieee:rpc}.

In the present analysis, the TPC provides the measurement for the
pattern recognition to find the particle tracks, and to measure their
momentum through the curvature of their trajectory. 
It also provides PID using the measurement of energy deposition.
The RPC system is used in this analysis to provide a calibration of
the PID capabilities of the TPC.
%The performance of the time-of-flight measurements with the RPCs is
%reported in ~\cite{ref:ieee:rpc}.

%====================================================================================
%summarize!
%\subsubsection{Operation and calibration of the TPC}
%\label{sec:tpc-operation}

In addition to the usual need for calibration of the detector, a number of
%difficulties 
hardware shortfalls, discovered mainly after the end of data-taking,
had to be overcome to use the TPC data reliably in the analysis.
The TPC is affected by a relatively large number of dead or noisy 
pads and static and dynamic distortions of the reconstructed trajectories.
Static distortions are caused by the inhomogeneity of the electric field,
due to an accidental mismatch between the inner and outer field cage 
(powered by two distinct HV supplies) and 
other sources.
%a partial "transparency" of the cathode wire grid. 
Dynamic distortions are caused instead by the
build-up of ion-charge density in the drift volume during the 400~ms 
long beam spill. 
All these effects were fully studied and available corrections are described 
in detail in Ref.~\cite{ref:harp:tantalum}. While methods to correct the dynamic 
distortions of the TPC tracks are being implemented, a pragmatic approach has been
followed in the present analysis. Only the events corresponding to the
early part of the spill, where the effects of the dynamic distortions are
still small, are used~\footnote{this translates into a cut on the maximum 
number of events ($N_{evt}$) to be retained}. The time interval between spills 
is large enough to drain all charges in the TPC related to the effect of the beam.
The combined effect of the distortions on the kinematic quantities used
in the analysis has been studied in detail and only that part of the data
for which the systematic errors can be assessed with physical 
benchmarks was used, as explained in \cite{ref:harp:tantalum}. More than 40\% of the
recorded data can be used on average in the current analysis.

The absolute scale of the momentum determination is determined using
elastic scattering data off a hydrogen target.
The angle of the forward scattered particle (pion or proton) is used to
give an absolute prediction for the momentum of the recoil proton.
This prediction is compared with the measurement in the TPC.
To study the stability of this measurement protons are selected in a
narrow band with a relatively large \dedx where the \dedx depends
strongly on momentum.  
The average momentum for the protons selected in this band remains
stable within 3\% as a function of time--in--spill over the part of the
spill used for the analysis. 

\section{Data selection and analysis}
\label{sec:selection}

The beam of positive particles used for this measurement contains mainly 
positrons, pions and protons, with small components of kaons,
deuterons and heavier ions.
Its composition depends on the selected beam momentum.
The analysis proceeds by first selecting a beam proton hitting the
target, not accompanied by other tracks. 
Then an event is required to be triggered by the ITC in order to be
retained. 
After the event selection the sample of tracks to be used for analysis
is defined.
Tracks are only considered if they contain at least twelve space
points out of a maximum of twenty.
This cut is applied to ensure a good measurement of the track
parameters and of the \dedx.
Furthermore, a quality requirement is applied on the fit to the
helix. 
The latter requirement introduces a very small loss of efficiency.
For tracks satisfying these conditions, a cut is made on \dzeroprime,
the distance 
of closest approach to the extrapolated trajectory of the incoming
beam particle in the plane perpendicular to the beam direction and
\zzeroprime, the 
$z$-coordinate where the distance of the secondary track and the beam
track is minimal.  
Finally, only tracks with momentum in the range between 100~\MeVc and
800~\MeVc are accepted. 
In addition, particles with transverse momentum below 55~\MeVc are
removed. 

Table~\ref{tab:events} shows the number of events and tracks at
various stages of the selection.
The total number of events taken by the data acquisition (``Total
 DAQ events'') includes triggers of all types as well as
calibration events; the number of ``Protons on target'' represents the
count of the incoming beam trigger after off-line selection of accepted
protons multiplied by the down-scale factor 64.
The number of accepted events for this analysis  (``Accepted protons
with LAI (Large Angle Interaction)'') is obtained using the same selection of
incoming protons in coincidence with a trigger in the ITC.
The large difference between the 
%first and seventh 
rows ``Total DAQ events'' and ``Accepted protons with LAI'' 
is due to the relatively large fraction of pions in the
beam and to the larger number of triggers taken for the measurements
with the forward spectrometer.
These data will be the subject of other publications. 
The line ``Maximum \evtspill'' refers to the 
%first 
last
number of
events \evtspill in spill used to avoid dynamic distortion corrections, with the
corresponding number of interaction triggers used in the analysis (``LAI
in accepted spill part'') and the fraction of the data used given under
``Fraction of triggers used''.
The lines ``Accepted momentum determination'' and ``In kinematic region
and originating from target'' give the number of {\em tracks} passing
the momentum fit quality requirements and the selection of tracks
originating in the target region.
Finally, the rows ``Negative particles'', ``Positive particles '',
``$\bfpim$ selected with PID'' and ``$\bfpip$ selected with PID'' show
the number of accepted tracks with negative and positive charge and the ones
passing in addition the pion PID criteria, respectively.

To give an impression of the complexity of the events, one can define an
`average multiplicity' as the ratio of the number of tracks with at
least twelve hits in the TPC (regardless of their momentum, angle or
spatial position) and the number of events accepted by the selection
criteria with at least one such track. 
With this definition, the average multiplicity is 2.2, 2.6, 3.1 and 3.4
for the 3~\GeVc, 5~\GeVc, 8~\GeVc and 12~\GeVc beams in p--C data, respectively.
%%% also Cu Sn
 
\input{C5-Cu5-Sn5_table1}

The double-differential cross-section for the production of a particle of 
type $\alpha$ can be expressed in the laboratory system as:

\begin{equation}
{\frac{{\mathrm{d}^2 \sigma_{\alpha}}}{{\mathrm{d}p_i \mathrm{d}\theta_j }}} =
\frac{1}{{N_{\mathrm{pot}} }}\frac{A}{{N_A \rho t}}
 \sum_{i',j',\alpha'} M_{ij\alpha i'j' \alpha'}^{-1} \cdot
{N_{i'j'}^{\alpha'} } 
\ ,
\label{eq:cross}
\end{equation}

where $\frac{{\mathrm{d}^2 \sigma_{\alpha}}}{{\mathrm{d}p_i \mathrm{d}\theta_j }}$
is expressed in bins of true momentum ($p_i$), angle ($\theta_j$) and
particle type ($\alpha$).
%% The terms on the right-hand side of the equation are as follows.

%\begin{itemize}

The factor  $\frac{A}{{N_A \rho t}}$ 
is the inverse of the number of target nuclei per unit area
($A$ is the atomic mass,
$N_A$ is the Avogadro number, $\rho$ and $t$ are the target density
and thickness)\footnote{We do not make a correction for the attenuation
of the proton beam in the target, so that strictly speaking the
cross-sections are valid for a $\lambda_{\mathrm{I}}=5\%$ target.}. 
The result is normalized to the number of incident protons on target
$N_{\mathrm{pot}}$. 

%\item 
The `raw yield' $N_{i'j'}^{\alpha'}$ 
is the number of particles of observed type $\alpha'$ in bins of reconstructed
momentum ($p_{i'}$) and  angle ($\theta_{j'}$). 
These particles must satisfy the event, track and PID 
selection criteria.
Although, owing to the stringent PID selection,  the background from
misidentified protons in the pion sample is small, the pion and proton
raw yields ($N_{i'j'}^{\alpha'}$, for 
$\alpha'=\pim, \pip, \mathrm{p}$) have been measured simultaneously. 
This makes it possible to correct for the small remaining proton
background in the pion data without prior assumptions concerning the
proton production cross-section.
%Figure~\ref{fig:raw-yield} shows the $p$ and \tht distribution for the
%momentum bins and angular bins chosen in the analysis.

The matrix $ M_{ij\alpha i'j' \alpha'}^{-1}$ 
corrects for the  efficiency and resolution of the detector. 
It unfolds the true variables $ij\alpha$ from the reconstructed
variables $i'j'\alpha'$  with a Bayesian technique~\cite{dagostini} 
and corrects  
the observed number of particles to take into account effects such as 
trigger efficiency, reconstruction efficiency, acceptance, absorption,
pion decay, tertiary production, 
PID efficiency, PID misidentification and electron background. 
The method used to correct for the various effects is  described in
more detail in Ref.~\cite{ref:harp:tantalum}.

In order to predict the population of the migration matrix element 
$M_{ij\alpha i'j'\alpha'}$, the resolution, efficiency
and acceptance of the detector are obtained from the Monte Carlo.
This is accurate provided the Monte Carlo
simulation describes these quantities correctly. 
Where some deviations
from the control samples measured from the data are found, 
the data are used to introduce (small) {\em ad hoc} corrections to the
Monte Carlo. 

Using the unfolding approach, possible known biases in the measurements
are taken into account automatically as long as they are described by
the Monte Carlo.
For example the energy-loss of particles inside the target and
material around the inner field cage is expressed as an average shift
of the measured momentum distribution compared to the physical
momentum. 
Known biases are therefore treated in the same way as resolution
effects. 
In the experiment simulation, which is based on the GEANT4
toolkit~\cite{ref:geant4}, the materials in the beam-line and the 
detector are accurately described as well as
the relevant features of the detector response and 
the digitization process.
%The experiment simulation is based on the GEANT4
%toolkit~\cite{ref:geant4}.
%The materials in the beam line and the detector are accurately
%reproduced in this simulation, as well as the relevant features of the
%detector response and the digitization process. 
The Monte Carlo
simulation compares well with data, as shown in Ref.~\cite{ref:harp:tantalum}.

The absolute normalization of the result is calculated in first
instance relative to the number of incident beam particles accepted by
the selection. 
After unfolding, the factor  $\frac{A}{{N_A \rho t}}$ is applied.
The beam normalization using down-scaled incident-proton triggers 
has uncertainties smaller than 2\%
  for all beam momentum settings.

The background due to interactions of the primary
protons outside the target (called `Empty target background') is
measured using data taken without the target mounted in the target
holder.
Owing to the selection criteria which only accept events from the
target region and the good definition of the interaction point this
background is negligible\footnote{The background of
interactions of the primary proton outside the target can be suppressed
for large angle tracks measured in the TPC owing to the good resolution
in $z$.  This is contrary to the situation in the forward spectrometer
where an interaction in the target cannot be distinguished from an
interaction in upstream or downstream
material~\cite{ref:alPaper,ref:bePaper}.} ($< 10^{-5}$).

The effects of these uncertainties on the final results are estimated
by repeating the analysis with the relevant input modified within the
estimated uncertainty intervals.
In many cases this procedure requires the construction of a set of
different migration matrices.
The correlations of the variations between the cross-section bins are
evaluated and expressed in the covariance matrix.
Each systematic error source is represented by its own covariance
matrix.
The sum of these matrices describes the total systematic error.

\section{Results}
\label{sec:results}

\begin{figure}[tbp]
\begin{center}
\epsfig{figure=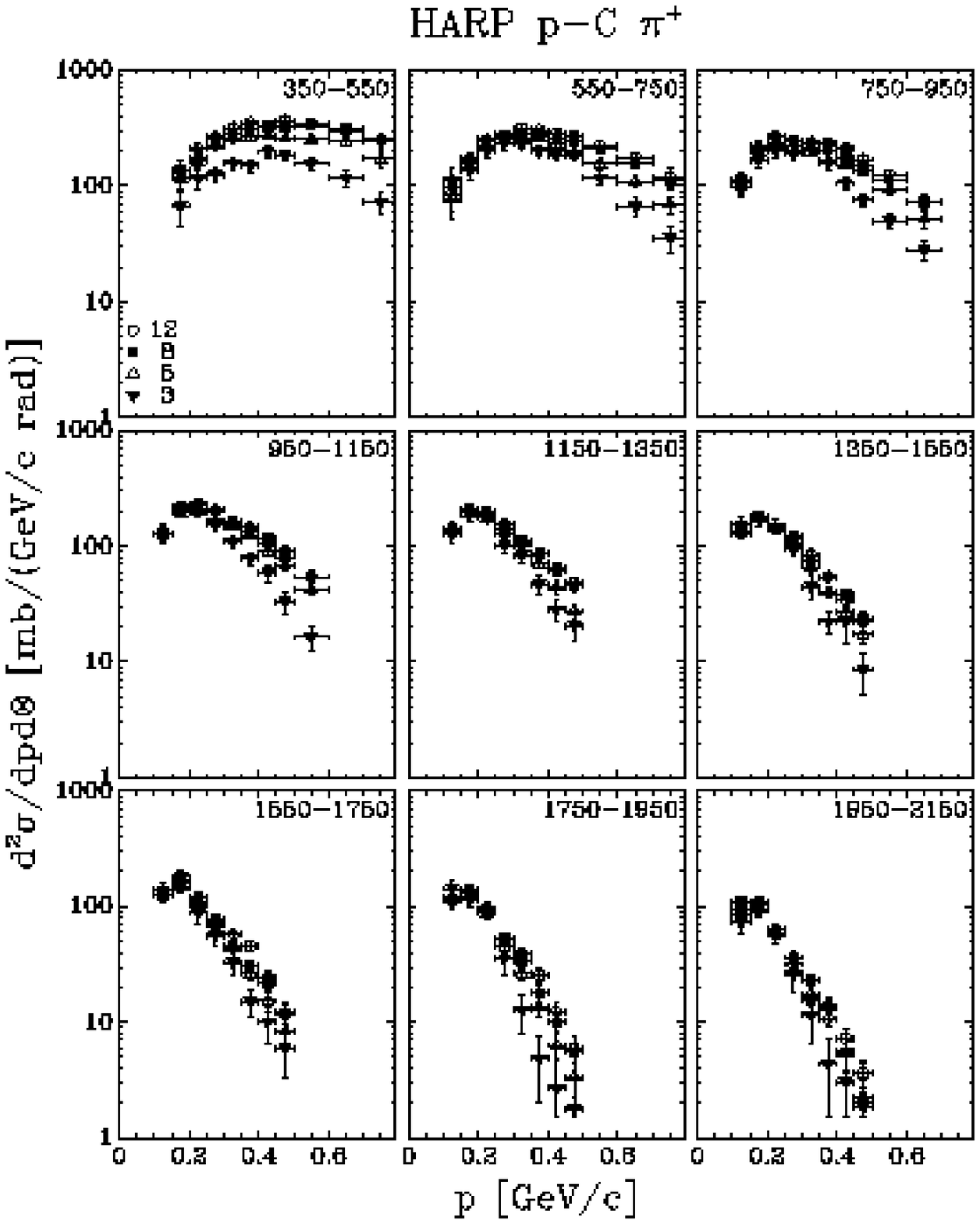,width=0.49\textwidth}
\epsfig{figure=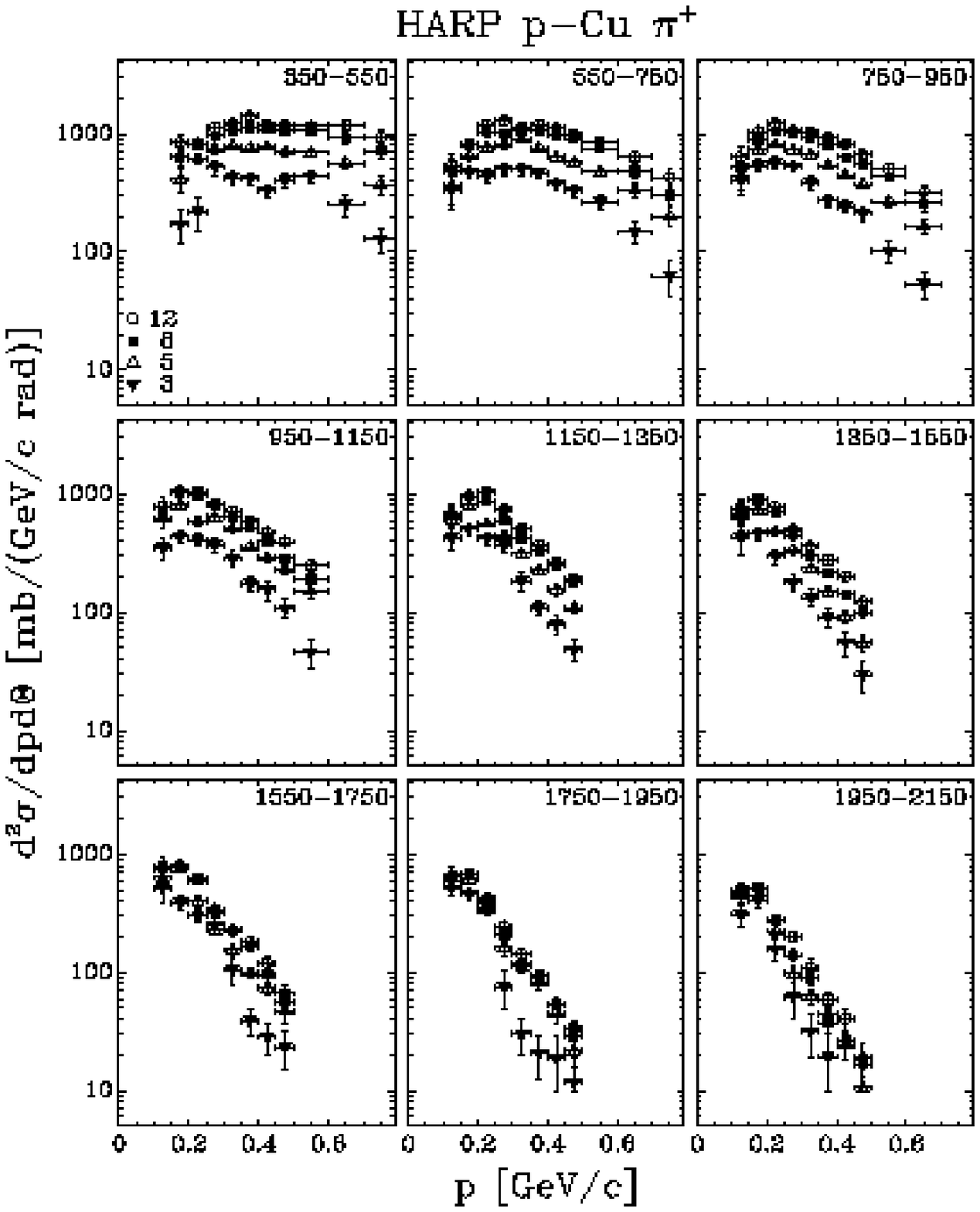,width=0.49\textwidth}
\epsfig{figure=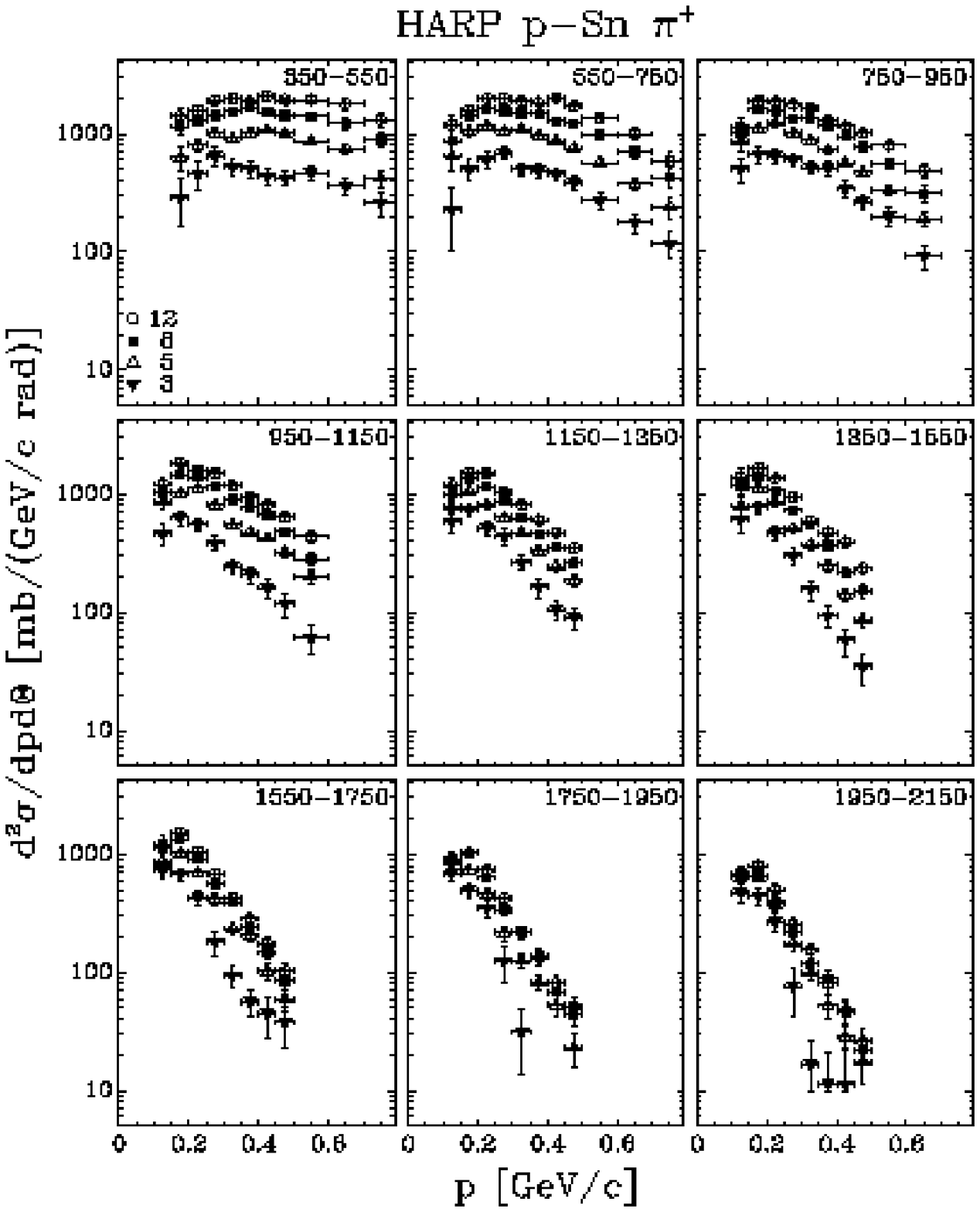,width=0.49\textwidth}
\caption{
Double-differential cross-sections for \pip production in
p--C, p--Cu and p--Sn interactions as a function of momentum displayed in different
angular bins (shown in \mrad in the panels).
The results are given for four incident beam momenta (filled triangles:
3~\GeVc; open triangles: 5~\GeVc; filled rectangles: 8~\GeVc; open
circles: 12~\GeVc).
The error bars represent the combination of statistical and systematic
 uncertainties.
}
\label{fig:xs-p-th-pbeam-plus}
\end{center}
\end{figure}

\begin{figure}[tbp]
\begin{center}
\epsfig{figure=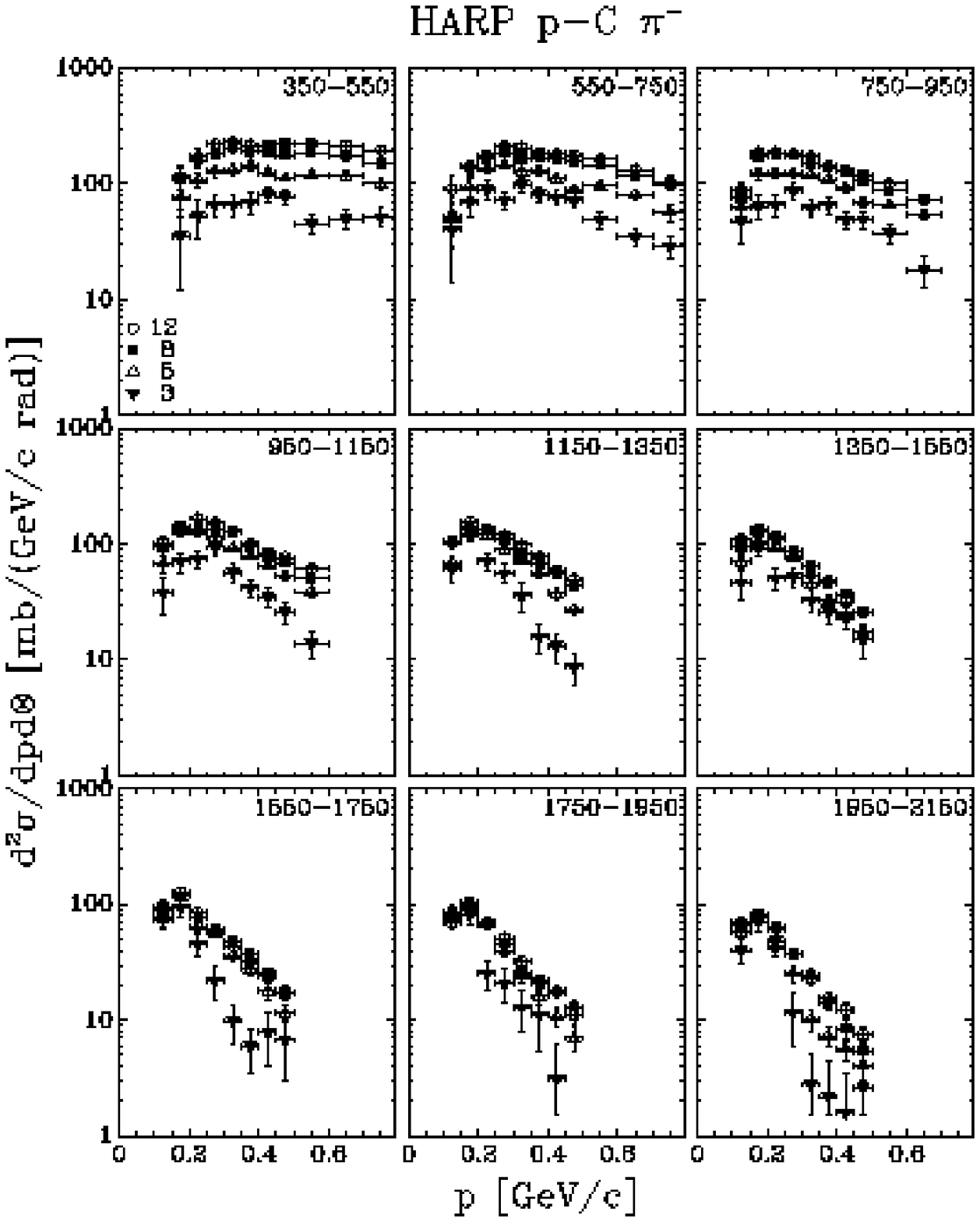,width=0.49\textwidth}
\epsfig{figure=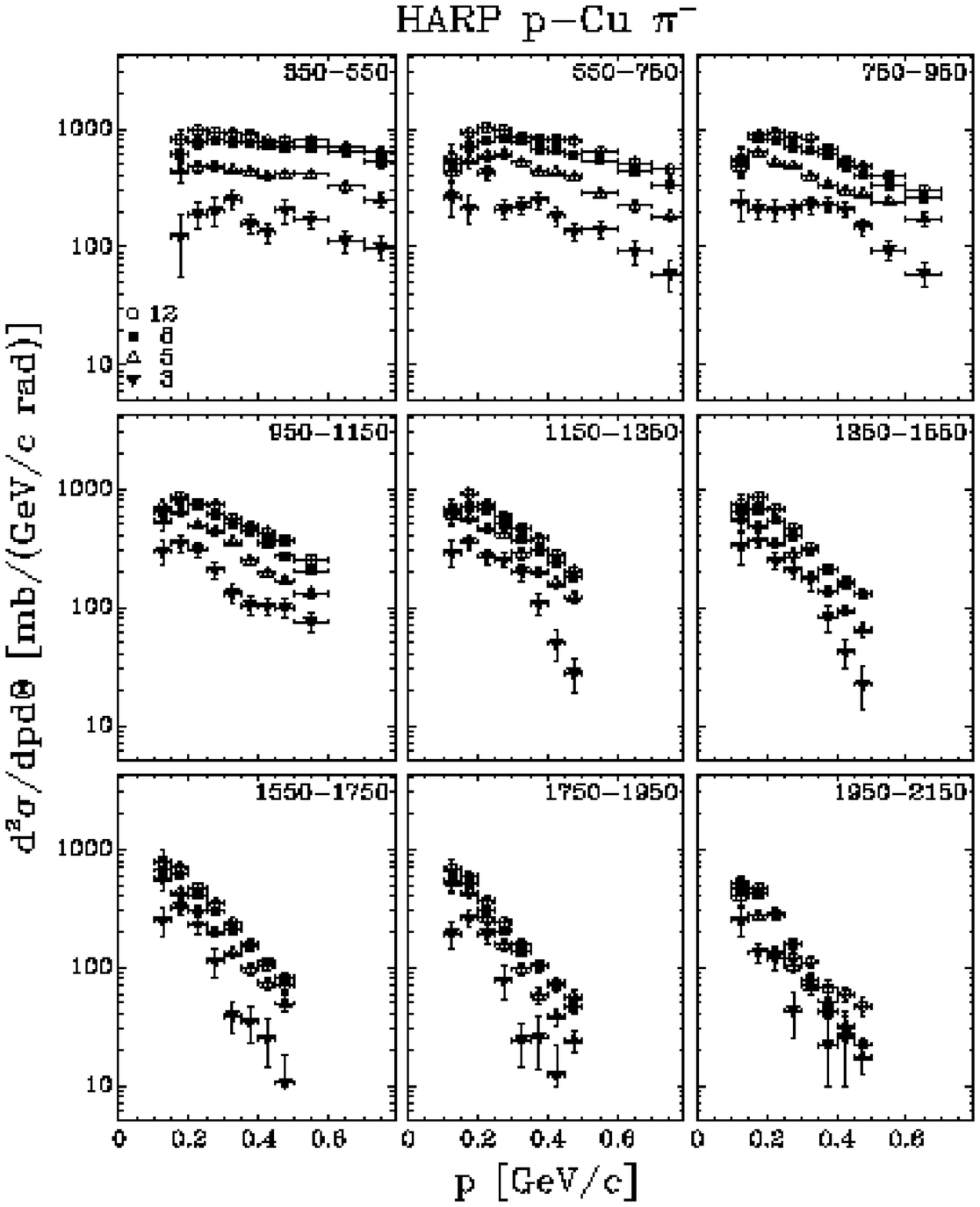,width=0.49\textwidth}
\epsfig{figure=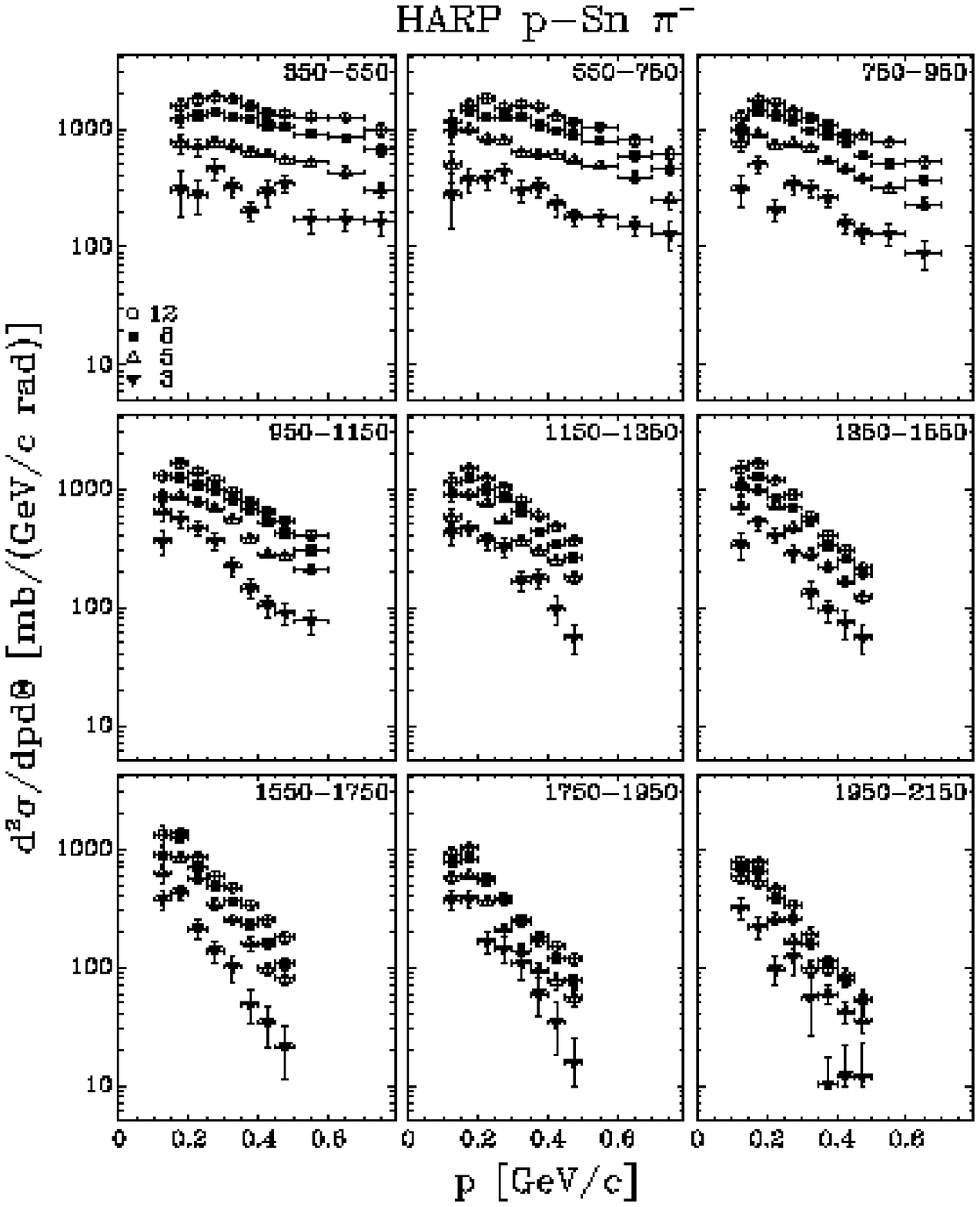,width=0.49\textwidth}
\caption{
Double-differential cross-sections for \pim production in
p--C, p--Cu  and p--Sn  interactions as a function of momentum displayed in different
angular bins (shown in \mrad in the panels).
The results are given for four incident beam momenta (filled triangles:
3~\GeVc; open triangles: 5~\GeVc; filled rectangles: 8~\GeVc; open
circles: 12~\GeVc).
The error bars represent the combination of statistical and systematic
 uncertainties.
}
\label{fig:xs-p-th-pbeam-minus}
\end{center}
\end{figure}

The measured double-differential cross-sections for the 
production of \pip and \pim in the laboratory system as a function of
the momentum and the polar angle for each incident beam momentum are
shown in Fig.~~\ref{fig:xs-p-th-pbeam-plus} and
\ref{fig:xs-p-th-pbeam-minus}, respectively.
The error bars  shown are the
square-roots of the diagonal elements in the covariance matrix,
where statistical and systematic uncertainties are combined
in quadrature.
Correlations cannot be shown in the figures.
The correlation of the statistical errors (introduced by the unfolding
procedure) are typically smaller than 20\% for adjacent momentum bins and
smaller for adjacent angular bins.
The correlations of the systematic errors are larger, typically 80\% for
adjacent bins.
Tables with the results of this analysis are also given in Appendix A.
A discussion of the error evaluation is given below. 
The overall scale error ($< 2\%$) is not shown.
The measurements for the different beam momenta are overlaid in the
same figure.
For the 3~\GeVc data the point--to--point
statistical error is larger than the systematic error, except for the
lowest secondary momentum bin. 
Especially in the middle of the range (around 400~\MeVc), the systematic
error is small.
Thus the fluctuations between the points are expected to be of
statistical nature.
In the first angular bins the momentum resolution is relatively large
compared to the bin size such that the unfolding procedure tends to
display statistical fluctuations over two bins.
Since the treatment of the data sets taken with different beam momenta
is identical, structures visible in the spectra at 3~\GeVc and not
visible in the other data sets are not likely to be artefacts of the
efficiency corrections.
Overall trends in the shapes, {\em i.e.} structures extending over more
than two bins are, however, to be considered significant.

\begin{figure}[tbp]
  \epsfig{figure=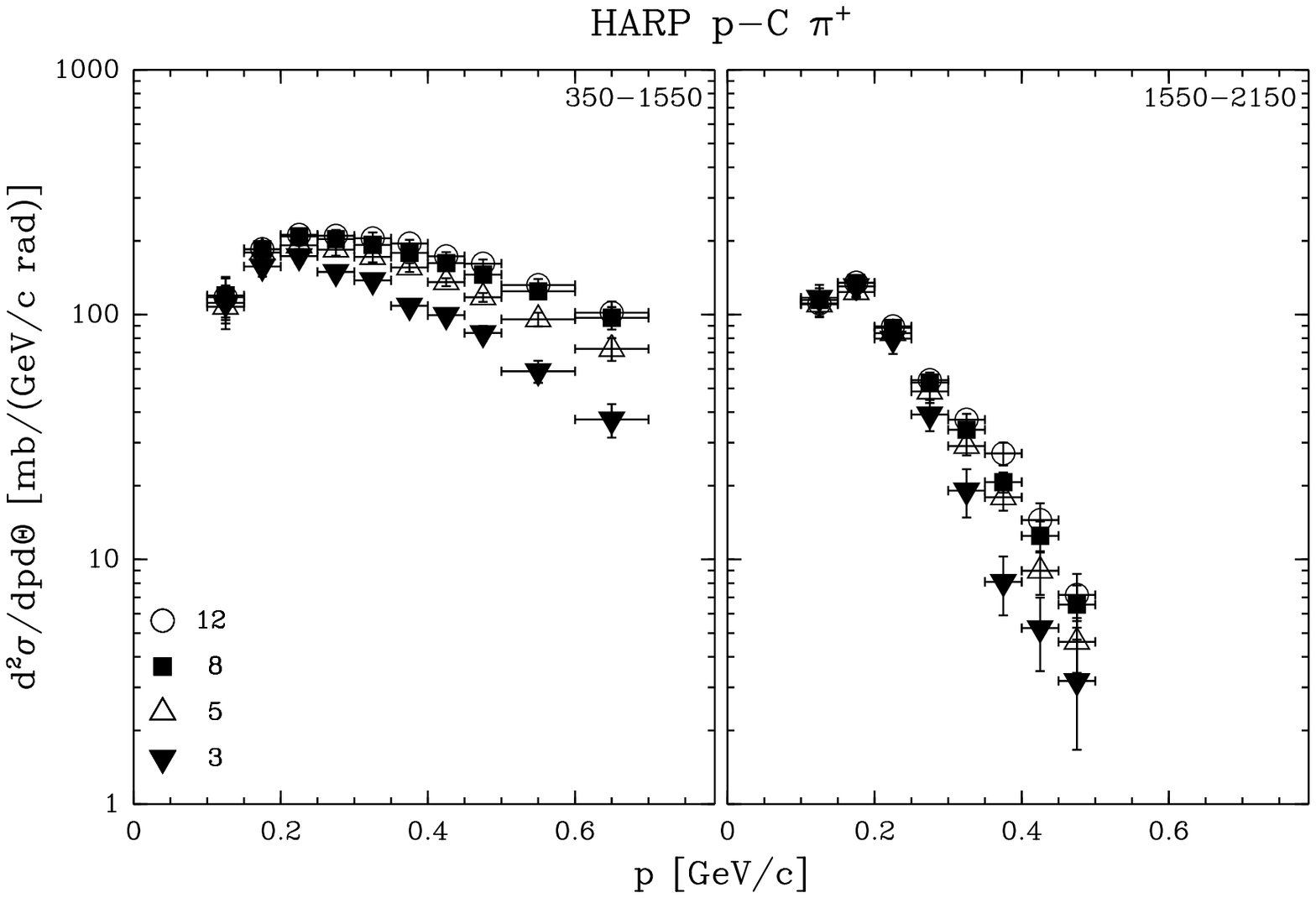,width=0.49\textwidth}
  \epsfig{figure=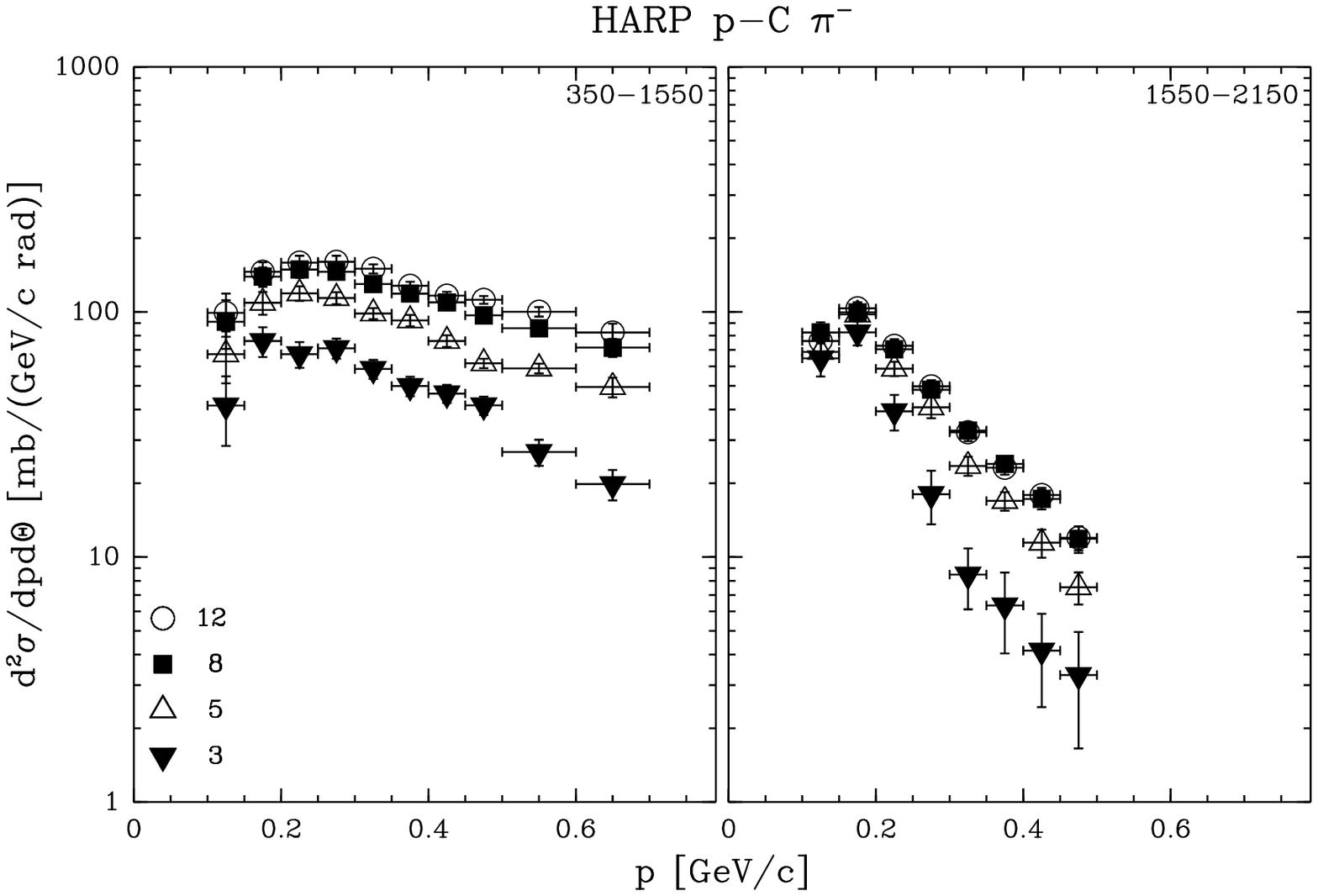,width=0.49\textwidth}
  \epsfig{figure=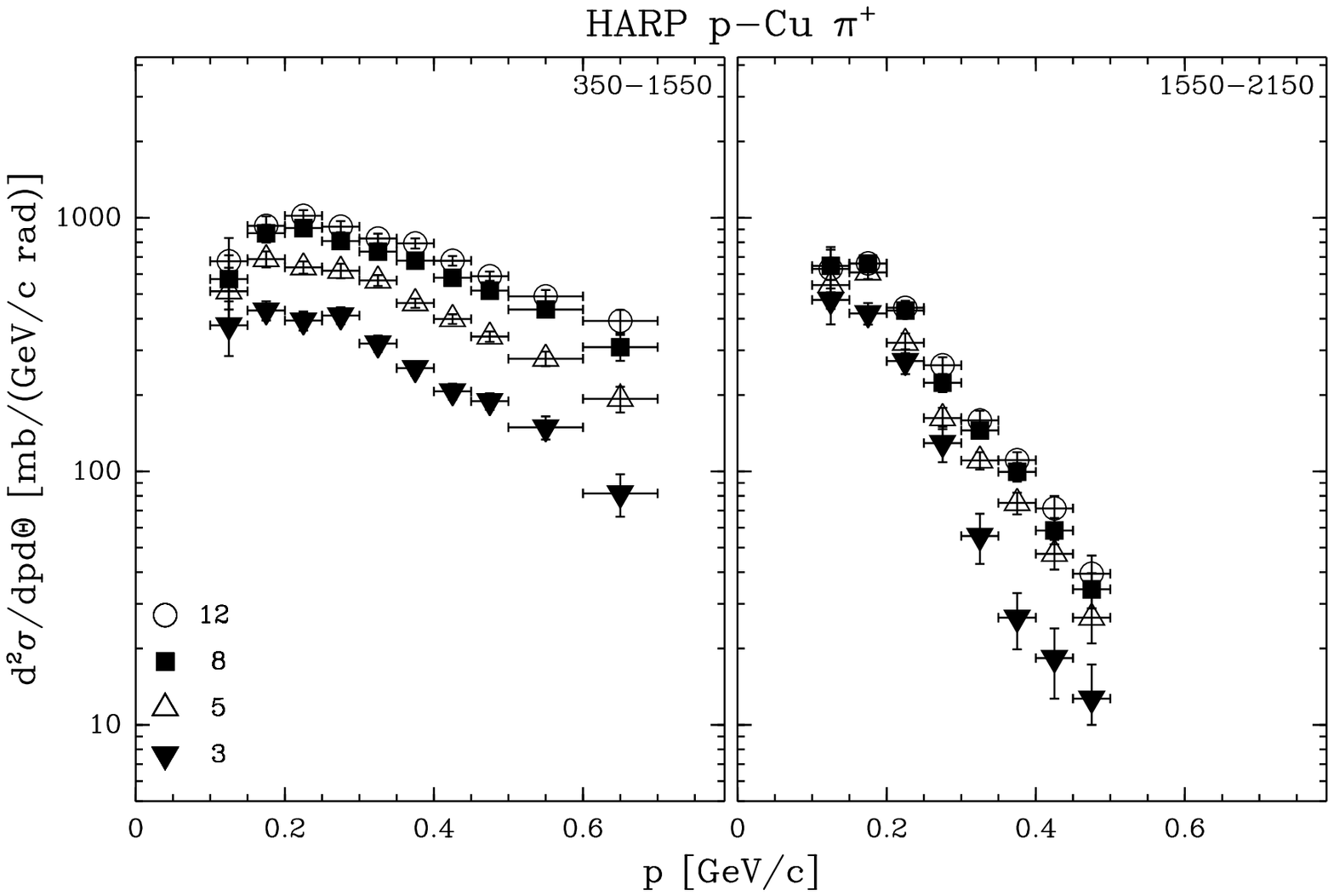,width=0.49\textwidth}
  \epsfig{figure=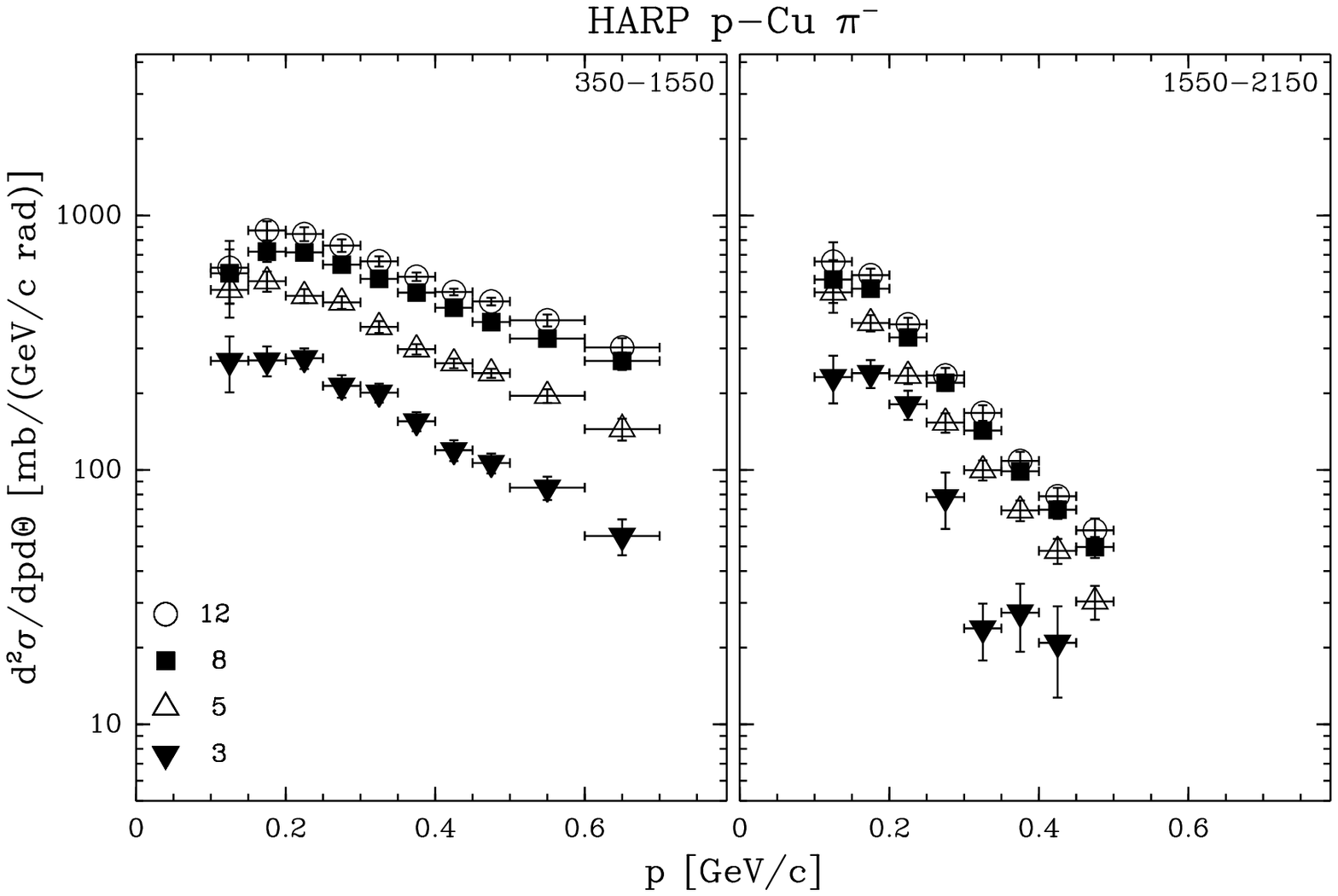,width=0.49\textwidth}
  \epsfig{figure=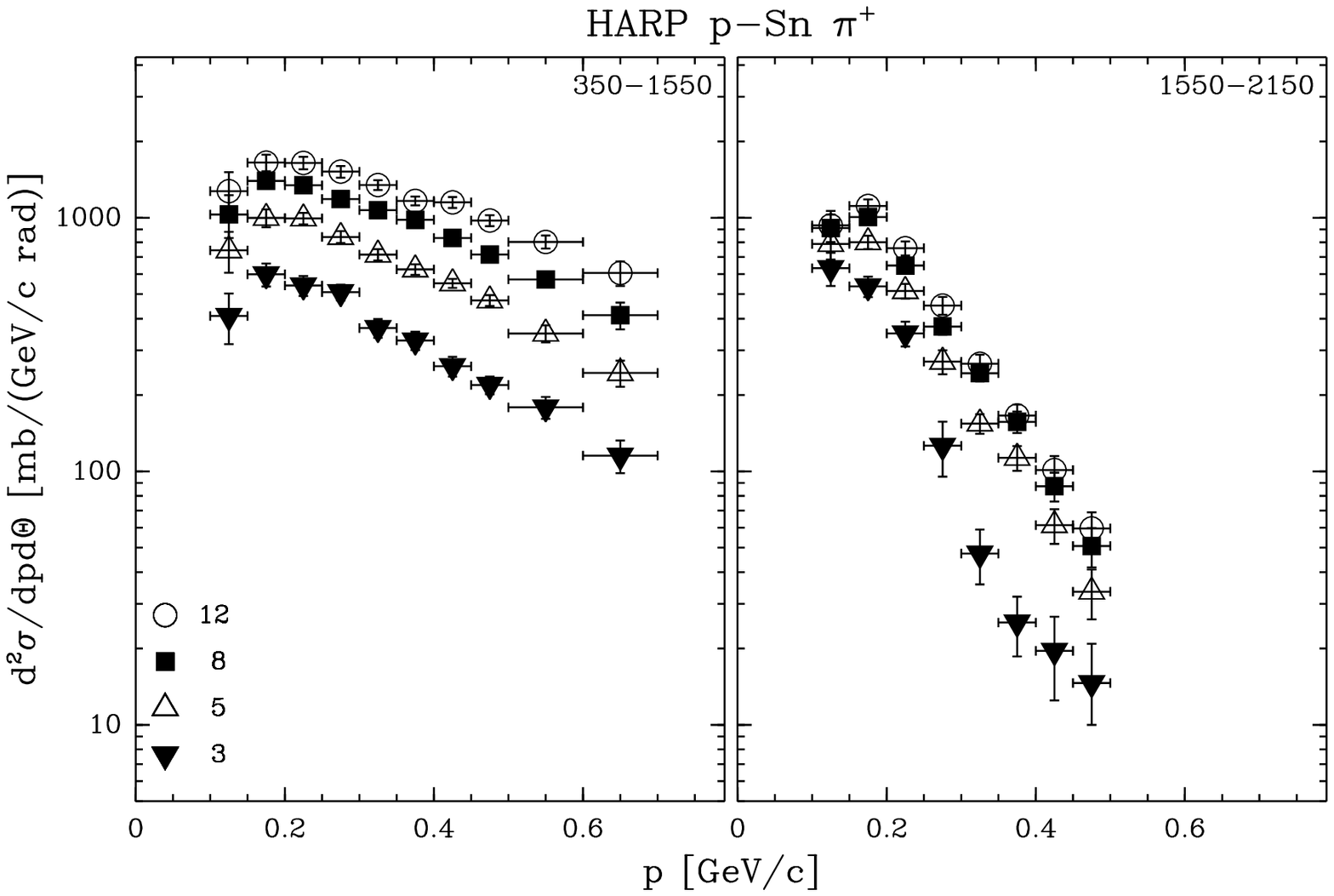,width=0.49\textwidth}
  \epsfig{figure=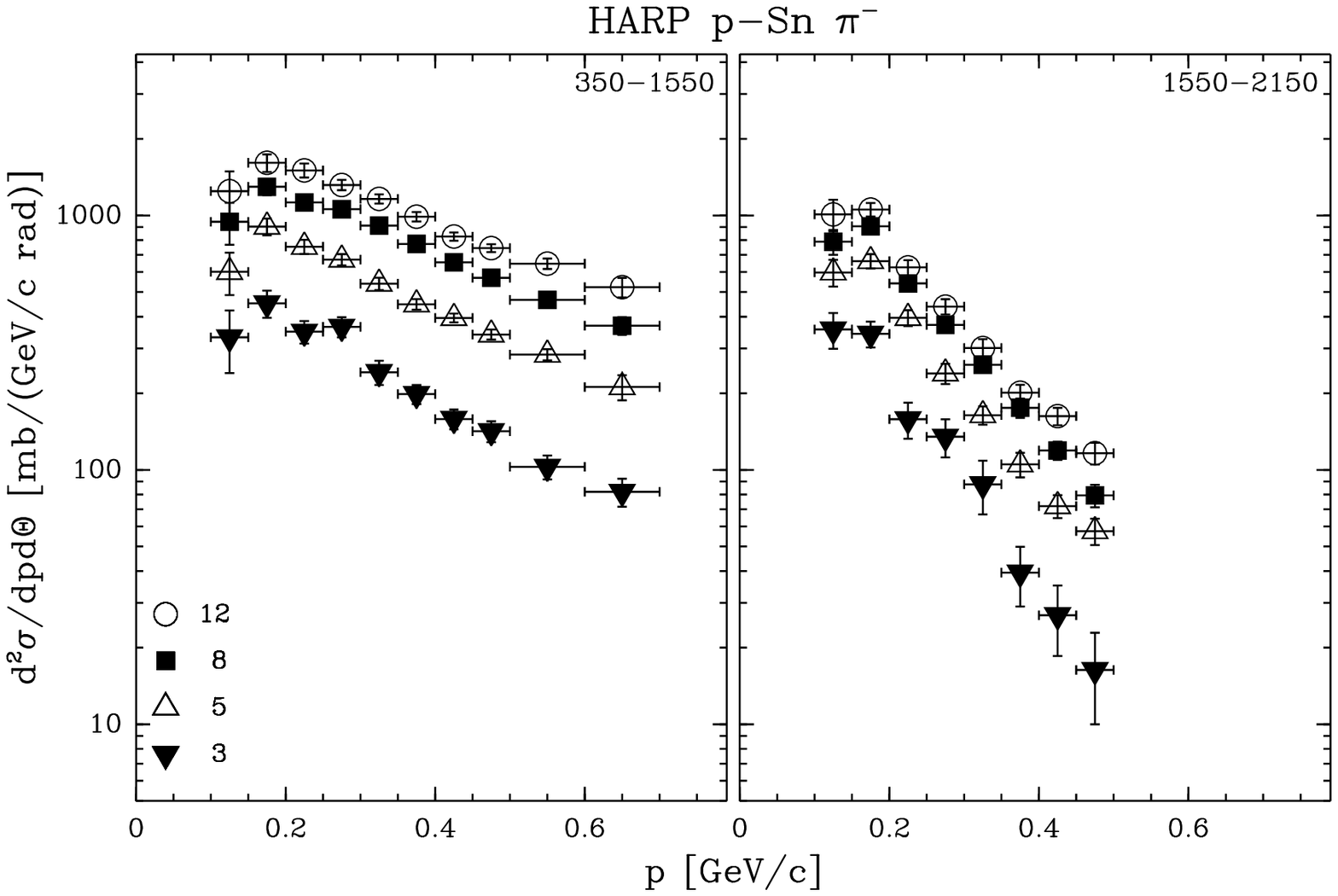,width=0.49\textwidth}
\caption{
Double-differential cross-sections for \pip  and \pim production in
p--C, p--Cu  and p--Sn  interactions as a function of momentum averaged over the
angular region covered by this experiment (shown in mrad).
The left panel of each pair shows forward production
(350~\mrad  $\le \theta <$ 1550~\mrad), while
the right panel of each pair shows backward production
(1550~\mrad  $\le \theta <$ 2150~\mrad).
The results are given for four incident beam momenta (filled triangles:
3~\GeVc; open triangles: 5~\GeVc; filled rectangles: 8~\GeVc; open
circles: 12~\GeVc).
The error bars obtained after summing the bins of the
 double-differential cross-sections take into account the correlations
 of the statistical and systematic uncertainties.
}
\label{fig:xs-p-pbeam}
\end{figure}

To better visualize the dependence on the incoming beam momentum, the
same data averaged over the angular range (separately for the
forward going and backward going tracks) covered by the analysis are shown
separately for \pip and \pim in Fig.~\ref{fig:xs-p-pbeam}.
The spectrum of pions produced in the backward direction is much
steeper than that in the forward direction.

The increase of the pion yield per proton is visible in 
addition to a change of spectrum towards higher momentum of the
secondaries produced by higher momentum beams in the forward
direction. 

\begin{figure}[tbp]
\begin{center}
  \epsfig{figure=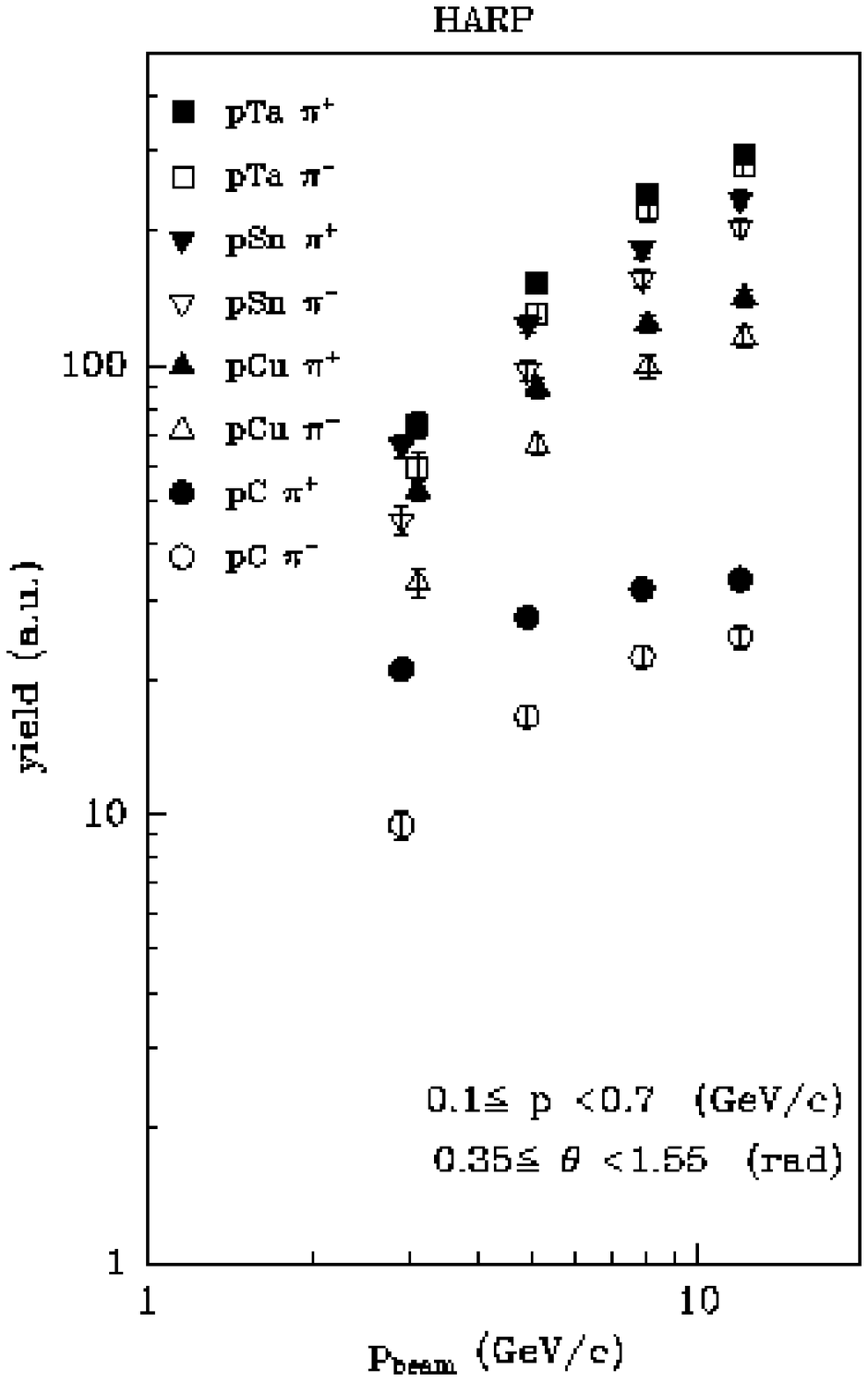,width=
0.425\textwidth}
 ~
  \epsfig{figure=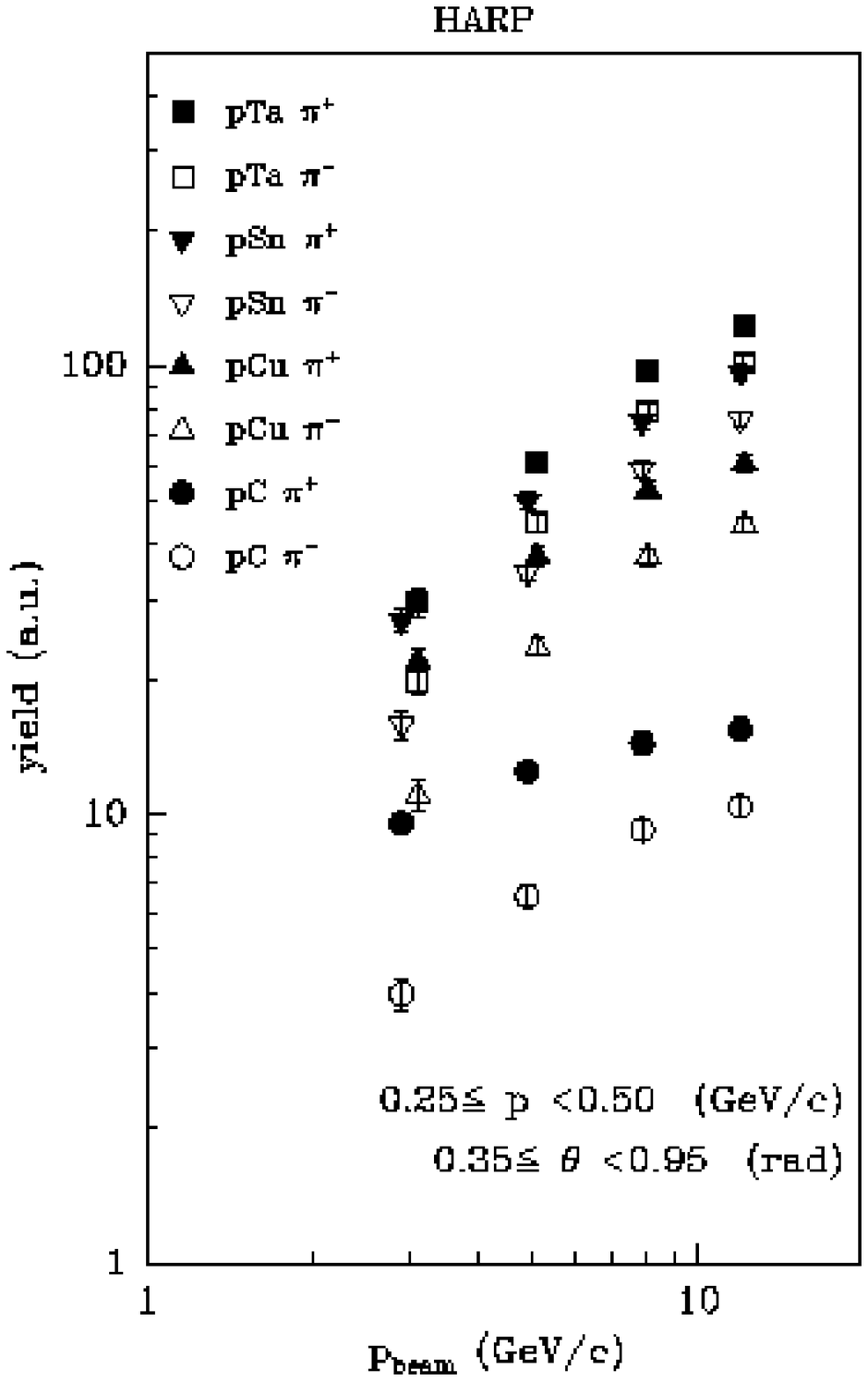,width=0.425\textwidth}
\end{center}
\caption{
Left: The dependence on the beam momentum of the pion production yields
 in p--C, p--Cu, p--Sn, p--Ta interactions
 interactions integrated over the forward angular region
 ($0.350~\rad \leq \theta < 1.550~\rad$) and momentum ($100~\MeVc \leq p < 700~\MeVc$).
 Right: The dependence on the beam momentum of the pion production yields
 integrated over the region
 ($0.350~\rad \leq \theta < 0.950~\rad$ and $250~\MeVc \leq p < 500~\MeVc$)
 with the same meaning of the symbols.
 The results are given in arbitrary units, with a consistent scale
 between the left and right panel.
Although the units are indicated as ``arbitrary'',
for the largest region (left panel), the yield is expressed as
${{\mathrm{d}^2 \sigma}}/{{\mathrm{d}p\mathrm{d}\Omega }}$ in
mb/(\GeVc~sr).
For the smaller region (left panel) the same normalization is chosen, but now scaled with the
relative bin size to show visually the correct ratio of number of pions
produced in this kinematical region with respect to the yield in the
 larger kinematical region.
Data points for different target nuclei and equal momenta are slightly
 shifted horizontally with respect to each other to increase the visibility.
}
\label{fig:xs-trend}
\end{figure}

The dependence of the integrated pion yields on the incident beam
momentum is shown in Fig.~\ref{fig:xs-trend} and compared with the p--Ta
data taken with the same apparatus (Ref.~\cite{ref:harp:tantalum}). 
The \pip and \pim yields integrated over the region 
$0.350~\rad \le \theta < 1.550~\rad$ and $100~\MeVc \le p < 700~\MeVc$ are
shown in the left panel and the data integrated over the region
$0.350~\rad \le \theta < 0.950~\rad$ and $250~\MeVc \le p < 500~\MeVc$ in
the right panel.
The beam energy dependence of the yields is clearly different in the
p--C data compared to the p--Ta data.
The dependence in the p--C data is much more flat with a saturation of
the yield between 8~\GeVc and 12~\GeVc (in both integration regions).
The \pip and \pim production yields exhibit a different behaviour.
The p--Cu and p--Sn data are more similar to the p--Ta data than the
p--C data indicating a smooth transition between light and heavy target
nuclei. 

\begin{figure}[tbp]
\begin{center}
  \epsfig{figure=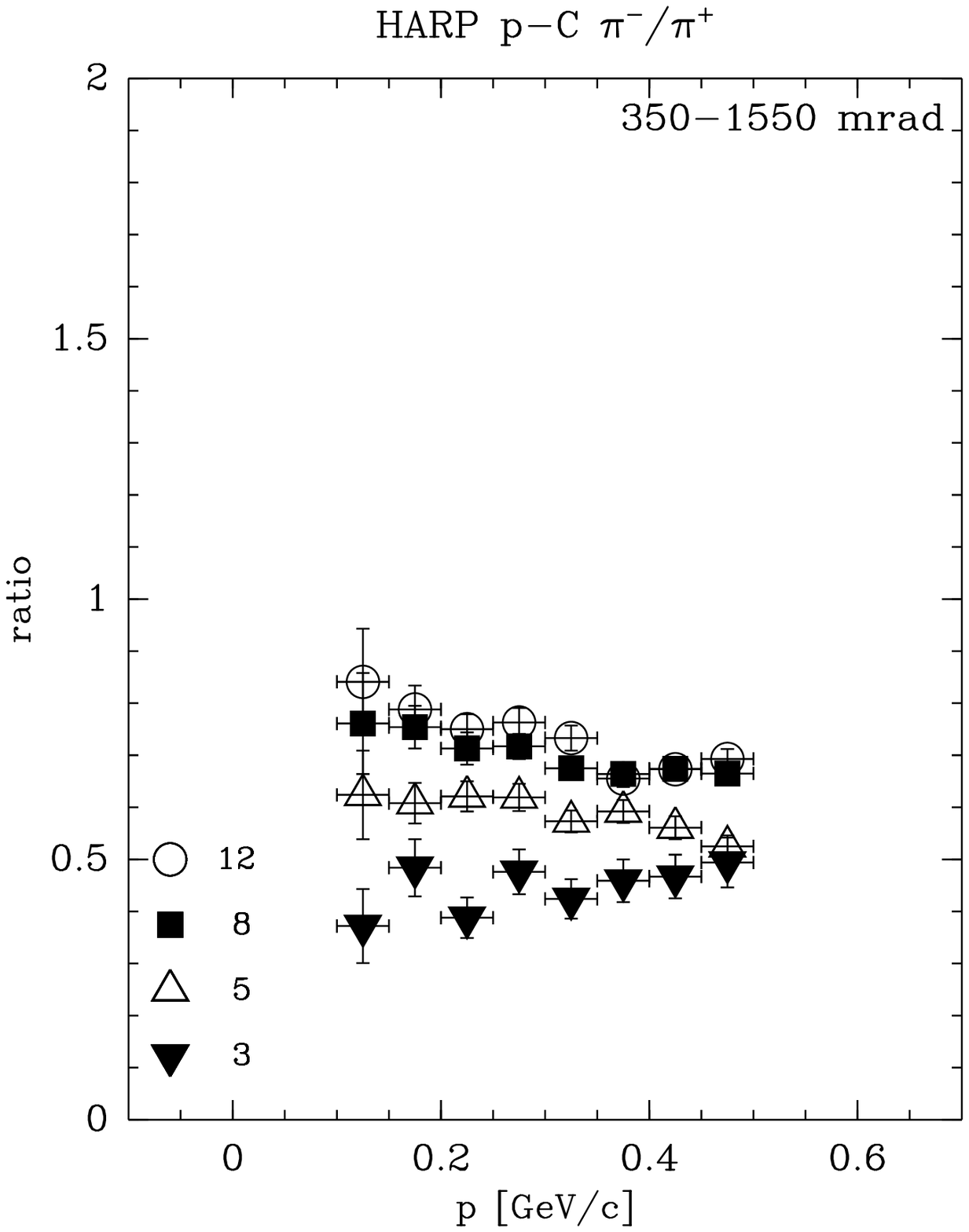,width=0.425\textwidth}
  \epsfig{figure=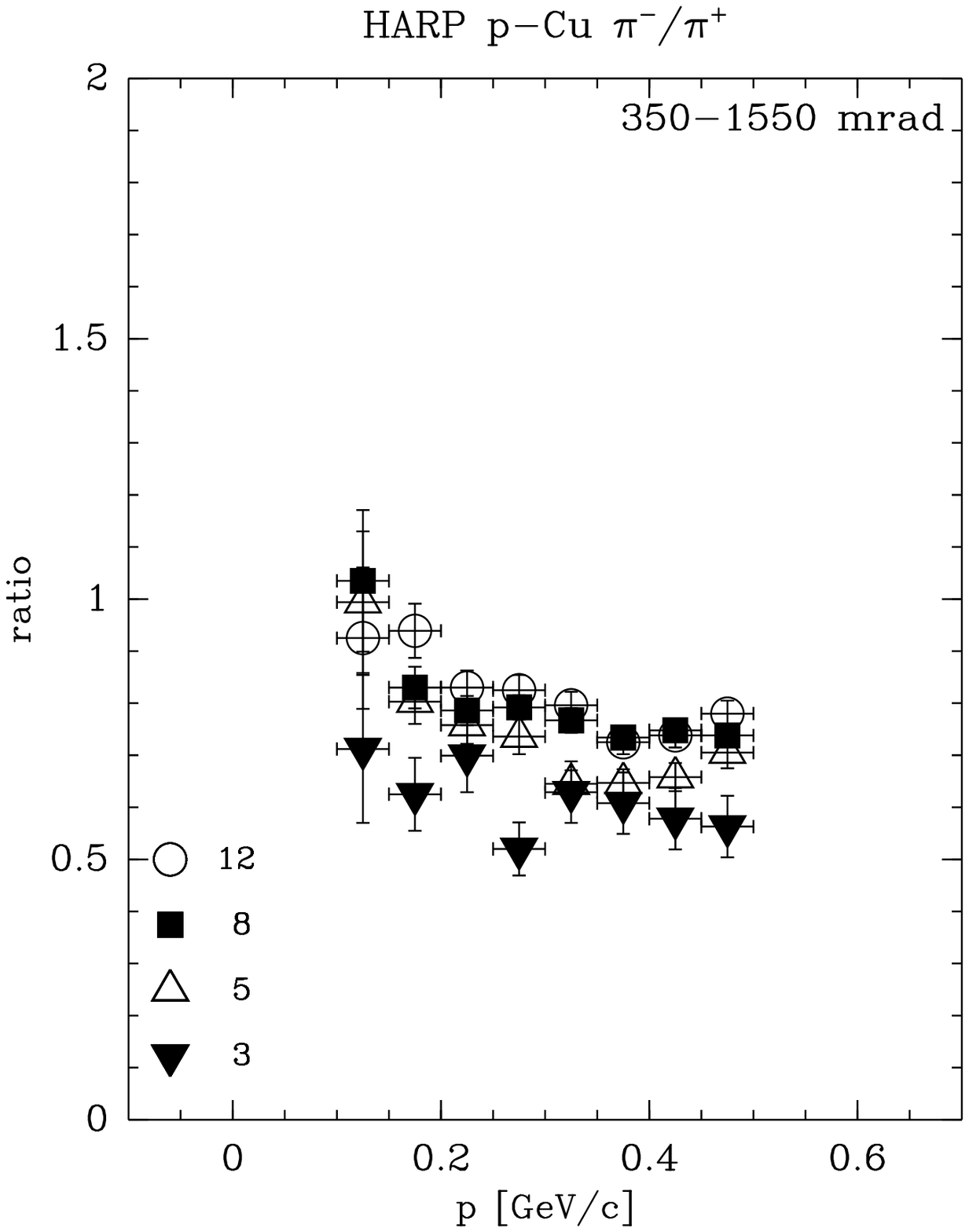,width=0.425\textwidth}
  \epsfig{figure=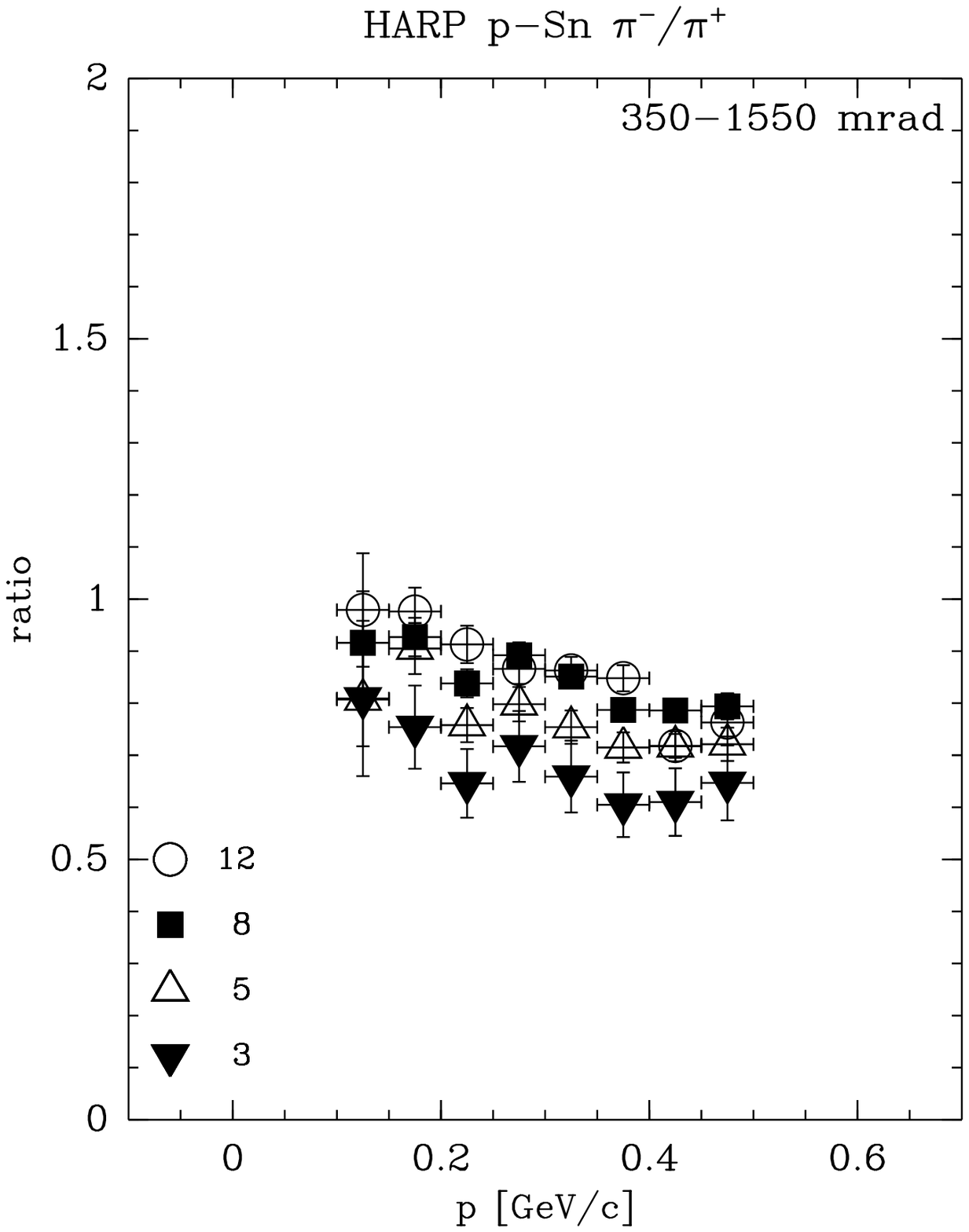,width=0.425\textwidth}
  \epsfig{figure=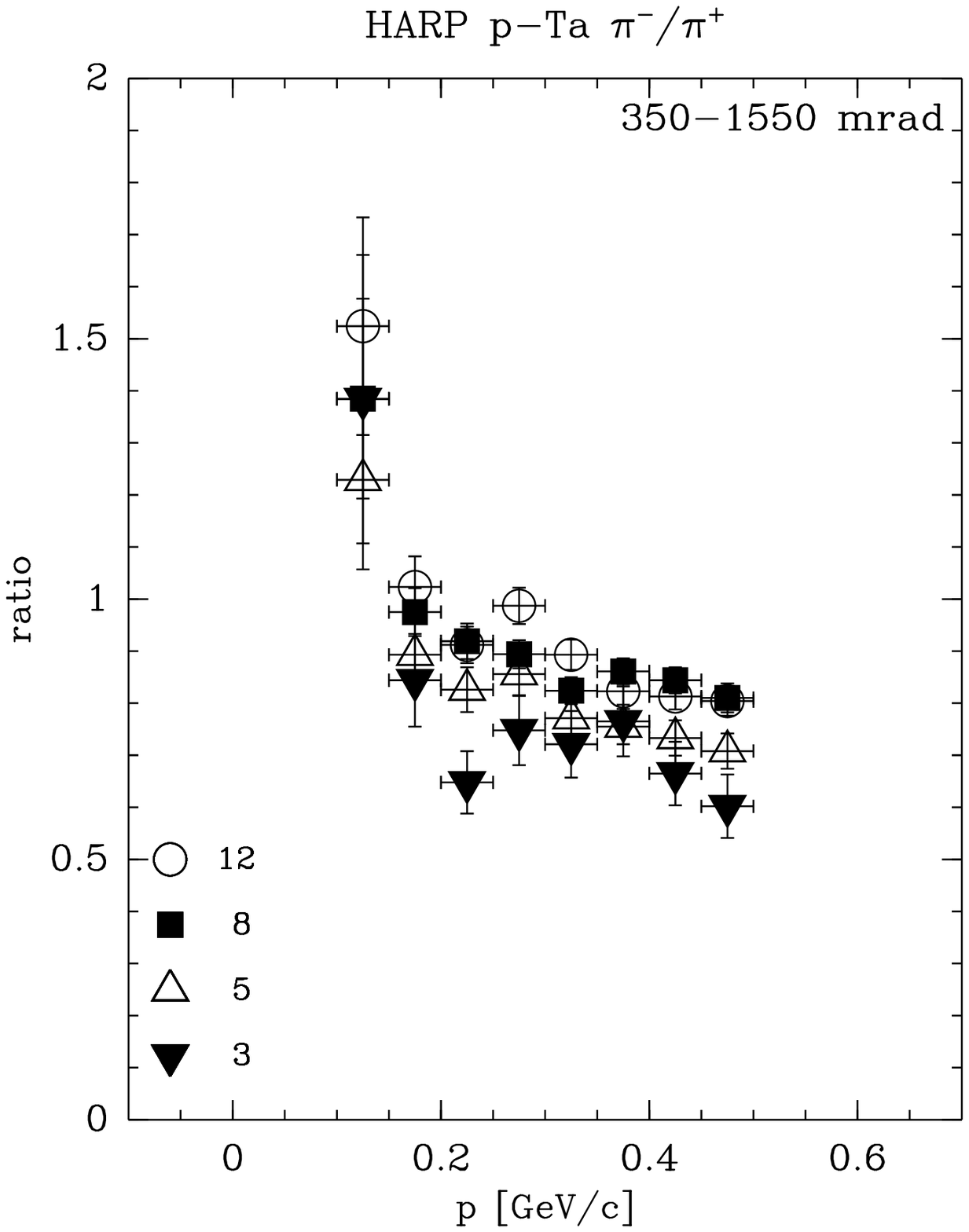,width=0.425\textwidth}
\end{center}
\caption{
The ratio of the differential cross-sections for \pim and \pip
 production in
p--C, p--Cu, p--Sn and p--Ta   interactions as a function of secondary
 momentum integrated over the 
forward angular region (shown in mrad).
The results are given for four incident beam momenta (filled triangles:
3~\GeVc; open triangles: 5~\GeVc; filled rectangles: 8~\GeVc; open
circles: 12~\GeVc).
}
\label{fig:xs-ratio}
\end{figure}

The integrated \pim/\pip ratio in the forward direction is displayed in
Fig.~\ref{fig:xs-ratio} as a function of secondary momentum. 
The previously published p--Ta data are reproduced in addition to the
measurements on the three target nuclei presented in this paper. 
In the covered  part of the momentum range more \pip's are produced than
\pim's.
The \pim/\pip ratio increases with increasing beam momentum and,
depending on the beam momentum, a change
of the sign of the slope of the ratio as a function of secondary momentum
is visible in the p--C data.
The latter feature is not present in the p--Cu, p--Sn and
p--Ta~\cite{ref:harp:tantalum} data. 
The ratio is closer to unity for the heavier target nuclei and a smaller
variation with beam momentum is observed.  
%% Especially the high \pim/\pip ratio observed in
%% the lowest secondary momentum bin in the p--Ta data is not present in
%% the p--C data.

\begin{figure}[tbp]
\begin{center}
  \epsfig{figure=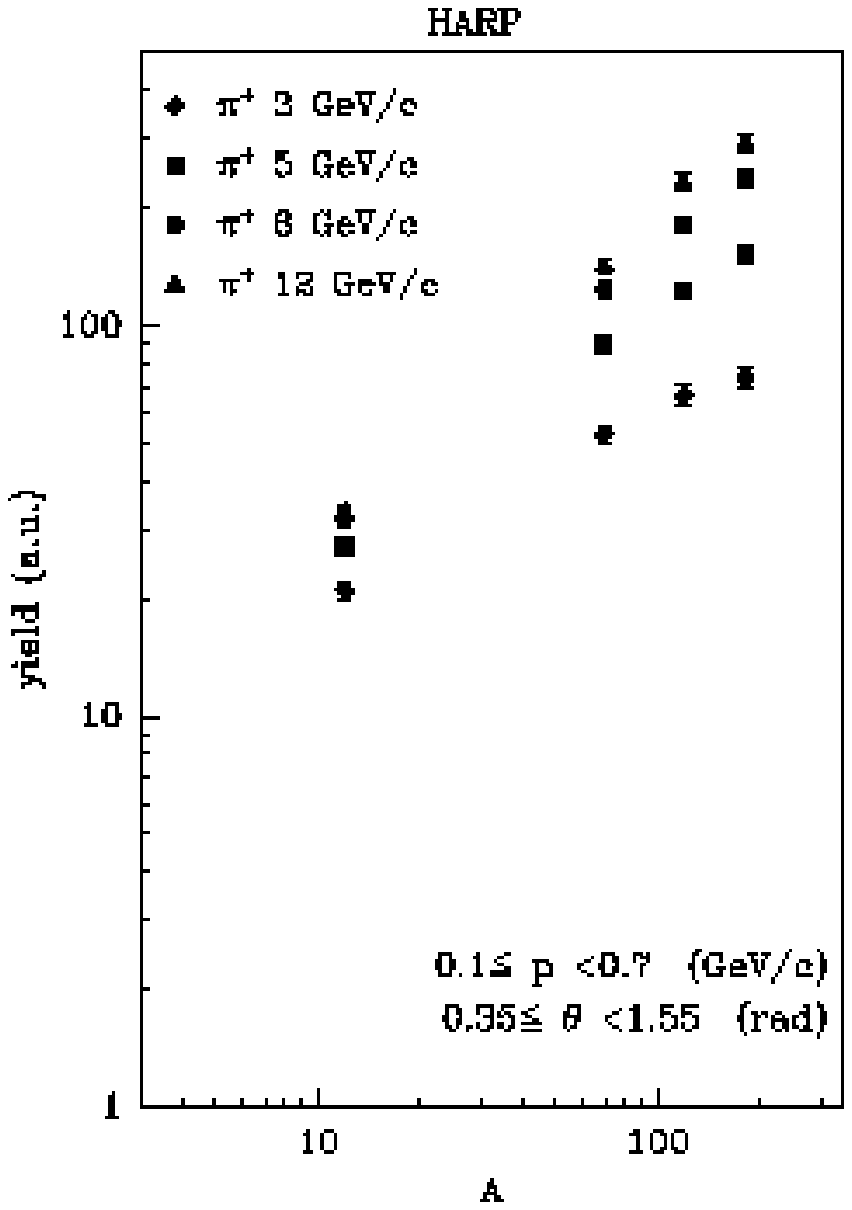,width=0.425\textwidth}
 ~
  \epsfig{figure=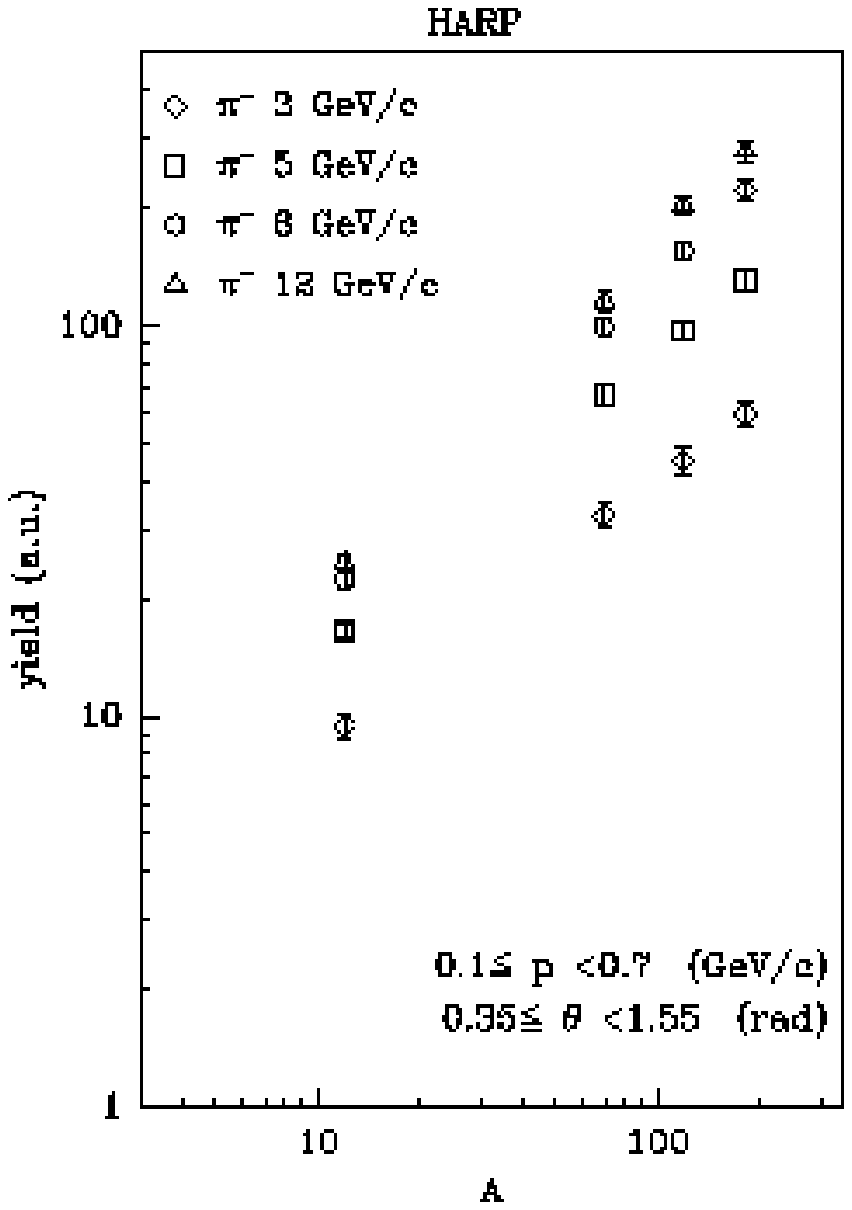,width=0.425\textwidth}
\end{center}
\caption{
 The dependence on the atomic number $A$ of the pion production yields
 in p--C, p--Cu, p--Sn, p--Ta interactions
 interactions integrated over the forward angular region
 ($0.350~\rad \leq \theta < 1.550~\rad$) and momentum ($100~\MeVc \leq p < 700~\MeV/c$).
 The results are given in arbitrary units, with a consistent scale
 between the left and right panel.
 The vertical scale used in this figure is consistent with the one in
 Fig.~\ref{fig:xs-trend}.
}
\label{fig:xs-a-dep}
\end{figure}

The dependence of the integrated pion yields on the atomic number $A$ is
shown in Fig.~\ref{fig:xs-a-dep} combining the results with the p--Ta
data (Ref.~\cite{ref:harp:tantalum}) taken with the same apparatus and analysed
using the same methods.
The \pip yields integrated over the region
$0.350~\rad \leq \theta < 1.550~\rad$ and $100~\MeVc \leq p < 700~\MeVc$ are
shown in the left panel and the \pim data integrated over the same region
in the right panel for four different beam momenta.
One observes a smooth behaviour of the integrated yields.
The $A$-dependence is slightly different for \pim and \pip production,
the latter saturating earlier, especially at lower beam momenta.

\subsection{Systematic errors}
\label{sec:syst}

The uncertainties are reported in some detail in
Table~\ref{tab:errors-3} for the carbon target data 
and summarized for the copper and tin target data in
Table~\ref{tab:errors-4}.   
One observes that only for the 3~\GeVc beam is the statistical error 
similar in magnitude to the systematic error, while the statistical
error is negligible for the 8~\GeVc and 12~\GeVc beams.
The statistical error is calculated by error propagation as part of the
unfolding procedure. 
It takes into account that the unfolding matrix is obtained from the
data themselves\footnote{The migration matrix is calculated without
prior knowledge of the cross-sections, while the unfolding procedure
determined the unfolding matrix from the migration matrix and the
distributions found in the data.} and hence contributes also to the
statistical error. 
This procedure almost doubles the statistical error, but avoids an important 
systematic error which would otherwise be introduced by assuming a
cross-section model {\em a priori} to calculate the corrections. 

The largest systematic error corresponds to the uncertainty in the
absolute momentum scale, which was estimated to be around 3\% using elastic
scattering (see detailed discussion in~\cite{ref:harp:tantalum}).
It is difficult to better constrain the absolute momentum scale, since
it depends on 
the knowledge of the beam momentum (known to 1\%) and the measurement
of the forward scattering angle in the elastic scattering interaction.
At low momentum in the relatively small angle forward
direction the uncertainty in the subtraction of the electron and
positron background due to \piz production is dominant.
This uncertainty is split between the variation in the shape of the
\piz spectrum and the normalization using the recognized electrons. 
The assumption is made that the \piz spectrum is similar to the
spectrum of charged pions.
Initial \pim and \pip spectra are obtained in an analysis without \piz
subtraction. 
The \pim spectra are then used in the MC for the \piz distributions. 
A full simulation of the production and decay into $\gamma$'s with
subsequent conversion in the detector materials is used to predict the
background electron and positron tracks.
In the region below 120~\MeVc a large fraction of the electrons can be
unambiguously identified.
These tracks are used as relative normalization between
data and MC.
The remaining background is then estimated from the distributions of
the simulated electron and positron tracks which are accepted as pion
tracks with the same criteria as used to select the data.
These normalized distributions are subtracted from the data
before the unfolding procedure is applied.
Uncertainties in the assumption of the \piz spectrum are taken into
account by an alternative assumption that their spectrum follows the
average of the \pim and \pip distribution.
An additional systematic error of 10\% is assigned to the
normalization of the \piz subtraction using the identified electrons
and positrons.

The target region definition  and
the uncertainty in the PID efficiency and background from tertiaries
are of similar size and are not negligible.
Relatively small errors are introduced by the uncertainties in
the absorption correction, absolute knowledge of the angular and the
momentum resolution.
The correction for tertiaries (particles produced in secondary
interactions) is relatively large at low momenta and large angles. 
As expected, this region is most affected by this component.

As already mentioned above, the overall normalization has an uncertainty
of 2\%, and is not reported in the table.
It is mainly due to the uncertainty in the efficiency that beam protons
counted in the normalization actually hit the target, with smaller
components from the target density and beam particle counting procedure.

\input{C5-syst-table}
\input{Cu5-Sn5_syst-table}

\subsection{Comparisons with earlier data}
\label{sec:compare}

\begin{figure}[tbp]
\epsfig{figure=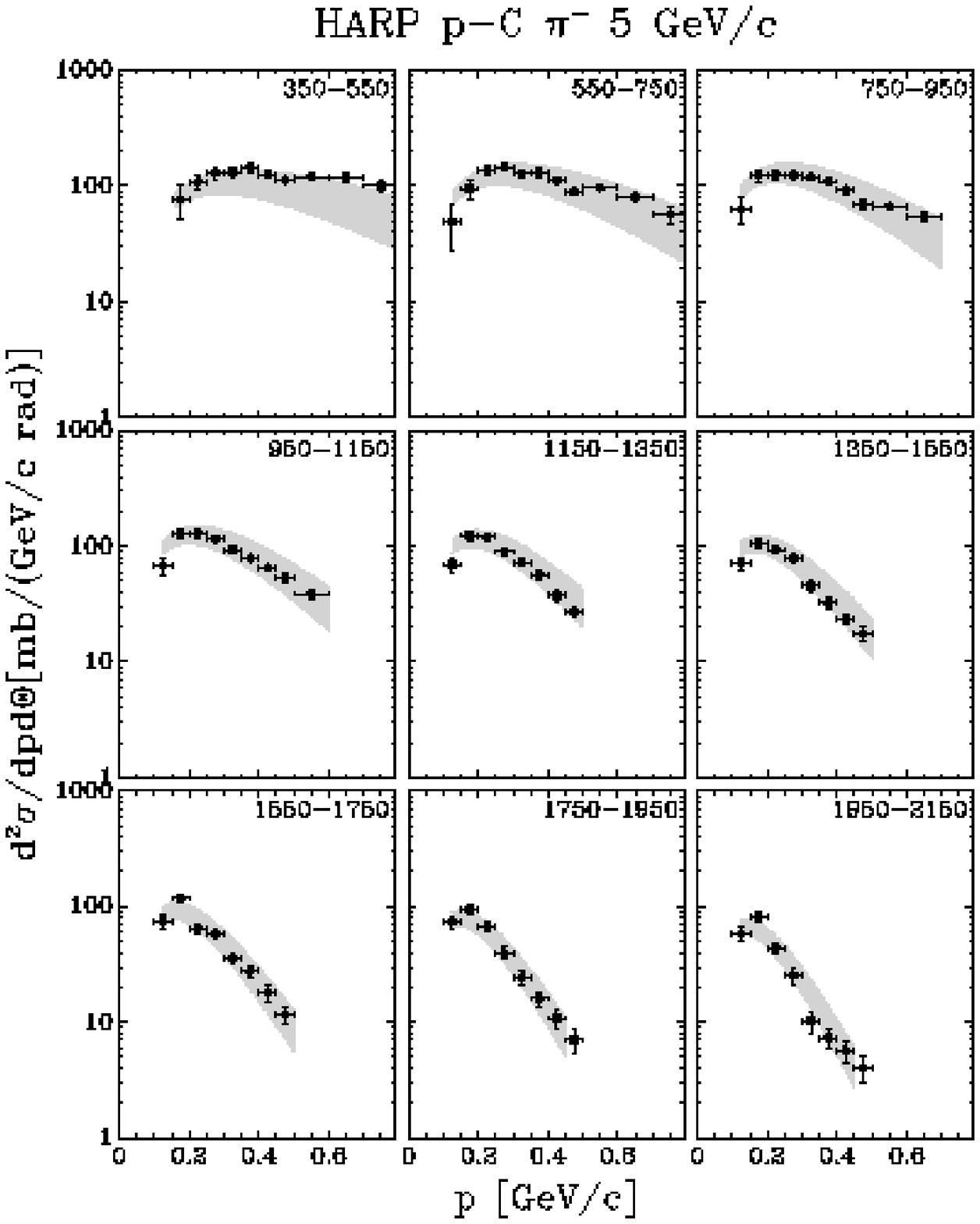,width=0.49\textwidth}
%~
%\epsfig{figure=plots_la72-CTHETA8-piminus_jinr_m_1-p_B.ps,width=0.49\textwidth}
~
\epsfig{figure=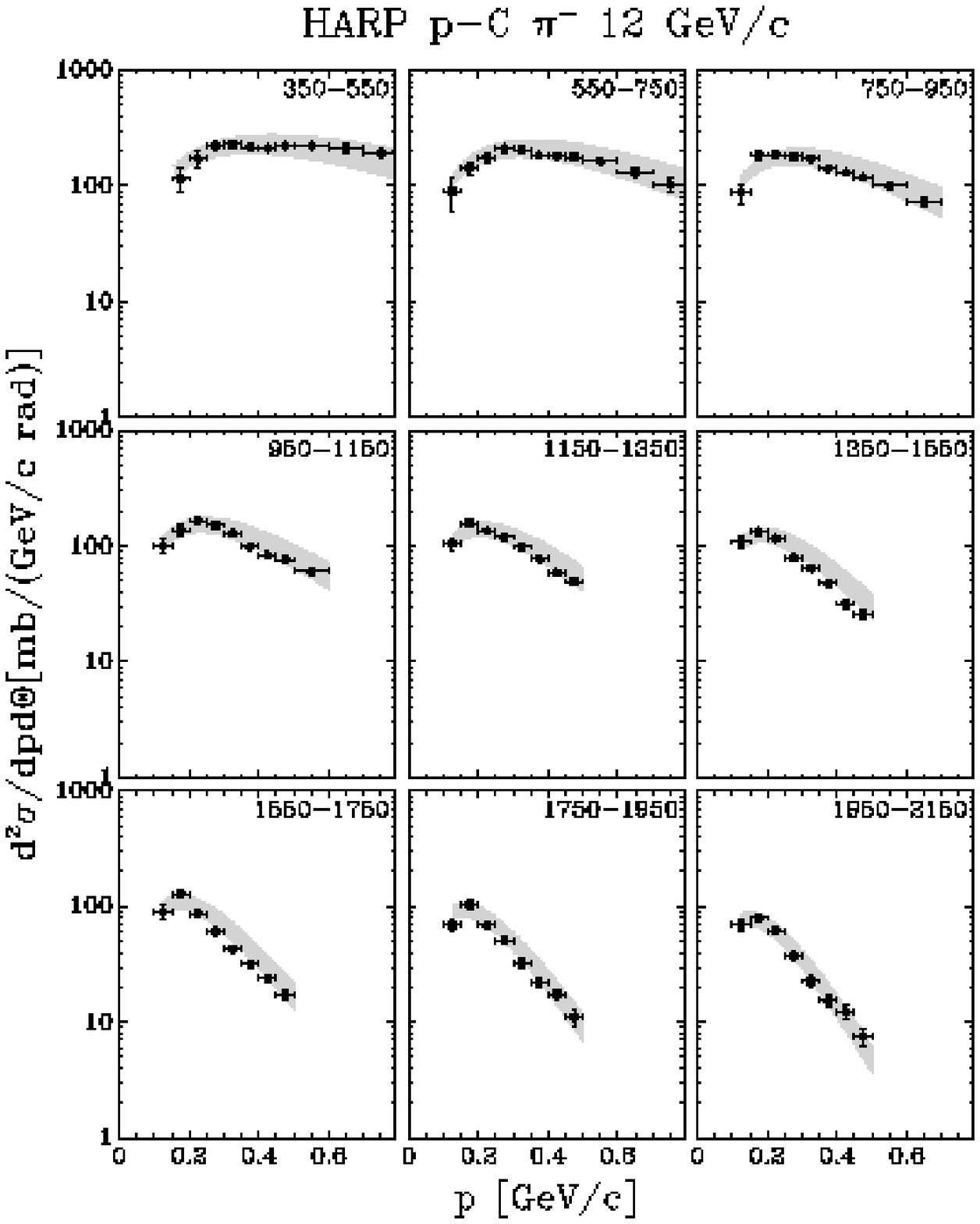,width=0.49\textwidth}
\caption{
Comparison of the HARP p--C results with data from
Ref.~\cite{ref:agakishiev} and \cite{ref:armutliiski}.
The left panel shows the comparison of the parametrization of
the 4.2~\GeVc data of Ref.~\cite{ref:agakishiev} with the 5~\GeVc data
reported here; the right panel shows the comparison of
the 10~\GeVc parametrization of \cite{ref:armutliiski} with the 12~\GeVc
data. 
The absolute normalization of the parametrization was fixed to the
data in both cases.
The band shows the range allowed by varying the slope parameters given
by \cite{ref:agakishiev} and \cite{ref:armutliiski} with two standard
 deviation and a 10\% variation on the absolute scale.
The angular ranges are shown in mrad in the panels.
}
\label{fig:compare}
\end{figure}

Very few pion production data sets are available in the literature for
p--C, p--Cu and p--Sn interactions in this energy region. 
Our data can be compared with results from
Ref.~\cite{ref:agakishiev} and \cite{ref:armutliiski}
where measurements of 
\pim production are reported in 4.2~\GeVc and 10~\GeVc p--C
interactions, respectively. 
The total number of \pim observed in the above references is about
1300 (5650) in the 4.2(10)~\GeVc data. 
In the papers cited above no tables of the double differential
cross-sections were provided, the measurements being given in
parametrized and graphical form only.
The authors of Ref.~\cite{ref:agakishiev} and \cite{ref:armutliiski}
give the results as a 
simple exponential in the invariant cross-section:
$\frac{E}{A} {\frac{{\mathrm{d}^3 \sigma}}{{\mathrm{d}p^3 }}}$,
where $E$ and $p$ are the energy and momentum of the produced
particle, respectively, and $A$ the atomic number of the target
nucleus\footnote{their spectra are parametrized 
in each angular bin with a function of
the form 
$f_{\pi^-}= c \ \exp{(-T/T_0)}$,
where $T$ is the kinetic energy of the produced particle and $T_0$ is
given by $T_0=T'/(1-\beta \ \cos{\theta})$.
For the 4.2~\GeVc data the values of the parameters are 
$T'=(0.089\pm0.006) \ \GeVc$ and $\beta=0.77\pm0.04$ and 
$T'=(0.100\pm0.002) \ \GeVc$ and $\beta=0.81\pm0.02$ for the 10~\GeVc
data.}. 
Unfortunately, no absolute normalization is given numerically.
To provide a comparison with these data, the parametrization was
integrated over the angular bins used in our analysis and with an
arbitrary overall normalization overlaid to our 
%8~\GeVc and 
%12~\GeVc 
results. 
We compare the 4.2~\GeVc parametrization of Ref.~\cite{ref:agakishiev} with our
5~\GeVc data and the Ref.~\cite{ref:armutliiski} parametrization with our 12~\GeVc
data. 
In the comparison with the 4.2~\GeVc parametrization the normalization $c$
is a simple constant, while for the 10~\GeVc parametrization a smooth
$\theta$-dependence consistent with a graphical analysis of
Ref.~\cite{ref:armutliiski} was used.
Thus only the comparison of the slopes with secondary momentum can be
considered significant.  
Since the 8~\GeVc and 12~\GeVc p--C results are very similar, the lack
of data with an exactly equal beam momentum does not play an important role.
The results of this comparison are shown in Fig.~\ref{fig:compare}.
The shaded band gives the excursion of the parametrization due to the
error in the slope parameters ($\pm 2\sigma$) with an additional assumed
10\% error on the absolute scale.
The latter additional error takes into account the fact that 
the errors on the slopes fitted to the individual angular
bins in the cited data are at least a factor of two larger than in the
exponential slope obtained from their global parametrization.
The agreement of our data with the simple parametrization is good.
To judge the quality of the comparison, one should keep in mind that the
statistics of Ref.~\cite{ref:agakishiev} and \cite{ref:armutliiski} is
much smaller (1300 \pim and 5650 \pim, respectively) 
than the statistics of the \pim samples in our 5~\GeVc and 12~\GeVc
data (18,000 and 43,000 \pim, respectively).  
The errors on the slopes fitted to the individual angular
bins in the cited data are at least a factor of two larger than in the
exponential slope obtained from their global parametrization.
The bands in the figure extend over the region where data from 
Ref.~\cite{ref:agakishiev} and Ref.~\cite{ref:armutliiski} are available.
%
% discussion about norm(theta) missing
%

\begin{figure}[tbp]
\epsfig{figure=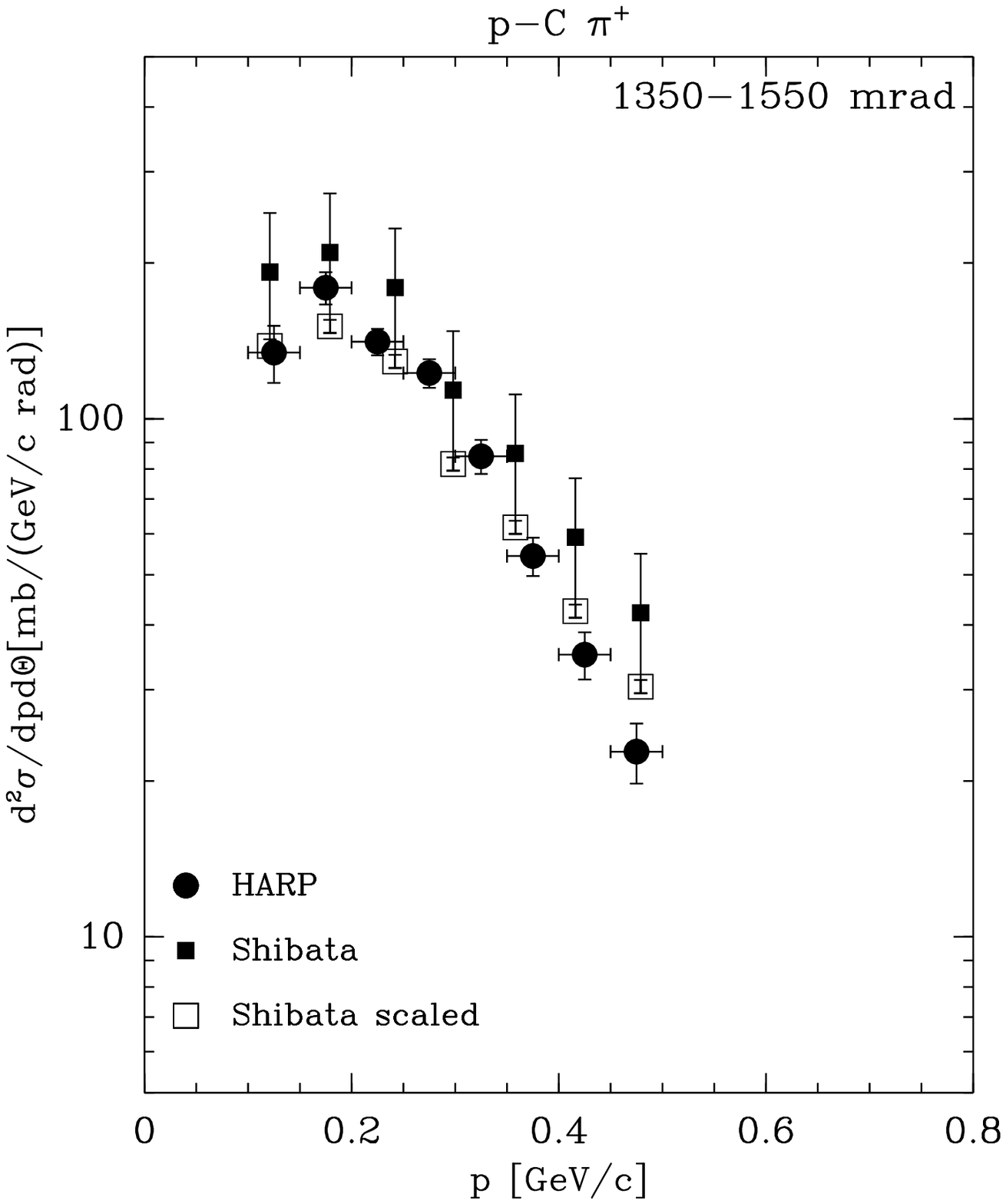,width=0.47\textwidth}
~
\epsfig{figure=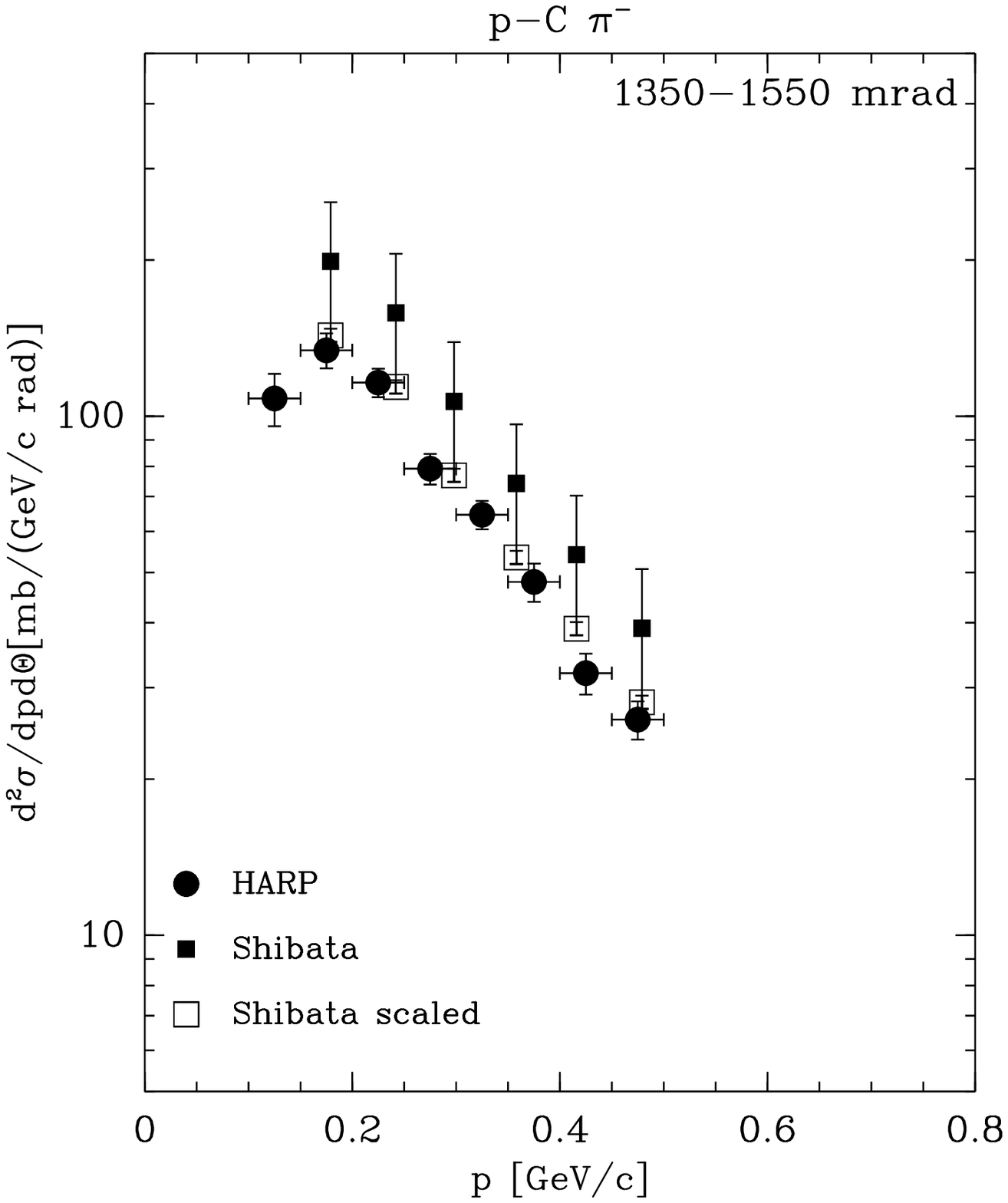,width=0.47\textwidth}
\caption{
Comparison of the HARP results with \pip and \pim production data at 90
 degrees from Ref.~\cite{ref:shibata} taken with 12~\GeVc protons.
The left panel shows the comparison of the \pip production
 data of Ref.~\cite{ref:shibata} with the data
reported here; the right panel shows the comparison with the \pim production
data. 
The smaller filled boxes show the data directly from Ref.~\cite{ref:shibata},
while the open boxes are scaled as explained in the text.
The latter data set compares well with the data described in this paper
(filled circles) in the angular region 1.35~\rad~$\le \ \theta \
<$1.55~\rad. 
}
\label{fig:shibata-c}
\end{figure}

Our p--C and p--Cu data can also be compared with \pip and \pim production measurements
taken with 12~\GeVc incident protons from Ref.~\cite{ref:shibata}.
These data were taken with a magnetic spectrometer and only measurements
at 90 degrees from the initial proton direction are available.
The statistical point--to--point errors are quoted to be 3\%, while the
overall normalization has a 30\% uncertainty due to the knowledge of the
acceptance.   
In Fig.~\ref{fig:shibata-c} their p--C data are shown together with the p--C data
reported in this paper.  
The filled boxes show the data directly from Ref.~\cite{ref:shibata},
while the open boxes are scaled with a factor 0.72.
This factor was defined by scaling the average of the \pim and \pip data
from Ref.~\cite{ref:shibata} at 179~\MeVc and 242~\MeVc to the HARP data
averaged over the same region.
The scale factor is within one standard deviation of the systematic
normalization uncertainty of Ref.~\cite{ref:shibata}. 
The latter data set compares well with the data described in this paper
(filled circles) in the angular region 1.35~\rad~$\le \ \theta \
<$1.55~\rad at the same proton beam momentum. 
 
\begin{figure}[tbp]
\epsfig{figure=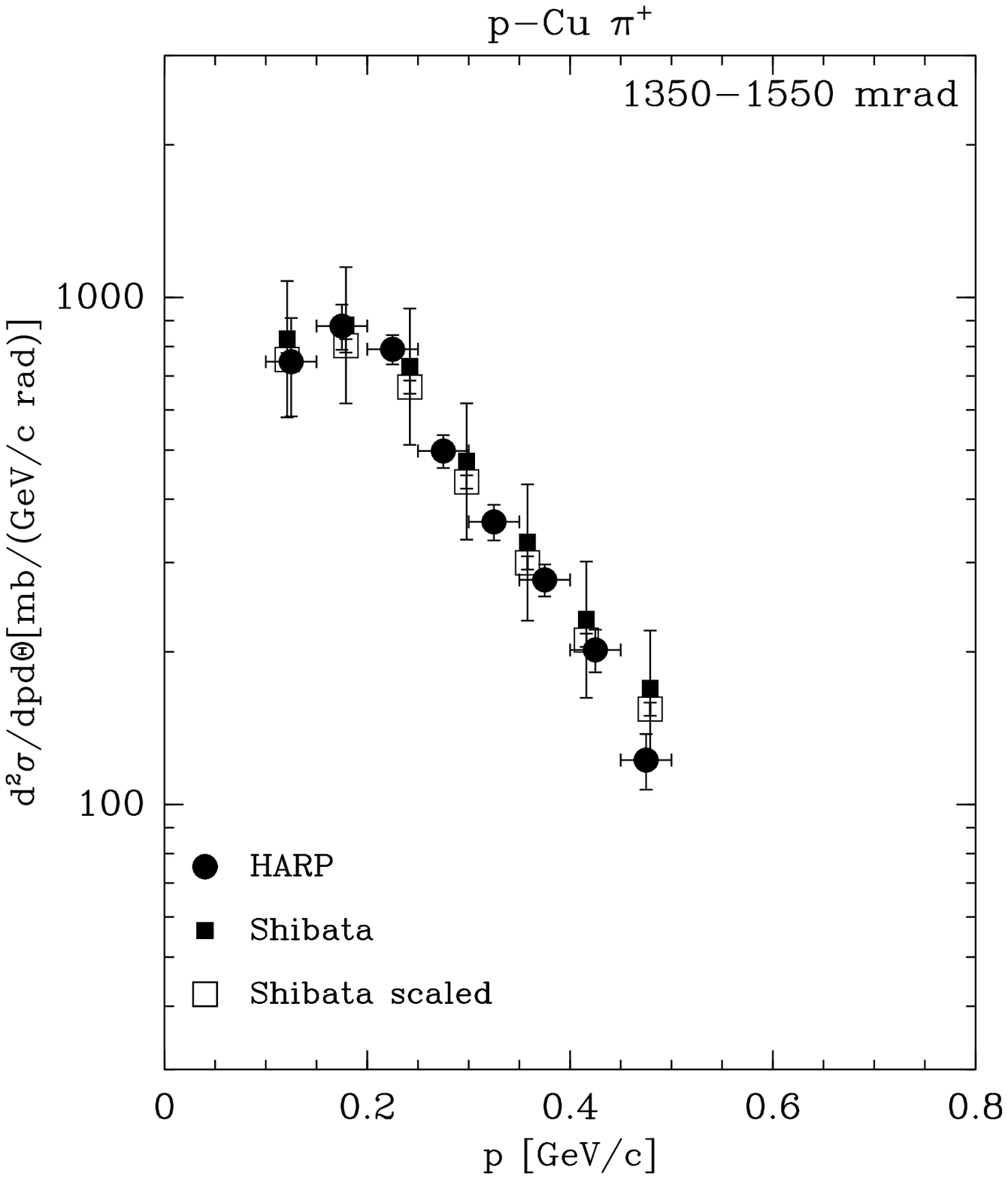,width=0.47\textwidth}
~
\epsfig{figure=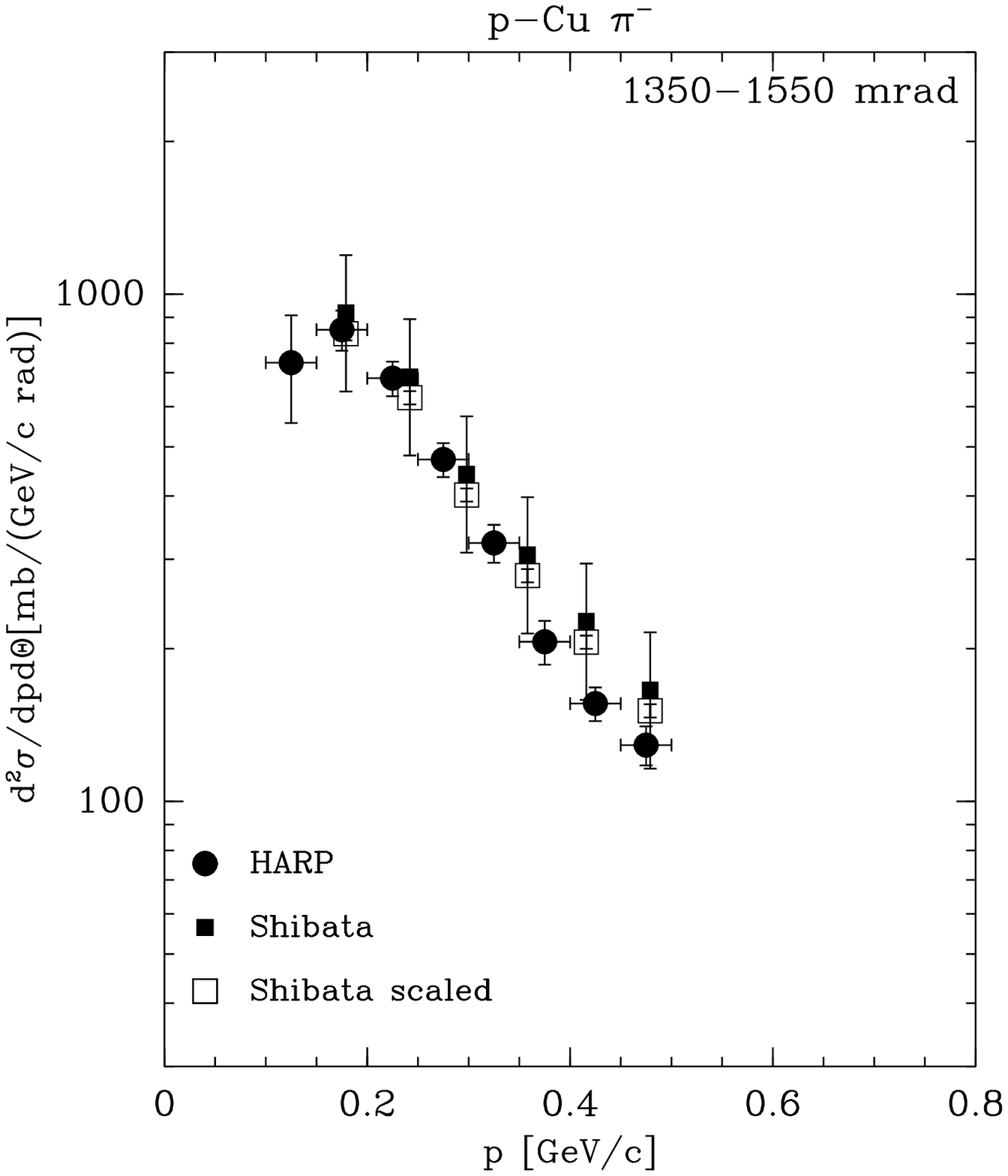,width=0.47\textwidth}
\caption{
Comparison of the HARP data with \pip and \pim production data at 90
 degrees from Ref.~\cite{ref:shibata} taken with 12~\GeVc protons.
The left panel shows the comparison of the \pip production
 data of Ref.~\cite{ref:shibata} with the data
reported here; the right panel shows the comparison with the \pim production
data.
The smaller filled boxes show the data directly from Ref.~\cite{ref:shibata},
while the open boxes are scaled as explained in the text.
The latter data set compares well with the data described in this paper
(filled circles) in the angular region 1.35~\rad~$\leq \ \theta \
<$1.55~\rad.
}
\label{fig:shibata-cu}
\end{figure}

In Fig.~\ref{fig:shibata-cu} the p--Cu data of Ref.~\cite{ref:shibata}
are shown together with the p--Cu results reported in this paper.
The filled boxes show the data directly from Ref.~\cite{ref:shibata},
while the open boxes are scaled with a factor 0.91.
This factor was defined in a similar procedure as described for the p--C
data. 
The scale factor is very close to unity and well within one standard
deviation of the systematic 
normalization uncertainty of Ref.~\cite{ref:shibata}.
Like for the p--C data, the full error bars including
the 30\% scale uncertainty of Ref.~\cite{ref:shibata} are drawn, for the
scaled data only their quoted statistical error of 3\% is shown.
The agreement of the two data sets is excellent.
The fact that the two scale factors are different may be due to the fact
that the scale uncertainty in Ref.~\cite{ref:shibata} holds separately
for data sets taken with different target nuclei.

\begin{figure}[tbp]
\begin{center}
   \begin{minipage}[b]{0.47\textwidth}
    \begin{center}
     \epsfig{figure=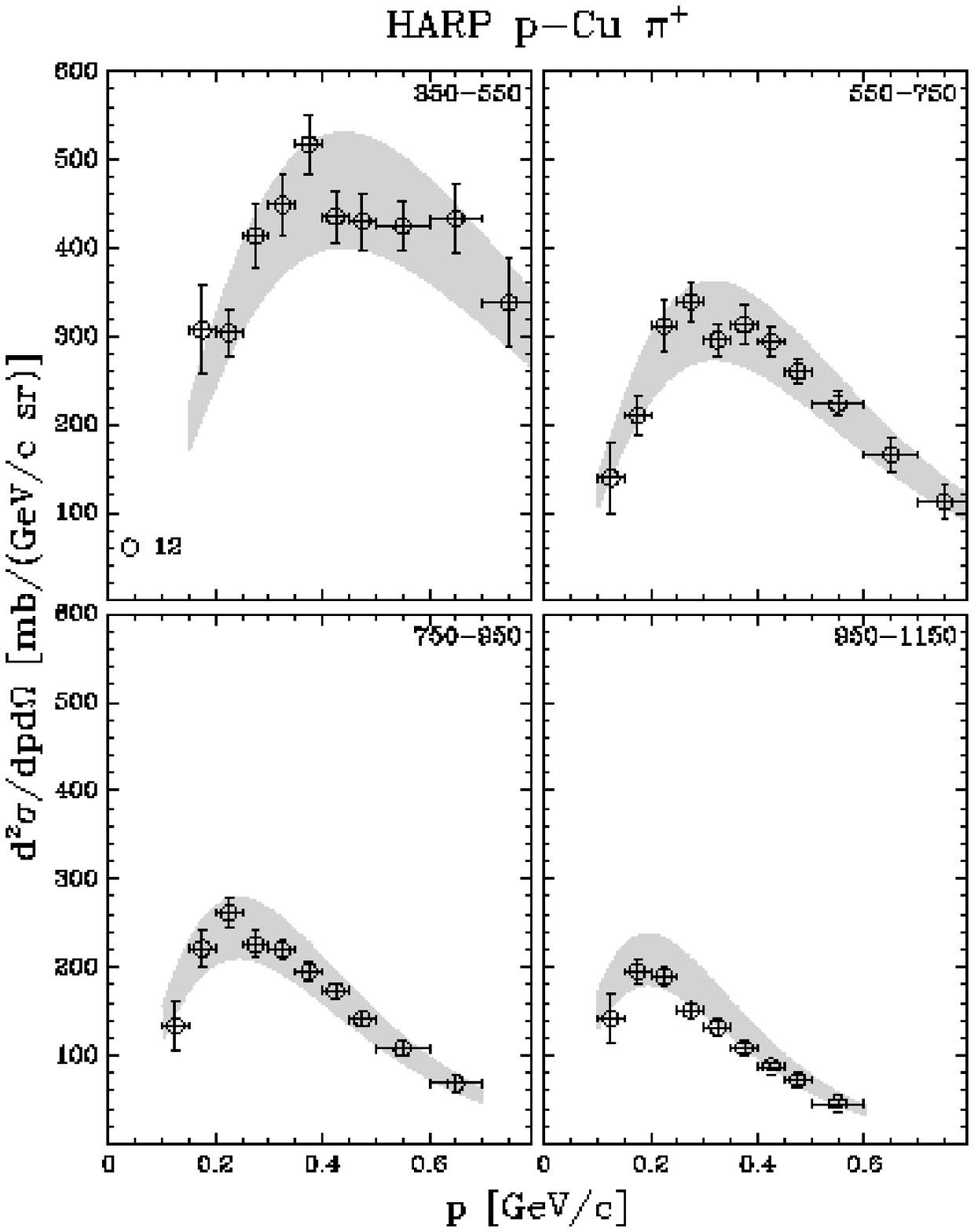,width=0.9\textwidth}
    \end{center}
   \end{minipage}
 ~
   \begin{minipage}[b]{0.47\textwidth}
    \begin{center}
     \epsfig{figure=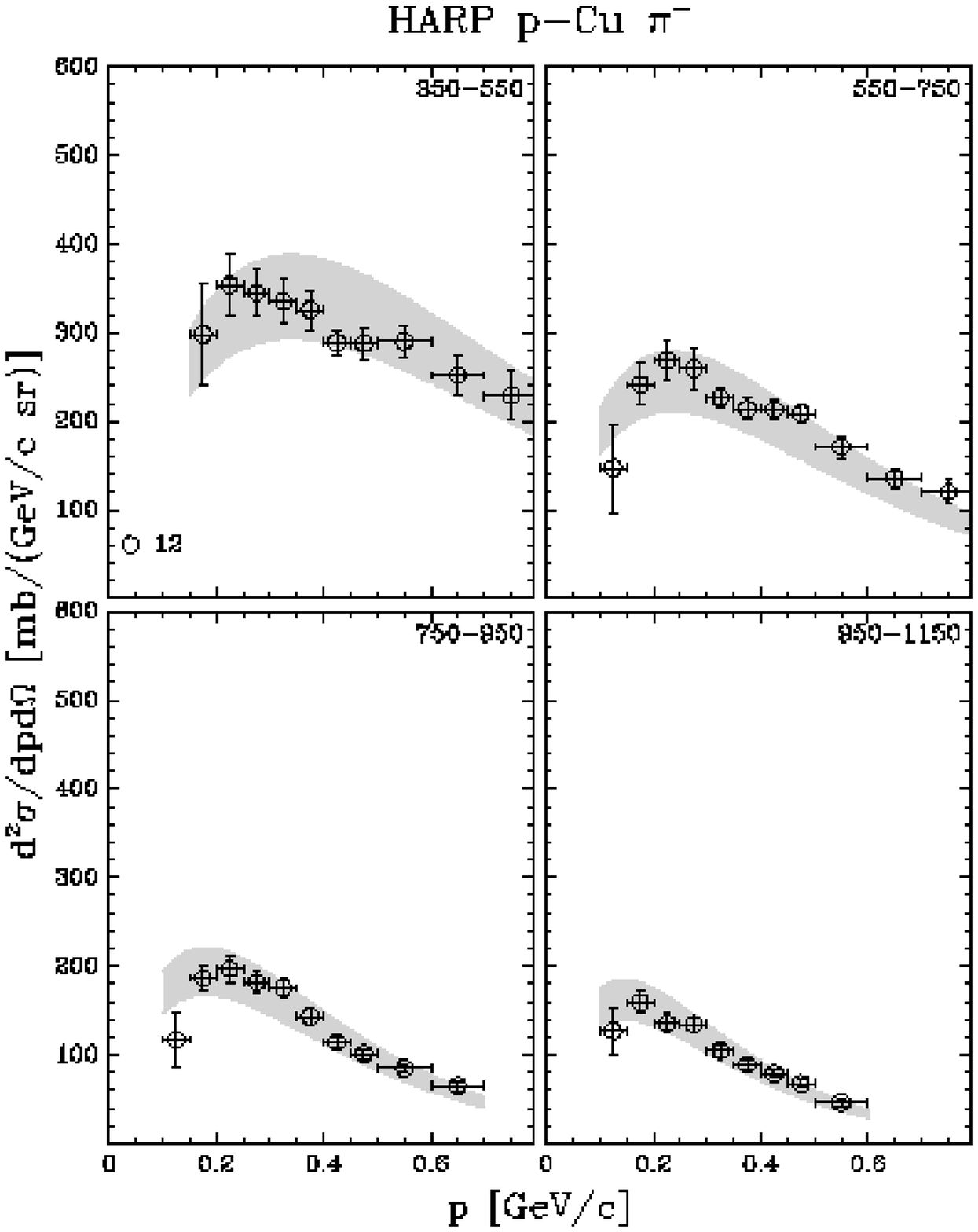,width=0.9\textwidth}
    \end{center}
   \end{minipage}

\end{center}
 \begin{center}
  \epsfig{figure=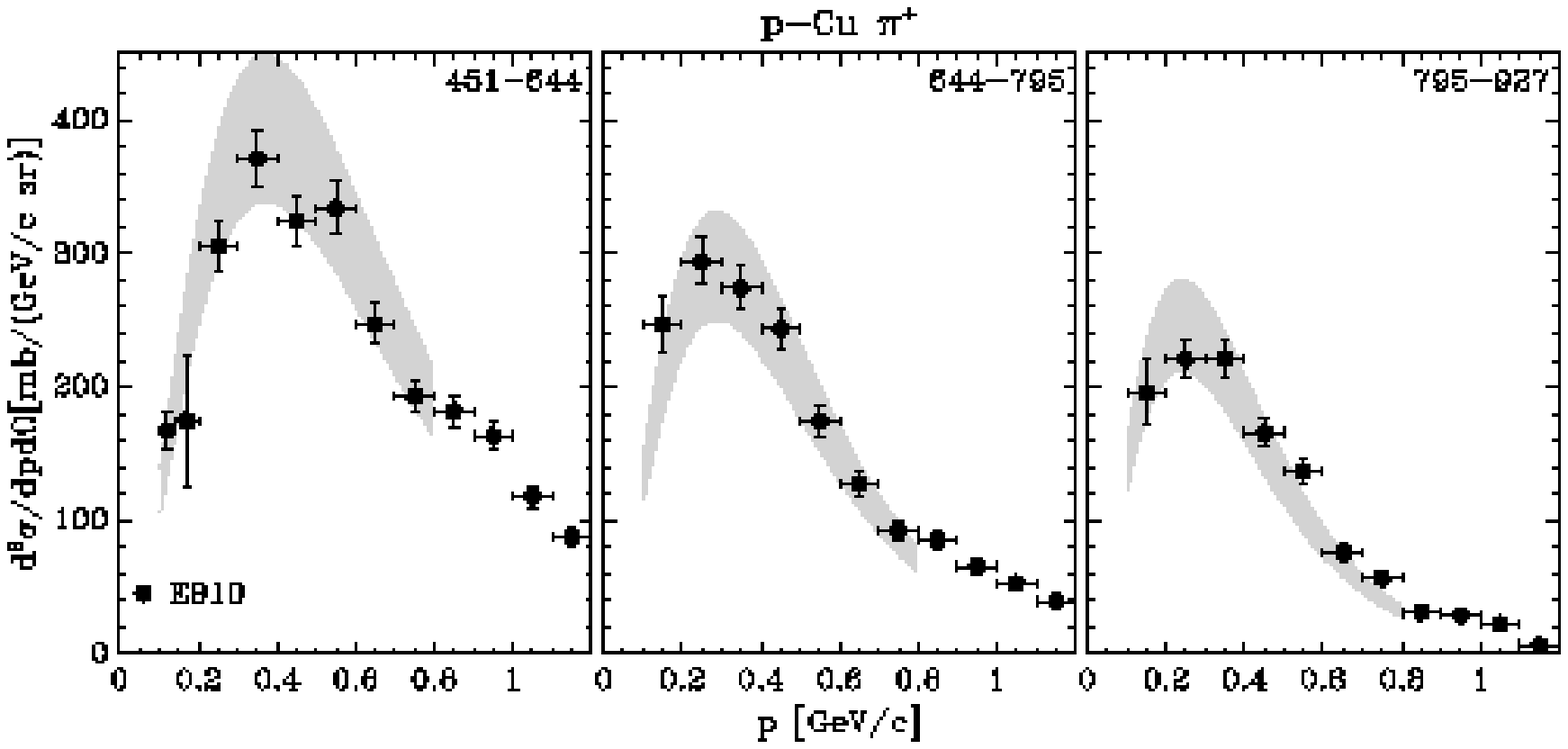,width=0.48\textwidth}
  ~
  \epsfig{figure=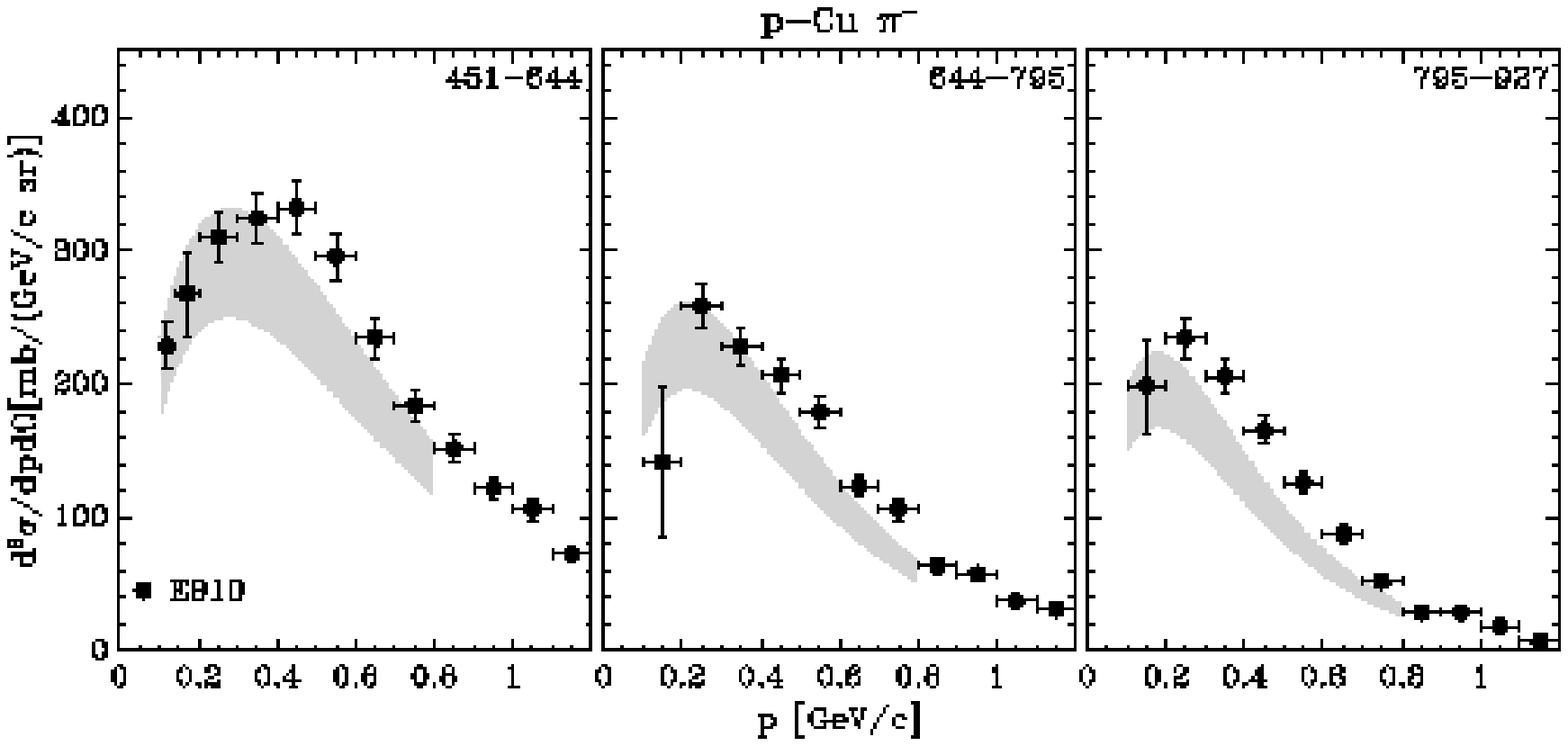,width=0.48\textwidth}
 \end{center}
\caption{
Comparison of the HARP data with \pip and \pim production data from
 Ref.~\cite{ref:E910} taken with 12.3~\GeVc protons.
The top panels show a parametrization of the \pip (left) and \pim(right)
 production  data described in this paper.
The data have been normalized to represent
$
{{\mathrm{d}^2 \sigma^{\pi}}}/{{\mathrm{d}p\mathrm{d}\Omega }} \ .
$
The shaded band represents the area between two parametrization which
 contain the data points.
The bottom panels show the comparison of the same parametrization,
 now binned according to the E910 data.  The bottom left (right) panel shows the
 \pip (\pim) production  data of Ref.~\cite{ref:E910}.
The angular regions are indicated in \mrad in the upper right-hand corner of each plot.
}
\label{fig:e910}
\end{figure}

Available data at 12.3 GeV/c from the E910 experiment \cite{ref:E910} are in
reasonable agreement with our p--Cu results as shown in
Fig.~\ref{fig:e910}.
In order to take into account the different angular binnings which
prevent a direct comparison, a Sanford-Wang parametrization is
fitted to our data.
The fit is performed to the data redefined as
$
{{\mathrm{d}^2 \sigma^{\pi}}}/{{\mathrm{d}p\mathrm{d}\Omega }} \ .
$
An area between two parametrizations is defined which contains our data
points as shown in Fig.~\ref{fig:e910} (top panels).
It is visible that the parametrization is not a perfect description to
our data. 
Therefore, we define a band of $\pm 15\%$ around the best fit which
contains almost all the HARP data points. 
The same parametrizations are then displayed in the binning of E910.
While the shape of the distributions are similar for both \pip and \pim
in HARP and E910 data sets, the absolute normalizations disagree by
5\%--10\%. 
For the individual data sets the systematic errors are between 5\% and
10\% depending on the range of secondary momentum.
Since these errors are correlated between bins, the discrepancy in the
\pip and \pim data separately are of the order of one standard
deviation. 
However, the effects are opposite in \pip and \pim, giving a 15\%
difference in the \pip/\pim ratio between two experiments which is of
the order of two standard deviations. 
This effect may point to an
underestimation of systematic effects on the absolute normalization in
one of the experiments or in the PID efficiency.
Part of the difference is also due to an imperfect parametrization
of our data sample.
Owing to the symmetry of the HARP TPC, including its trigger counter, we
do not expect a large systematic error in the HARP data between \pip and
\pim production cross-sections.

\section{Summary and Conclusions}
\label{sec:summary}

An analysis of the production of pions at
large angles with respect to the beam direction for protons of
3~\GeVc, 5~\GeVc, 8~\GeVc and 12~\GeVc impinging on  thin
(5\% interaction length) carbon, copper and tin targets is described.
The secondary pion yield is measured in a large angular and momentum
range and double-differential cross-sections are obtained.
A detailed error estimation has been discussed.
Results on the dependence of pion production on the target atomic number A
are also presented.

The use of a single detector for a range of beam momenta makes it
possible to measure the dependence of the pion yield on the secondary
particle momentum and emission angle $\theta$ with high precision.
The $A$ dependence of the cross-section can be studied using the
combination of the present data with the data obtained with 
tantalum~\cite{ref:harp:tantalum}.

Very few pion production measurements in this energy range are reported
in the literature.
The only comparable results found in the literature agrees with the
analysis described in this paper.
Hadronic production models describing this energy range can now be
compared with our new results and, if needed, improved.
Data taken with different target materials and beam momenta
will be presented in subsequent papers.

\section{Acknowledgements}

We gratefully acknowledge the help and support of the PS beam staff
and of the numerous technical collaborators who contributed to the
detector design, construction, commissioning and operation.  
In particular, we would like to thank
G.~Barichello,
R.~Brocard,
K.~Burin,
V.~Carassiti,
F.~Chignoli,
D.~Conventi,
G.~Decreuse,
M.~Delattre,
C.~Detraz,  
A.~Domeniconi,
M.~Dwuznik,   
F.~Evangelisti,
B.~Friend,
A.~Iaciofano,
I.~Krasin, 
D.~Lacroix,
J.-C.~Legrand,
M.~Lobello, 
M.~Lollo,
J.~Loquet,
F.~Marinilli,
J.~Mulon,
L.~Musa,
R.~Nicholson,
A.~Pepato,
P.~Petev, 
X.~Pons,
I.~Rusinov,
M.~Scandurra,
E.~Usenko,
and
R.~van der Vlugt,
for their support in the construction of the detector.
The collaboration acknowledges the major contributions and advice of
M.~Baldo-Ceolin, 
L.~Linssen, 
M.T.~Muciaccia and A. Pullia
during the construction of the experiment.
The collaboration is indebted to 
V.~Ableev,
%%%M.~Baldo~Ceolin,
F.~Bergsma,
P.~Binko,
E.~Boter,
M.~Calvi, 
C.~Cavion, 
A.~Chukanov,  
A.~De~Min, 
M.~Doucet,
D.~D\"{u}llmann,
V.~Ermilova, 
W.~Flegel,
Y.~Hayato,
A.~Ichikawa,
A.~Ivanchenko,
O.~Klimov,
T.~Kobayashi,
%S.~Kotov,
D.~Kustov,
M.~Laveder,  
%%%L.~Linssen,
M.~Mass,
H.~Meinhard,
%%%M.T.~Muciaccia, 
T.~Nakaya,
K.~Nishikawa,
M.~Pasquali,
M.~Placentino,
%I.~Potrap, 
%%%A.~Pullia,
S.~Simone,
S.~Troquereau,
S.~Ueda and A.~Valassi
for their contributions to the experiment.

We acknowledge the contributions of 
V.~Ammosov,
G.~Chelkov,
D.~Dedovich,
F.~Dydak,
M.~Gostkin,
A.~Guskov, 
D.~Khartchenko, 
V.~Koreshev,
Z.~Kroumchtein,
I.~Nefedov,
A.~Semak, 
J.~Wotschack,
V.~Zaets and
A.~Zhemchugov
to the work described in this paper.

 The experiment was made possible by grants from
the Institut Interuniversitaire des Sciences Nucl\'eair\-es and the
Interuniversitair Instituut voor Kernwetenschappen (Belgium), 
Ministerio de Educacion y Ciencia, Grant FPA2003-06921-c02-02 and
Generalitat Valenciana, grant GV00-054-1,
CERN (Geneva, Switzerland), 
the German Bundesministerium f\"ur Bildung und Forschung (Germany), 
the Istituto Na\-zio\-na\-le di Fisica Nucleare (Italy), 
INR RAS (Moscow) and the Particle Physics and Astronomy Research Council (UK).
We gratefully acknowledge their support.
This work was supported in part by the Swiss National Science Foundation
and the Swiss Agency for Development and Cooperation in the framework of
the programme SCOPES - Scientific co-operation between Eastern Europe
and Switzerland.

\clearpage

\begin{appendix}

\section{Cross-section data}\label{app:data}
\input{C5_result-table-scl}
\input{Cu5_result-table-scl}
\input{Sn5_result-table-scl}

%%%%%%%%%%%%%%%%%%%%%%%%%%%%%%%%%%

\end{appendix}

\clearpage

%\section*{References}

%%%%%%%%%%%

\end{document}

%% file: harpauthors.tex
\thispagestyle{plain}
\begin{center}
{\large HARP collaboration}\\
\newcommand{\afdoct}{{3}\xspace}
\vspace{0.1cm}
{\small
%\begin{center}
M.G.~Catanesi, 
%M.T.~Muciaccia, 
E.~Radicioni%,
%S.~Simone
\\ 
{\bf Universit\`{a} degli Studi e Sezione INFN, Bari, Italy} %done
\\
R.~Edgecock, 
M.~Ellis$^{1}$,          %, OK
S.~Robbins$^{2,3}$,      %, OK
F.J.P.~Soler$^{4}$
\\
{\bf Rutherford Appleton Laboratory, Chilton, Didcot, UK} %done
\\
C.~G\"{o}\ss ling %,
%M.~Mass  % one paper (same as roma iii student)
\\
{\bf Institut f\"{u}r Physik, Universit\"{a}t Dortmund, Germany} %done
\\
%I.~Boyko,          % not member of HARP
S.~Bunyatov, 
%G.~Chelkov,        % pending explicit request
%%A.~Chukanov,      % removed by Boris (only shifts)
%%D.~Dedovitch,     % pending explicit request
%A.~Elagin,         % not member of HARP
%%M.~Gostkin,       % pending explicit request 
%%A.~Guskov,        % pending explicit request, 
%D.~Khartchenko,    % removed by Boris (only shifts)
%%O.~Klimov,        % removed by Boris (only shifts) 
%S.~Kotov,          % (only shifts) 
%%I.~Krasin,        % techn.paper only (technician)
A.~Krasnoperov, 
%Z.~Kroumchtein,    % pending explicit request
%%D.~Kustov,        % removed by Boris (only shifts) 
%D.~Naumov,         % late request
%Y.~Nefedov,        % pending explicit request
%K.~Nikolaev,       % not member of HARP
B.~Popov$^5$, 
%I.~Potrap,,        % (only shifts) 
V.~Serdiouk,        % TPC only - propose to keep on this paper
V.~Tereschenko %, 
%A.~Zhemchugov      % pending explicit request
\\
{\bf Joint Institute for Nuclear Research, JINR Dubna, Russia} %later
\\
E.~Di~Capua, 
G.~Vidal--Sitjes$^{6}$  % now at ..
%tech: V. Carassiti
%tech: F. Evangelisti
\\
{\bf Universit\`{a} degli Studi e Sezione INFN, Ferrara, Italy}  %done
\\
A.~Artamonov$^7$,   % and ITEP, Moscow
P.~Arce$^8$,        % and univ. 
%F.~Dydak, 
S.~Giani, 
S.~Gilardoni,       %$^3$, %supported by DOCT 
P.~Gorbunov$^{7}$,  %,                new footnote
A.~Grant,  
A.~Grossheim$^{10}$, %$^1$, %supported by DOCT 
P.~Gruber$^{11}$,    %supported by DOCT 
V.~Ivanchenko$^{12}$,  %,                new footnote 
%J.-C.~Legrand,     % only technical paper, or first TPC paper?
A.~Kayis-Topaksu$^{13}$,
%L.~Linssen,  %% asked to be removed
J.~Panman, 
I.~Papadopoulos,  
J.~Pasternak, %$^{1,4}$,  %supported by DOCT 
E.~Tcherniaev, 
I.~Tsukerman$^7$,   % and ITEP, Moscow
R.~Veenhof, 
C.~Wiebusch$^{14}$,    % now at ..
%J.~Wotschack,
P.~Zucchelli$^{9,15}$ %on leave of absence from INFN-Ferrara
\\
{\bf CERN, Geneva, Switzerland} 
\\
A.~Blondel, 
S.~Borghi$^{16}$,  % new footnote or CERN
M.~Campanelli,       % ???????????????
M.C.~Morone$^{17}$, 
G.~Prior$^{18}$,   %supported by DOCT 
R.~Schroeter
\\
{\bf Section de Physique, Universit\'{e} de Gen\`{e}ve, Switzerland} %done
\\
R.~Engel,
C.~Meurer
\\
{\bf Institut f\"{u}r Physik, Forschungszentrum Karlsruhe, Germany}
\\
\newcommand{\afkyot}{{19}\xspace}
I.~Kato$^{10,\afkyot}$ %,                new footnote
%T.~Nakaya$^{\afkyot}$,
%K.~Nishikawa$^{\afkyot}$,
%S.~Ueda$^{\afkyot}$
\\
{\bf University of Kyoto, Japan} % 
\\
%V.~Ableev, % one or few papers
%C.~Cavion, 
U.~Gastaldi%, 
%M.~Placentino
\\
{\bf Laboratori Nazionali di Legnaro dell' INFN, Legnaro, Italy} %done
\\
\newcommand{\aflanl}{{20}\xspace}
G.~B.~Mills$^{\aflanl}$  
\\
{\bf Los Alamos National Laboratory, Los Alamos, USA} % 
\\
J.S.~Graulich$^{21}$, 
G.~Gr\'{e}goire 
\\
{\bf Institut de Physique Nucl\'{e}aire, UCL, Louvain-la-Neuve,
  Belgium} %ok
\\
M.~Bonesini, 
%%M.~Calvi,         % one physics paper
%%A.~De~Min,          % ??????
F.~Ferri,           % ?
M.~Paganoni,        % ?????? 
F.~Paleari          % ?
% for technical paper:  Francesco Chignoli 
\\
{\bf Universit\`{a} degli Studi e Sezione INFN Milano Bicocca, Milano, Italy} %done
\\
%S.~Gninenko, 
M.~Kirsanov
%Yu.~Musienko, 
%A.~Poljarush, 
%A.~Toropin
%V.Postoev % for technical paper
\\
{\bf Institute for Nuclear Research, Moscow, Russia} %done (?)
\\
A. Bagulya, 
%V.~Chechin, 
% acknowledgements, tech: V.~Ermilova$^{\dagger}$, 
V.~Grichine,  % also supported Geneva
N.~Polukhina%, 
%N.~Starkov
\\
{\bf P. N. Lebedev Institute of Physics (FIAN), Russian Academy of
Sciences, Moscow, Russia} %done
\\
V.~Palladino
\\
{\bf Universit\`{a} ``Federico II'' e Sezione INFN, Napoli, Italy} % ok 
\\
\newcommand{\afclmb}{{20}\xspace}
L.~Coney$^{\afclmb}$, 
D.~Schmitz$^{\afclmb}$
\\
{\bf Columbia University, New York, USA} % 
\\
G.~Barr, 
A.~De~Santo$^{22}$, % now at ..
C.~Pattison, 
K.~Zuber$^{23}$  % now at ..
\\
{\bf Nuclear and Astrophysics Laboratory, University of Oxford, UK} % ok
\\
%M.~Baldo~Ceolin,  preferes not to sign
% tech: G.~Barichello, 
F.~Bobisut, 
D.~Gibin,
A.~Guglielmi, 
%M.~Laveder, 
%A.~Menegolli, 
M.~Mezzetto
% tech: A. Pepato
%, M.~Vascon
\\
{\bf Universit\`{a} degli Studi e Sezione INFN, Padova, Italy} % done
\\
J.~Dumarchez, 
%S.~Troquereau, % one paper
F.~Vannucci 
\\
{\bf LPNHE, Universit\'{e}s de Paris VI et VII, Paris, France} % done
\\
%%V.~Ammosov,        %pending explicit request 
%V.~Gapienko,  %shifts?  reduce for next paper
%%V.~Koreshev,        %pending explicit request 
%%A.~Semak,   % ???,        %pending explicit request
%Yu.~Sviridov,  %shifts?  reduce for next paper
%tech: E.~Usenko, now at INR Troitsk, and LA paper (RPC)
%%V.~Zaets    % ???,        %pending explicit request
%%\\
%%{\bf Institute for High Energy Physics, Protvino, Russia} %done
%%\\
U.~Dore
\\
{\bf Universit\`{a} ``La Sapienza'' e Sezione INFN Roma I, Roma,
  Italy} % ok 
\\
D.~Orestano, 
%M.~Pasquali,  %student worked on calorimeter - one paper
F.~Pastore, 
A.~Tonazzo, 
L.~Tortora
% tech: Alfredo Iaciofano, Marco Lobello, Franco Marinilli
\\
{\bf Universit\`{a} degli Studi e Sezione INFN Roma III, Roma, Italy}
% done
\\
C.~Booth, 
C.~Buttar$^{4}$,  %"now at the University of Glasgow".
P.~Hodgson, 
L.~Howlett
%tech:  R. Nicholson
\\
{\bf Dept. of Physics, University of Sheffield, UK} %done
\\
M.~Bogomilov, 
% tech:  K.Burin
M.~Chizhov, 
D.~Kolev, 
% tech: P.~Petev, I.~Rusinov, 
R.~Tsenov
\\
{\bf Faculty of Physics, St. Kliment Ohridski University, Sofia,
  Bulgaria} %done
\\
%G.~Maneva, 
S.~Piperov, 
%S.~Stoykova, 
P.~Temnikov
\\
{\bf Institute for Nuclear Research and Nuclear Energy, 
Academy of Sciences, Sofia, Bulgaria} % done
\\
M.~Apollonio, 
P.~Chimenti,  % also supported Geneva
G.~Giannini, 
G.~Santin$^{24}$  % presently (now) at ESA / ESTEC, Noordwijk, The Netherlands , also supported Geneva, remove after one
\\
{\bf Universit\`{a} degli Studi e Sezione INFN, Trieste, Italy} % done
\\
%Y.~Hayato$^{\afkek}$, 
%A.~Ichikawa$^{\afkek}$, 
%T.~Kobayashi$^{\afkek}$
%\\
%{\bf KEK, Tsukuba, Japan} %
%\\
J.~Burguet--Castell, 
A.~Cervera--Villanueva, 
J.J.~G\'{o}mez--Cadenas, % also supported Geneva
J. Mart\'{i}n--Albo,
P.~Novella,
M.~Sorel,
A.~Tornero
\\
{\bf  Instituto de F\'{i}sica Corpuscular, IFIC, CSIC and Universidad de Valencia,
Spain} % done
}
\end{center}
\thispagestyle{plain}
\vfill
%\newpage
%\vspace{1cm}
\rule{0.3\textwidth}{0.4mm}
\newline
\newpage
$^{~1}${Now at FNAL, Batavia, Illinois, USA.}
\newline
$^{~2}$Jointly appointed by Nuclear and Astrophysics Laboratory,
            University of Oxford, UK.
\newline
$^{~3}${Now at Codian Ltd., Langley, Slough, UK.}
%Now at Bergische Universit\"{a}t Wuppertal, Germany.}
\newline
$^{~4}${Now at University of Glasgow, UK.}
\newline
$^{~5}${Also supported by LPNHE, Paris, France.}
\newline
%$^{~6}${Supported by the CERN Doctoral Student Programme.}
%\newline
%
$^{~6}${Now at Imperial College, University of London, UK.}
\newline
$^{~7}${ITEP, Moscow, Russian Federation.}
\newline
$^{~8}${Permanently at Instituto de F\'{\i}sica de Cantabria,
            Univ. de Cantabria, Santander, Spain.} 
\newline
$^{~9}${Now at SpinX Technologies, Geneva, Switzerland.}
\newline
$^{10}${Now at TRIUMF, Vancouver, Canada.}
\newline
$^{11}${Now at University of St. Gallen, Switzerland.}
\newline
$^{12}${On leave of absence from Ecoanalitica, Moscow State University,
Moscow, Russia.}
%short: EMSU, 119899, Moscow, Russia 
%the Budker Institute for Nuclear Physics, Novosibirsk, Russia.
\newline
$^{13}${Now at \c{C}ukurova University, Adana, Turkey.}
\newline
$^{14}${Now at III Phys. Inst. B, RWTH Aachen, Aachen, Germany.}
\newline
$^{15}$On leave of absence from INFN, Sezione di Ferrara, Italy.
\newline
$^{16}${Now at CERN, Geneva, Switzerland.}
\newline
$^{17}${Now at Univerity of Rome Tor Vergata, Italy.}
\newline
$^{18}${Now at Lawrence Berkeley National Laboratory, Berkeley, California, USA.}
\newline
$^{19}${K2K Collaboration.}
\newline
$^{20}${MiniBooNE Collaboration.}
\newline
$^{21}${Now at Section de Physique, Universit\'{e} de Gen\`{e}ve, Switzerland, Switzerland.}
\newline
$^{22}${Now at Royal Holloway, University of London, UK.}
\newline
$^{23}${Now at University of Sussex, Brighton, UK.}
\newline
$^{24}${Now at ESA/ESTEC, Noordwijk, The Netherlands.}

%% file: C5-Cu5-Sn5_table1.tex
% copper (tin) 
\begin{table}[tbp!] 
\caption{Total number of events and tracks used in the carbon, copper and tin 
  5\%~$\lambda_{\mathrm{I}}$ target data sets, and the number of
  protons on target as calculated from the pre-scaled trigger count.} 
\label{tab:events}
{\small
\begin{center}
\begin{tabular}{ l  l r  r r r} \hline
\bf{Data set}& &\bf{3 \bfGeVc}&\bf{5 \bfGeVc}&\bf{8 \bfGeVc}&\bf{12 \bfGeVc}\\ \hline
    Total DAQ events   &  (C) & 1304255 & 2648351 & 1878590 & 1875610 \\
                       &  (Cu)& 992549  & 2166883 & 2599056 & 748123  \\
                       &  (Sn)& 1636933 & 2827930 & 2780036 & 950582  \\ 
    Protons on target  &  (C) & 1107456 & 4872896 & 6143552 & 7393024 \\
     (selected min. bias$\times$64)  & (Cu) & 971840 &  3626048 & 7606272& 2990656 \\
                                     & (Sn) & 1379008&  4598848 & 8260544& 3842112 \\   
    Acc. protons with LAI        &  (C) & 56712 & 255922 & 337150 & 509713 \\
                                 &  (Cu)& 59873 & 237894 & 541852 & 226250 \\
                                 &  (Sn)& 83549 & 304949 & 600581 & 295053 \\     
    Maximum \evtspill            &  (C) & 140   & 140    &  170   & 150    \\
                                 &  (Cu)& 130   & 120    &  120   & 130    \\
                                 &  (Sn)& 110   & 110    &  120   & 110    \\    
    LAI in accepted spill part   &  (C) & 26231 & 108215 & 161331 & 217899 \\
                                 &  (Cu)& 27287 & 87974  & 175770 & 90752  \\
                                 &  (Sn)& 30029 & 98078  & 199209 & 100872 \\   
    Fraction of triggers used    &  (C) & 46 \% & 42 \%  & 48 \%  & 43 \%  \\
                                 &  (Cu)& 46 \% & 37 \%  & 32 \%  & 40 \%  \\
                                 &  (Sn)& 36 \% & 32 \%  & 33 \%  & 34 \%  \\  
    Accepted momentum            &  (C) & 32483 & 154984 & 258338 & 304993 \\
    determination                &  (Cu)& 37681 & 156847 & 374701 & 209043 \\
                                 &  (Sn)& 42949 & 188994 & 481436 & 274700 \\  
    In kinematic region and      &  (C) & 20508 & 95999  & 150444 & 173077 \\
    originating from target      &  (Cu)& 23896 & 99652  & 229002 & 122273 \\
                                 &  (Sn)& 29090 & 125864 & 305214 & 167137 \\  
    Negative particles           &  (C) & 2873 & 20328  & 38892 & 48699 \\
                                 &  (Cu)& 3016 & 18242  &  52447 & 31239 \\
                                 &  (Sn)& 3352 & 20721  &  63846 & 39395 \\    
    Positive particles           &  (C) & 17635  & 75671  & 111552  & 124378  \\
                                 &  (Cu)& 20880  & 81410  & 176555  & 91034  \\
                                 &  (Sn)& 25738  & 105143 & 241368  & 127742 \\  
    \bf{$\bfpim$ selected with PID} & (C) & 2661 & 18513 & 35115 & 42994 \\
                                    & (Cu)& 2728 & 16451 & 46820 & 27636 \\
                                    & (Sn)& 3100 & 18762 & 56697 & 34595 \\
    \bf{$\bfpip$ selected with PID} & (C) & 5554 & 28446 & 47165 & 54481 \\
                                    & (Cu)& 4403 & 22087 & 57218 & 32234  \\
                                    & (Sn)& 4439 & 23402 & 64410 & 38000 \\   
\end{tabular}
\end{center}
}
\end{table}
% counters of target.py used:
% #11
% #41 x64
% #21
% #22
% nevt
% #23
% #23/#22
% #72
% mult (mean of trk_mult1 in ..target.aida)
% #84
% #87
% #91
% #92
% #88
% #89

%% file: C5-syst-table.tex
\begin{table}[tbp!] 
\small{
\begin{center}
\caption{Contributions to the experimental uncertainties for the thin carbon
 target data. The numbers
 represent the uncertainty in percent of 
 the cross-section integrated over the angle and momentum region
 indicated.
 The overall normalization has an uncertainty
 of 2\%, and is not reported in the table.} 
\label{tab:errors-3}
\vspace{2mm}
\begin{tabular}{ l rrr | rrr | rr} \hline
\bf{Momentum range  (\MeVc)}&\multicolumn{3}{c|}{100 -- 300}
                            &\multicolumn{3}{c|}{300 -- 500}
                            &\multicolumn{2}{c}{500 -- 700} \\
\hline
\bf{Angle range: from (\rad)}&0.35--&0.95--&1.55--
                       &0.35--&0.95--&1.55--
                       &0.35--&0.95-- \\
\hspace{2.2cm} \bf{to (\rad)}      &0.95&1.55&2.15
                       &0.95&1.55&2.15
                       &0.95&1.55 \\
%3
\hline
\bf{Error source}&\multicolumn{8}{c}{\bf{3 \GeVc beam}}\\
\hline
Absorption               &   1.0 &  0.7 &  0.6 &  0.5 &  0.3 &  0.1 &  0.3 &  0.5 \\
Tertiaries               &   2.9 &  2.2 &  1.3 &  2.6 &  2.2 &  0.9 &  0.4 &  0.0 \\
Target region cut        &   2.3 &  0.6 &  0.4 &  1.2 &  0.6 &  0.6 &  1.1 &  0.3 \\
Efficiency               &   1.5 &  1.8 &  1.5 &  1.4 &  2.4 &  2.6 &  1.7 &  2.8 \\
Shape of $\pi^0$         &   8.1 &  2.1 &  0.8 &  0.2 &  0.0 &  0.0 &  0.0 &  0.0 \\
Normalization of $\pi^0$ &   1.5 &  0.4 &  0.2 &  0.1 &  0.0 &  0.0 &  0.0 &  0.0 \\
Particle ID              &   0.1 &  0.1 &  0.0 &  1.2 &  1.0 &  0.6 &  5.7 &  5.0 \\
Momentum resolution      &   2.0 &  0.3 &  1.0 &  0.2 &  0.3 &  0.5 &  0.7 &  1.9 \\
Momentum scale           &   6.7 &  2.8 &  0.5 &  1.3 &  6.5 & 10.0 &  7.4 & 15.0 \\
Angle bias               &   0.8 &  0.2 &  0.5 &  0.0 &  1.5 &  1.6 &  0.9 &  2.7 \\
\bf{Total systematics}   &  11.6 &  4.6 &  2.6 &  3.6 &  7.5 & 10.5 &  9.6 & 16.4 \\
\bf{Statistics}          &   4.5 &  3.8 &  4.9 &  3.5 &  5.9 & 16.0 &  4.7 & 13.0 \\
%5
\hline
&\multicolumn{8}{c}{\bf{5 \GeVc beam}}\\
\hline
Absorption               &   1.0 &  0.6 &  0.5 &  0.5 &  0.2 &  0.0 &  0.3 &  0.6 \\
Tertiaries               &   2.8 &  1.9 &  0.9 &  2.5 &  2.2 &  1.5 &  0.3 &  0.3 \\
Target region cut        &   1.7 &  0.9 &  0.8 &  2.1 &  0.1 &  0.4 &  1.1 &  1.0 \\
Efficiency               &   1.7 &  2.1 &  1.7 &  1.2 &  2.1 &  3.0 &  1.2 &  2.5 \\
Shape of $\pi^0$         &   6.8 &  1.4 &  0.4 &  0.2 &  0.0 &  0.1 &  0.0 &  0.0 \\
Normalization of $\pi^0$ &   2.0 &  0.7 &  0.4 &  0.1 &  0.0 &  0.0 &  0.0 &  0.0 \\
Particle ID              &   0.1 &  0.0 &  0.0 &  1.0 &  0.7 &  0.8 &  5.2 &  4.6 \\
Momentum resolution      &   2.4 &  0.0 &  0.7 &  0.0 &  0.2 &  0.6 &  0.4 &  0.5 \\
Momentum scale           &   7.2 &  3.2 &  1.4 &  1.4 &  3.4 &  6.9 &  3.3 &  9.8 \\
Angle bias               &   0.8 &  0.3 &  0.4 &  0.2 &  1.4 &  1.0 &  1.1 &  2.0 \\
\bf{Total systematics}   &  11.0 &  4.7 &  2.7 &  3.9 &  4.8 &  7.8 &  6.5 & 11.4 \\
\bf{Statistics        }  &   2.2 &  1.8 &  2.3 &  1.4 &  2.2 &  4.5 &  1.7 &  3.5 \\
%8
\hline
&\multicolumn{8}{c}{\bf{8 \GeVc beam}}\\
\hline
Absorption               &   1.0 &  0.6 &  0.4 &  0.5 &  0.2 &  0.0 &  0.3 &  0.7 \\
Tertiaries               &   2.8 &  1.8 &  0.5 &  2.5 &  2.1 &  1.4 &  0.5 &  0.4 \\
Target region cut        &   4.1 &  2.7 &  1.8 &  3.6 &  1.3 &  1.0 &  2.5 &  1.9 \\
Efficiency               &   1.5 &  2.1 &  1.4 &  1.1 &  1.8 &  2.3 &  1.2 &  2.1 \\
Shape of $\pi^0$         &   4.6 &  0.8 &  0.1 &  0.1 &  0.0 &  0.1 &  0.0 &  0.0 \\
Normalization of $\pi^0$ &   2.2 &  0.7 &  0.4 &  0.1 &  0.0 &  0.0 &  0.0 &  0.0 \\
Particle ID              &   0.0 &  0.0 &  0.1 &  1.0 &  0.6 &  0.5 &  5.2 &  4.2 \\
Momentum resolution      &   2.5 &  0.2 &  0.5 &  0.2 &  0.0 &  0.3 &  0.8 &  0.2 \\
Momentum scale           &   7.5 &  3.3 &  1.8 &  2.1 &  2.1 &  6.0 &  2.9 &  9.9 \\
Angle bias               &   0.6 &  0.4 &  0.5 &  0.4 &  1.1 &  1.3 &  0.9 &  2.0 \\
\bf{Total systematics}   &  10.8 &  5.2 &  3.1 &  5.1 &  3.9 &  6.8 &  6.7 & 11.3 \\
\bf{Statistics}          &   1.7 &  1.4 &  1.8 &  1.1 &  1.7 &  3.2 &  1.2 &  2.5 \\
%12
\hline
&\multicolumn{8}{c}{\bf{12 \GeVc beam}}\\
\hline
Absorption               &   1.0 &  0.6 &  0.3 &  0.5 &  0.2 &  0.0 &  0.2 &  0.6 \\
Tertiaries               &   1.6 &  1.2 &  0.1 &  1.8 &  1.3 &  0.7 &  0.2 &  0.8 \\
Target region cut        &   3.8 &  1.9 &  1.4 &  2.4 &  1.0 &  0.0 &  1.6 &  0.3 \\
Efficiency               &   1.4 &  1.9 &  1.4 &  1.1 &  1.7 &  2.4 &  1.0 &  2.1 \\
Shape of $\pi^0$         &   4.6 &  0.9 &  0.1 &  0.3 &  0.1 &  0.0 &  0.1 &  0.1 \\
Normalization of $\pi^0$ &   2.8 &  0.8 &  0.5 &  0.2 &  0.1 &  0.1 &  0.1 &  0.1 \\
Particle ID              &   0.1 &  0.1 &  0.1 &  1.1 &  0.7 &  0.4 &  5.2 &  4.4 \\
Momentum resolution      &   2.6 &  0.4 &  0.2 &  0.4 &  0.5 &  0.8 &  0.6 &  0.1 \\
Momentum scale           &   7.7 &  3.8 &  1.9 &  2.4 &  2.4 &  6.1 &  3.4 &  9.7 \\
Angle bias               &   0.6 &  0.4 &  0.4 &  0.4 &  1.4 &  0.9 &  1.0 &  1.9 \\
\bf{Total systematics}   &  10.7 &  5.1 &  2.9 &  4.3 &  3.8 &  6.7 &  6.6 & 11.1 \\
\bf{Statistics}          &   1.6 &  1.4 &  1.7 &  1.0 &  1.5 &  2.9 &  1.1 &  2.2 \\
\end{tabular}
\end{center}
}
\end{table}

%% file: Cu5-Sn5_syst-table.tex
% copper 
\begin{table}[tbp!] 
\small{
\begin{center}
\caption{Summary of experimental uncertainties for the copper (tin) analysis. The numbers
 represent the uncertainty in percent of 
 the cross-section integrated over the angle and momentum region indicated. 
 The overall normalization has an uncertainty
 of 2\%, and is not reported in the table.} 
\label{tab:errors-4}
\vspace{2mm}
\begin{tabular}{ l rrr | rrr | rr} \hline
\bf{p (\GeVc) }&\multicolumn{3}{c|}{0.1 -- 0.3}
                            &\multicolumn{3}{c|}{0.3 -- 0.5}
                            &\multicolumn{2}{c}{0.5 -- 0.7} \\
\hline
\bf{Angle}&350--&950--&1550--
           &350--&950--&1550--
                       &350--&950-- \\
\bf{(\mrad)}    &950&1550&2150
                       &950&1550&2150
                       &950&1550 \\
%3
\hline
\bf{3 \GeVc }&&&&&&&&\\
\hline
\bf{Total syst.}    &10.7 (13.7)& 7.8 (6.2) & 6.2(4.6) & 3.5 (3.4)& 7.3 (6.3) & 9.3 (11.7) &10.0 (7.7)&12.8 (12.4)\\
\bf{Statistics}           & 5.1 (4.9)& 4.4 (4.1)& 5.2 (5.0)& 4.0 (4.3) & 6.1 (6.5)&15.1 (14.1)& 5.3 (5.6)&12.4 (12.0)\\
%5
\hline
\bf{5 \GeVc }&&&&&&&&\\
\hline
\bf{Total syst.}    &10.5 (9.7) & 7.9 (6.0) & 6.8 (5.0) & 3.6 (3.5) & 5.3 (4.8) & 6.5 (7.0) & 7.0 (7.2) &12.6 (13.1)\\
\bf{Statistics        }   & 2.2 (2.1) & 2.0 (1.8) & 2.5 (2.2) & 1.7 (1.7) & 2.4 (2.4) & 4.4 (4.3) & 2.0 (2.1) & 3.7 (3.6) \\
%8
\hline
\bf{8 \GeVc }&&&&&&&&\\
\hline
\bf{Total syst.}    &10.4 (9.3) & 7.6 (6.4) & 6.8 (4.9) & 4.1 (3.6) & 4.3 (4.9) & 5.8 (7.0) & 6.5 (6.9) &10.2 (10.7)\\
\bf{Statistics}           & 1.4 (1.3) & 1.3 (1.2) & 1.6 (1.4) & 1.0 (1.0) & 1.4 (1.4) & 2.6 (2.5) & 1.2 (1.2) & 2.1 (2.1)\\
%12
\hline
\bf{12 \GeVc }&&&&&&&&\\
\hline
\bf{Total syst.}    &10.6 (9.9)& 7.9 (6.5) & 7.0 (5.7) & 3.6 (3.5)& 4.1 (4.6) & 6.1 (6.8) & 6.6 (6.5) &11.3 (10.7)\\
\bf{Statistics}           & 1.7 (1.6) & 1.7 (1.5) & 2.1 (1.9) & 1.3 (1.2) & 1.9 (1.8) & 3.5 (3.3) & 1.4 (1.4) & 2.7 (2.5)\\
\end{tabular}
\end{center}
}
\end{table}

%% file: C5_result-table-scl.tex
\begin{table}[hp!] 
\begin{center}
  \caption{\label{tab:xsec-p}
    HARP results for the double-differential $\pi^+$ production
    cross-section in the laboratory system,
    $d^2\sigma^{\pi^+}/(dpd\theta)$ for carbon. Each row refers to a
    different $(p_{\hbox{\small min}} \le p<p_{\hbox{\small max}},
    \theta_{\hbox{\small min}} \le \theta<\theta_{\hbox{\small max}})$ bin,
    where $p$ and $\theta$ are the pion momentum and polar angle, respectively.
    The central value as well as the square-root of the diagonal elements
    of the covariance matrix are given.}
\vspace{2mm}
%\begin{tabular}{rrrr|r@{$\pm$}lr{$\pm$}lr{$\pm$}lr{$\pm$}l} 
\begin{tabular}{rrrr|r@{$\pm$}lr@{$\pm$}lr@{$\pm$}lr@{$\pm$}l} 
\hline
$\theta_{\hbox{\small min}}$ &
$\theta_{\hbox{\small max}}$ &
$p_{\hbox{\small min}}$ &
$p_{\hbox{\small max}}$ &
\multicolumn{8}{c}{$d^2\sigma^{\pi^+}/(dpd\theta)$} 
\\
(rad) & (rad) & (\GeVc) & (\GeVc) &
\multicolumn{8}{c}{(barn/(\GeVc rad))}
\\
  &  &  & 
&\multicolumn{2}{c}{$ \bf{3 \ \GeVc}$} 
&\multicolumn{2}{c}{$ \bf{5 \ \GeVc}$} 
&\multicolumn{2}{c}{$ \bf{8 \ \GeVc}$} 
&\multicolumn{2}{c}{$ \bf{12 \ \GeVc}$} 
\\ 
\hline
 0.35 & 0.55 & 0.10 & 0.15& 0.039 &  0.032& 0.06 &  0.04& 0.11 &  0.05& 0.12 &  0.05\\ 
      &      & 0.15 & 0.20& 0.068 &  0.024& 0.116 &  0.028& 0.138 &  0.026& 0.137 &  0.028\\ 
      &      & 0.20 & 0.25& 0.116 &  0.022& 0.165 &  0.021& 0.179 &  0.019& 0.209 &  0.023\\ 
      &      & 0.25 & 0.30& 0.125 &  0.019& 0.223 &  0.023& 0.257 &  0.026& 0.265 &  0.022\\ 
      &      & 0.30 & 0.35& 0.158 &  0.019& 0.258 &  0.021& 0.286 &  0.023& 0.320 &  0.032\\ 
      &      & 0.35 & 0.40& 0.150 &  0.020& 0.267 &  0.018& 0.310 &  0.028& 0.351 &  0.016\\ 
      &      & 0.40 & 0.45& 0.198 &  0.021& 0.270 &  0.015& 0.332 &  0.019& 0.325 &  0.023\\ 
      &      & 0.45 & 0.50& 0.185 &  0.019& 0.259 &  0.013& 0.329 &  0.020& 0.368 &  0.024\\ 
      &      & 0.50 & 0.60& 0.155 &  0.016& 0.256 &  0.015& 0.332 &  0.024& 0.347 &  0.020\\ 
      &      & 0.60 & 0.70& 0.117 &  0.019& 0.239 &  0.022& 0.301 &  0.027& 0.314 &  0.031\\ 
      &      & 0.70 & 0.80& 0.072 &  0.016& 0.172 &  0.028& 0.25 &  0.04& 0.25 &  0.04\\ 
\hline
 0.55 & 0.75 & 0.10 & 0.15& 0.112 &  0.034& 0.078 &  0.027& 0.098 &  0.026& 0.086 &  0.027\\ 
      &      & 0.15 & 0.20& 0.137 &  0.022& 0.168 &  0.019& 0.172 &  0.017& 0.161 &  0.017\\ 
      &      & 0.20 & 0.25& 0.198 &  0.027& 0.228 &  0.019& 0.253 &  0.022& 0.251 &  0.024\\ 
      &      & 0.25 & 0.30& 0.231 &  0.025& 0.254 &  0.021& 0.279 &  0.022& 0.264 &  0.017\\ 
      &      & 0.30 & 0.35& 0.230 &  0.022& 0.252 &  0.017& 0.286 &  0.022& 0.315 &  0.029\\ 
      &      & 0.35 & 0.40& 0.198 &  0.019& 0.268 &  0.019& 0.284 &  0.020& 0.304 &  0.014\\ 
      &      & 0.40 & 0.45& 0.185 &  0.017& 0.230 &  0.013& 0.261 &  0.016& 0.287 &  0.013\\ 
      &      & 0.45 & 0.50& 0.182 &  0.018& 0.198 &  0.011& 0.247 &  0.015& 0.270 &  0.013\\ 
      &      & 0.50 & 0.60& 0.118 &  0.017& 0.156 &  0.012& 0.208 &  0.013& 0.226 &  0.014\\ 
      &      & 0.60 & 0.70& 0.066 &  0.012& 0.109 &  0.013& 0.159 &  0.018& 0.177 &  0.018\\ 
      &      & 0.70 & 0.80& 0.036 &  0.009& 0.069 &  0.012& 0.112 &  0.021& 0.119 &  0.021\\ 
\hline
 0.75 & 0.95 & 0.10 & 0.15& 0.109 &  0.023& 0.099 &  0.020& 0.114 &  0.018& 0.106 &  0.019\\ 
      &      & 0.15 & 0.20& 0.168 &  0.024& 0.221 &  0.019& 0.207 &  0.018& 0.204 &  0.017\\ 
      &      & 0.20 & 0.25& 0.200 &  0.023& 0.235 &  0.017& 0.239 &  0.015& 0.268 &  0.019\\ 
      &      & 0.25 & 0.30& 0.189 &  0.020& 0.219 &  0.014& 0.244 &  0.018& 0.250 &  0.018\\ 
      &      & 0.30 & 0.35& 0.202 &  0.021& 0.201 &  0.014& 0.239 &  0.014& 0.238 &  0.013\\ 
      &      & 0.35 & 0.40& 0.156 &  0.018& 0.166 &  0.009& 0.204 &  0.012& 0.233 &  0.012\\ 
      &      & 0.40 & 0.45& 0.105 &  0.013& 0.154 &  0.008& 0.175 &  0.009& 0.210 &  0.010\\ 
      &      & 0.45 & 0.50& 0.075 &  0.010& 0.135 &  0.008& 0.148 &  0.008& 0.170 &  0.010\\ 
      &      & 0.50 & 0.60& 0.050 &  0.008& 0.094 &  0.010& 0.114 &  0.009& 0.126 &  0.010\\ 
      &      & 0.60 & 0.70& 0.028 &  0.006& 0.052 &  0.009& 0.073 &  0.010& 0.073 &  0.012\\ 
%      &      & 0.70 & 0.80& 0.019 &  0.005& 0.031 &  0.007& 0.045 &  0.010& 0.043 &  0.010\\ 
\hline
 0.95 & 1.15 & 0.10 & 0.15& 0.130 &  0.025& 0.127 &  0.019& 0.126 &  0.018& 0.126 &  0.017\\ 
      &      & 0.15 & 0.20& 0.216 &  0.025& 0.194 &  0.015& 0.208 &  0.015& 0.220 &  0.019\\ 
      &      & 0.20 & 0.25& 0.197 &  0.021& 0.201 &  0.012& 0.236 &  0.018& 0.225 &  0.012\\ 
      &      & 0.25 & 0.30& 0.154 &  0.018& 0.168 &  0.011& 0.203 &  0.014& 0.205 &  0.011\\ 
      &      & 0.30 & 0.35& 0.110 &  0.014& 0.149 &  0.009& 0.159 &  0.009& 0.164 &  0.008\\ 
      &      & 0.35 & 0.40& 0.078 &  0.010& 0.122 &  0.008& 0.133 &  0.007& 0.146 &  0.007\\ 
      &      & 0.40 & 0.45& 0.059 &  0.010& 0.089 &  0.007& 0.107 &  0.006& 0.117 &  0.007\\ 
      &      & 0.45 & 0.50& 0.033 &  0.008& 0.068 &  0.007& 0.081 &  0.006& 0.090 &  0.007\\ 
      &      & 0.50 & 0.60& 0.016 &  0.004& 0.043 &  0.005& 0.055 &  0.006& 0.054 &  0.006\\ 
%      &      & 0.60 & 0.70& 0.006 &  0.002& 0.021 &  0.004& 0.032 &  0.005& 0.028 &  0.005\\ 
%      &      & 0.70 & 0.80& 0.003 &  0.002& 0.012 &  0.003& 0.016 &  0.004& 0.018 &  0.004\\ 
\hline
\end{tabular}
\end{center}
\end{table}

\begin{table}[hp!] 
\begin{center}
\begin{tabular}{rrrr|r@{$\pm$}lr@{$\pm$}lr@{$\pm$}lr@{$\pm$}l} 
\hline
$\theta_{\hbox{\small min}}$ &
$\theta_{\hbox{\small max}}$ &
$p_{\hbox{\small min}}$ &
$p_{\hbox{\small max}}$ &
\multicolumn{8}{c}{$d^2\sigma^{\pi^+}/(dpd\theta)$} 
\\
(rad) & (rad) & (\GeVc) & (\GeVc) &
\multicolumn{8}{c}{(barn/(\GeVc rad))}
\\
  &  &  & 
&\multicolumn{2}{c}{$ \bf{3 \ \GeVc}$} 
&\multicolumn{2}{c}{$ \bf{5 \ \GeVc}$} 
&\multicolumn{2}{c}{$ \bf{8 \ \GeVc}$} 
&\multicolumn{2}{c}{$ \bf{12 \ \GeVc}$} 
\\ 
\hline
 1.15 & 1.35 & 0.10 & 0.15& 0.127 &  0.023& 0.143 &  0.018& 0.143 &  0.021& 0.132 &  0.018\\ 
      &      & 0.15 & 0.20& 0.184 &  0.022& 0.197 &  0.015& 0.200 &  0.012& 0.209 &  0.014\\ 
      &      & 0.20 & 0.25& 0.180 &  0.020& 0.177 &  0.011& 0.203 &  0.011& 0.177 &  0.010\\ 
      &      & 0.25 & 0.30& 0.102 &  0.013& 0.128 &  0.009& 0.139 &  0.010& 0.154 &  0.009\\ 
      &      & 0.30 & 0.35& 0.084 &  0.012& 0.106 &  0.007& 0.111 &  0.007& 0.108 &  0.006\\ 
      &      & 0.35 & 0.40& 0.047 &  0.008& 0.071 &  0.006& 0.086 &  0.006& 0.085 &  0.005\\ 
      &      & 0.40 & 0.45& 0.028 &  0.006& 0.044 &  0.006& 0.061 &  0.005& 0.065 &  0.004\\ 
      &      & 0.45 & 0.50& 0.021 &  0.005& 0.028 &  0.004& 0.045 &  0.004& 0.048 &  0.005\\ 
%      &      & 0.50 & 0.60& 0.009 &  0.003& 0.016 &  0.002& 0.026 &  0.004& 0.026 &  0.004\\ 
%      &      & 0.60 & 0.70& 0.004 &  0.002& 0.008 &  0.002& 0.012 &  0.003& 0.013 &  0.002\\ 
%      &      & 0.70 & 0.80& 0.002 &  0.002& 0.004 &  0.002& 0.007 &  0.002& 0.007 &  0.002\\ 
\hline
 1.35 & 1.55 & 0.10 & 0.15& 0.153 &  0.024& 0.135 &  0.018& 0.129 &  0.016& 0.134 &  0.017\\ 
      &      & 0.15 & 0.20& 0.170 &  0.022& 0.179 &  0.013& 0.184 &  0.014& 0.179 &  0.013\\ 
      &      & 0.20 & 0.25& 0.149 &  0.019& 0.145 &  0.010& 0.144 &  0.010& 0.141 &  0.008\\ 
      &      & 0.25 & 0.30& 0.095 &  0.013& 0.114 &  0.010& 0.100 &  0.007& 0.123 &  0.008\\ 
      &      & 0.30 & 0.35& 0.045 &  0.011& 0.066 &  0.007& 0.075 &  0.005& 0.085 &  0.006\\ 
      &      & 0.35 & 0.40& 0.022 &  0.005& 0.040 &  0.004& 0.055 &  0.004& 0.054 &  0.005\\ 
      &      & 0.40 & 0.45& 0.022 &  0.008& 0.027 &  0.003& 0.039 &  0.004& 0.035 &  0.004\\ 
      &      & 0.45 & 0.50& 0.009 &  0.004& 0.017 &  0.003& 0.024 &  0.003& 0.023 &  0.003\\ 
%      &      & 0.50 & 0.60& 0.005 &  0.002& 0.009 &  0.002& 0.013 &  0.002& 0.012 &  0.002\\ 
%      &      & 0.60 & 0.70& 0.002 &  0.001& 0.005 &  0.001& 0.006 &  0.001& 0.005 &  0.001\\ 
%      &      & 0.70 & 0.80& 0.002 &  0.002& 0.003 &  0.001& 0.004 &  0.001& 0.004 &  0.001\\ 
\hline
 1.55 & 1.75 & 0.10 & 0.15& 0.137 &  0.022& 0.124 &  0.016& 0.123 &  0.016& 0.125 &  0.015\\ 
      &      & 0.15 & 0.20& 0.182 &  0.023& 0.146 &  0.011& 0.157 &  0.011& 0.166 &  0.010\\ 
      &      & 0.20 & 0.25& 0.088 &  0.019& 0.105 &  0.008& 0.120 &  0.008& 0.111 &  0.008\\ 
      &      & 0.25 & 0.30& 0.056 &  0.010& 0.073 &  0.008& 0.070 &  0.007& 0.078 &  0.006\\ 
      &      & 0.30 & 0.35& 0.033 &  0.008& 0.044 &  0.004& 0.045 &  0.004& 0.057 &  0.004\\ 
      &      & 0.35 & 0.40& 0.015 &  0.004& 0.026 &  0.004& 0.031 &  0.003& 0.045 &  0.005\\ 
      &      & 0.40 & 0.45& 0.010 &  0.003& 0.015 &  0.003& 0.022 &  0.003& 0.024 &  0.004\\ 
      &      & 0.45 & 0.50& 0.006 &  0.003& 0.008 &  0.002& 0.012 &  0.002& 0.012 &  0.002\\ 
%      &      & 0.50 & 0.60& 0.002 &  0.002& 0.004 &  0.001& 0.004 &  0.001& 0.006 &  0.001\\ 
%      &      & 0.60 & 0.70& 0.001 &  0.001& 0.002 &  0.001& 0.002 &  0.001& 0.003 &  0.001\\ 
%      &      & 0.70 & 0.80& 0.001 &  0.001& 0.001 &  0.000& 0.001 &  0.000& 0.002 &  0.001\\ 
\hline
 1.75 & 1.95 & 0.10 & 0.15& 0.142 &  0.023& 0.120 &  0.014& 0.111 &  0.015& 0.114 &  0.015\\ 
      &      & 0.15 & 0.20& 0.114 &  0.016& 0.131 &  0.009& 0.141 &  0.009& 0.131 &  0.008\\ 
      &      & 0.20 & 0.25& 0.093 &  0.014& 0.088 &  0.008& 0.088 &  0.008& 0.094 &  0.006\\ 
      &      & 0.25 & 0.30& 0.035 &  0.010& 0.045 &  0.005& 0.052 &  0.004& 0.053 &  0.006\\ 
      &      & 0.30 & 0.35& 0.013 &  0.005& 0.026 &  0.004& 0.034 &  0.004& 0.039 &  0.003\\ 
      &      & 0.35 & 0.40& 0.005 &  0.003& 0.014 &  0.003& 0.018 &  0.003& 0.025 &  0.004\\ 
      &      & 0.40 & 0.45& 0.003 &  0.002& 0.006 &  0.002& 0.010 &  0.002& 0.012 &  0.003\\ 
      &      & 0.45 & 0.50& 0.002 &  0.002& 0.003 &  0.001& 0.006 &  0.001& 0.006 &  0.002\\ 
%      &      & 0.50 & 0.60& 0.001 &  0.001& 0.001 &  0.001& 0.003 &  0.001& 0.002 &  0.001\\ 
%      &      & 0.60 & 0.70& 0.000 &  0.001& 0.001 &  0.000& 0.001 &  0.000& 0.001 &  0.000\\ 
%      &      & 0.70 & 0.80& 0.000 &  0.001& 0.000 &  0.000& 0.000 &  0.000& 0.001 &  0.000\\ 
\hline
 1.95 & 2.15 & 0.10 & 0.15& 0.073 &  0.015& 0.087 &  0.012& 0.110 &  0.013& 0.095 &  0.012\\ 
      &      & 0.15 & 0.20& 0.096 &  0.016& 0.094 &  0.008& 0.106 &  0.008& 0.108 &  0.008\\ 
      &      & 0.20 & 0.25& 0.059 &  0.012& 0.059 &  0.006& 0.057 &  0.005& 0.063 &  0.006\\ 
      &      & 0.25 & 0.30& 0.026 &  0.009& 0.027 &  0.004& 0.036 &  0.004& 0.032 &  0.004\\ 
      &      & 0.30 & 0.35& 0.012 &  0.005& 0.017 &  0.002& 0.023 &  0.002& 0.016 &  0.002\\ 
      &      & 0.35 & 0.40& 0.004 &  0.003& 0.014 &  0.002& 0.014 &  0.002& 0.011 &  0.002\\ 
      &      & 0.40 & 0.45& 0.003 &  0.003& 0.005 &  0.002& 0.006 &  0.002& 0.007 &  0.001\\ 
      &      & 0.45 & 0.50& 0.002 &  0.002& 0.002 &  0.001& 0.002 &  0.001& 0.004 &  0.001\\ 
%      &      & 0.50 & 0.60& 0.000 &  0.001& 0.001 &  0.000& 0.001 &  0.000& 0.001 &  0.000\\ 
%      &      & 0.60 & 0.70& 0.000 &  0.001& 0.000 &  0.000& 0.000 &  0.000& 0.000 &  0.000\\ 
%      &      & 0.70 & 0.80& 0.000 &  0.001& 0.000 &  0.000& 0.000 &  0.000& 0.000 &  0.000\\ 
 
\hline

\end{tabular}
\end{center}
\end{table}

\begin{table}[hp!] 
\begin{center}
  \caption{\label{tab:xsec-n}
    HARP results for the double-differential $\pi^-$ production
    cross-section in the laboratory system,
    $d^2\sigma^{\pi^-}/(dpd\theta)$ for carbon. Each row refers to a
    different $(p_{\hbox{\small min}} \le p<p_{\hbox{\small max}},
    \theta_{\hbox{\small min}} \le \theta<\theta_{\hbox{\small max}})$ bin,
    where $p$ and $\theta$ are the pion momentum and polar angle, respectively.
    The central value as well as the square-root of the diagonal elements
    of the covariance matrix are given.}
\vspace{2mm}
\begin{tabular}{rrrr|r@{$\pm$}lr@{$\pm$}lr@{$\pm$}lr@{$\pm$}l} 
\hline
$\theta_{\hbox{\small min}}$ &
$\theta_{\hbox{\small max}}$ &
$p_{\hbox{\small min}}$ &
$p_{\hbox{\small max}}$ &
\multicolumn{8}{c}{$d^2\sigma^{\pi^-}/(dpd\theta)$} 
\\
(rad) & (rad) & (\GeVc) & (\GeVc) &
\multicolumn{8}{c}{(barn/(\GeVc rad))}
\\
  &  &  & 
&\multicolumn{2}{c}{$ \bf{3 \ \GeVc}$} 
&\multicolumn{2}{c}{$ \bf{5 \ \GeVc}$} 
&\multicolumn{2}{c}{$ \bf{8 \ \GeVc}$} 
&\multicolumn{2}{c}{$ \bf{12 \ \GeVc}$} 
\\ 
\hline
 0.35 & 0.55 & 0.10 & 0.15& 0.015 &  0.020& 0.08 &  0.04& 0.13 &  0.06& 0.10 &  0.05\\ 
      &      & 0.15 & 0.20& 0.036 &  0.023& 0.076 &  0.025& 0.113 &  0.025& 0.115 &  0.028\\ 
      &      & 0.20 & 0.25& 0.053 &  0.020& 0.109 &  0.018& 0.162 &  0.023& 0.172 &  0.029\\ 
      &      & 0.25 & 0.30& 0.066 &  0.015& 0.130 &  0.016& 0.179 &  0.017& 0.220 &  0.020\\ 
      &      & 0.30 & 0.35& 0.066 &  0.014& 0.133 &  0.013& 0.204 &  0.019& 0.229 &  0.016\\ 
      &      & 0.35 & 0.40& 0.069 &  0.014& 0.146 &  0.013& 0.193 &  0.013& 0.216 &  0.012\\ 
      &      & 0.40 & 0.45& 0.082 &  0.012& 0.126 &  0.009& 0.193 &  0.014& 0.212 &  0.014\\ 
      &      & 0.45 & 0.50& 0.078 &  0.012& 0.113 &  0.007& 0.177 &  0.011& 0.223 &  0.012\\ 
      &      & 0.50 & 0.60& 0.045 &  0.008& 0.120 &  0.008& 0.186 &  0.012& 0.222 &  0.012\\ 
      &      & 0.60 & 0.70& 0.050 &  0.009& 0.118 &  0.010& 0.176 &  0.015& 0.212 &  0.017\\ 
      &      & 0.70 & 0.80& 0.052 &  0.010& 0.100 &  0.011& 0.152 &  0.017& 0.194 &  0.020\\ 
\hline
 0.55 & 0.75 & 0.10 & 0.15& 0.040 &  0.026& 0.049 &  0.022& 0.053 &  0.024& 0.090 &  0.030\\ 
      &      & 0.15 & 0.20& 0.070 &  0.018& 0.095 &  0.019& 0.140 &  0.019& 0.143 &  0.017\\ 
      &      & 0.20 & 0.25& 0.089 &  0.017& 0.136 &  0.014& 0.166 &  0.017& 0.177 &  0.015\\ 
      &      & 0.25 & 0.30& 0.072 &  0.013& 0.147 &  0.012& 0.187 &  0.013& 0.211 &  0.017\\ 
      &      & 0.30 & 0.35& 0.098 &  0.016& 0.128 &  0.009& 0.166 &  0.012& 0.206 &  0.012\\ 
      &      & 0.35 & 0.40& 0.081 &  0.013& 0.131 &  0.011& 0.170 &  0.010& 0.184 &  0.009\\ 
      &      & 0.40 & 0.45& 0.077 &  0.011& 0.111 &  0.009& 0.165 &  0.009& 0.181 &  0.009\\ 
      &      & 0.45 & 0.50& 0.072 &  0.010& 0.089 &  0.006& 0.158 &  0.009& 0.178 &  0.009\\ 
      &      & 0.50 & 0.60& 0.049 &  0.008& 0.097 &  0.006& 0.145 &  0.009& 0.164 &  0.009\\ 
      &      & 0.60 & 0.70& 0.035 &  0.007& 0.081 &  0.008& 0.119 &  0.010& 0.133 &  0.012\\ 
      &      & 0.70 & 0.80& 0.029 &  0.007& 0.057 &  0.010& 0.099 &  0.013& 0.104 &  0.014\\ 
\hline
 0.75 & 0.95 & 0.10 & 0.15& 0.048 &  0.018& 0.063 &  0.016& 0.076 &  0.016& 0.087 &  0.017\\ 
      &      & 0.15 & 0.20& 0.064 &  0.015& 0.124 &  0.014& 0.175 &  0.018& 0.187 &  0.019\\ 
      &      & 0.20 & 0.25& 0.065 &  0.014& 0.125 &  0.011& 0.184 &  0.015& 0.187 &  0.012\\ 
      &      & 0.25 & 0.30& 0.087 &  0.014& 0.124 &  0.011& 0.177 &  0.011& 0.180 &  0.013\\ 
      &      & 0.30 & 0.35& 0.062 &  0.010& 0.119 &  0.009& 0.148 &  0.009& 0.173 &  0.009\\ 
      &      & 0.35 & 0.40& 0.065 &  0.011& 0.110 &  0.008& 0.141 &  0.010& 0.142 &  0.007\\ 
      &      & 0.40 & 0.45& 0.049 &  0.007& 0.093 &  0.007& 0.128 &  0.007& 0.132 &  0.006\\ 
      &      & 0.45 & 0.50& 0.049 &  0.008& 0.070 &  0.006& 0.107 &  0.007& 0.119 &  0.005\\ 
      &      & 0.50 & 0.60& 0.038 &  0.007& 0.066 &  0.004& 0.087 &  0.005& 0.101 &  0.006\\ 
      &      & 0.60 & 0.70& 0.018 &  0.006& 0.055 &  0.006& 0.074 &  0.007& 0.074 &  0.007\\ 
%      &      & 0.70 & 0.80& 0.011 &  0.004& 0.039 &  0.007& 0.056 &  0.008& 0.059 &  0.008\\ 
\hline
 0.95 & 1.15 & 0.10 & 0.15& 0.038 &  0.013& 0.068 &  0.012& 0.094 &  0.016& 0.101 &  0.013\\ 
      &      & 0.15 & 0.20& 0.071 &  0.014& 0.129 &  0.011& 0.141 &  0.010& 0.138 &  0.014\\ 
      &      & 0.20 & 0.25& 0.074 &  0.013& 0.130 &  0.011& 0.137 &  0.012& 0.167 &  0.012\\ 
      &      & 0.25 & 0.30& 0.094 &  0.014& 0.116 &  0.008& 0.134 &  0.009& 0.152 &  0.009\\ 
      &      & 0.30 & 0.35& 0.057 &  0.011& 0.093 &  0.006& 0.126 &  0.010& 0.129 &  0.007\\ 
      &      & 0.35 & 0.40& 0.042 &  0.007& 0.079 &  0.005& 0.095 &  0.006& 0.099 &  0.006\\ 
      &      & 0.40 & 0.45& 0.035 &  0.007& 0.065 &  0.004& 0.078 &  0.004& 0.084 &  0.004\\ 
      &      & 0.45 & 0.50& 0.026 &  0.006& 0.053 &  0.004& 0.069 &  0.004& 0.077 &  0.004\\ 
      &      & 0.50 & 0.60& 0.014 &  0.004& 0.039 &  0.004& 0.051 &  0.004& 0.061 &  0.004\\ 
%      &      & 0.60 & 0.70& 0.009 &  0.003& 0.023 &  0.004& 0.034 &  0.004& 0.042 &  0.005\\ 
%      &      & 0.70 & 0.80& 0.008 &  0.003& 0.017 &  0.003& 0.026 &  0.004& 0.028 &  0.005\\ 
\hline
\end{tabular}
\end{center}
\end{table}

\begin{table}[hp!] 
\begin{center}
\begin{tabular}{rrrr|r@{$\pm$}lr@{$\pm$}lr@{$\pm$}lr@{$\pm$}l} 
\hline
$\theta_{\hbox{\small min}}$ &
$\theta_{\hbox{\small max}}$ &
$p_{\hbox{\small min}}$ &
$p_{\hbox{\small max}}$ &
\multicolumn{8}{c}{$d^2\sigma^{\pi^-}/(dpd\theta)$} 
\\
(rad) & (rad) & (\GeVc) & (\GeVc) &
\multicolumn{8}{c}{(barn/(\GeVc rad))}
\\
  &  &  & 
&\multicolumn{2}{c}{$ \bf{3 \ \GeVc}$} 
&\multicolumn{2}{c}{$ \bf{5 \ \GeVc}$} 
&\multicolumn{2}{c}{$ \bf{8 \ \GeVc}$} 
&\multicolumn{2}{c}{$ \bf{12 \ \GeVc}$} 
\\ 
\hline
 1.15 & 1.35 & 0.10 & 0.15& 0.061 &  0.015& 0.070 &  0.010& 0.103 &  0.012& 0.106 &  0.013\\ 
      &      & 0.15 & 0.20& 0.124 &  0.019& 0.125 &  0.013& 0.139 &  0.010& 0.158 &  0.011\\ 
      &      & 0.20 & 0.25& 0.070 &  0.012& 0.121 &  0.009& 0.135 &  0.010& 0.135 &  0.008\\ 
      &      & 0.25 & 0.30& 0.056 &  0.011& 0.089 &  0.007& 0.111 &  0.008& 0.119 &  0.007\\ 
      &      & 0.30 & 0.35& 0.036 &  0.010& 0.073 &  0.006& 0.082 &  0.005& 0.099 &  0.005\\ 
      &      & 0.35 & 0.40& 0.016 &  0.004& 0.057 &  0.005& 0.068 &  0.004& 0.078 &  0.004\\ 
      &      & 0.40 & 0.45& 0.013 &  0.004& 0.037 &  0.004& 0.056 &  0.004& 0.059 &  0.004\\ 
      &      & 0.45 & 0.50& 0.009 &  0.003& 0.027 &  0.003& 0.043 &  0.004& 0.050 &  0.004\\ 
%      &      & 0.50 & 0.60& 0.007 &  0.003& 0.020 &  0.002& 0.028 &  0.003& 0.034 &  0.003\\ 
%      &      & 0.60 & 0.70& 0.005 &  0.002& 0.014 &  0.002& 0.017 &  0.003& 0.022 &  0.003\\ 
%      &      & 0.70 & 0.80& 0.004 &  0.002& 0.011 &  0.002& 0.012 &  0.002& 0.016 &  0.003\\ 
\hline
 1.35 & 1.55 & 0.10 & 0.15& 0.046 &  0.013& 0.071 &  0.009& 0.095 &  0.011& 0.108 &  0.013\\ 
      &      & 0.15 & 0.20& 0.093 &  0.016& 0.105 &  0.010& 0.129 &  0.012& 0.134 &  0.010\\ 
      &      & 0.20 & 0.25& 0.051 &  0.011& 0.093 &  0.008& 0.110 &  0.007& 0.116 &  0.007\\ 
      &      & 0.25 & 0.30& 0.052 &  0.010& 0.078 &  0.007& 0.087 &  0.006& 0.079 &  0.005\\ 
      &      & 0.30 & 0.35& 0.033 &  0.007& 0.045 &  0.005& 0.054 &  0.005& 0.065 &  0.004\\ 
      &      & 0.35 & 0.40& 0.026 &  0.006& 0.032 &  0.003& 0.046 &  0.003& 0.048 &  0.004\\ 
      &      & 0.40 & 0.45& 0.024 &  0.006& 0.024 &  0.003& 0.037 &  0.003& 0.032 &  0.003\\ 
      &      & 0.45 & 0.50& 0.015 &  0.005& 0.018 &  0.002& 0.026 &  0.003& 0.026 &  0.002\\ 
%      &      & 0.50 & 0.60& 0.007 &  0.003& 0.011 &  0.002& 0.017 &  0.002& 0.019 &  0.002\\ 
%      &      & 0.60 & 0.70& 0.003 &  0.002& 0.006 &  0.001& 0.010 &  0.002& 0.012 &  0.002\\ 
%      &      & 0.70 & 0.80& 0.002 &  0.002& 0.005 &  0.001& 0.007 &  0.001& 0.009 &  0.002\\ 
\hline
 1.55 & 1.75 & 0.10 & 0.15& 0.077 &  0.016& 0.075 &  0.012& 0.095 &  0.012& 0.090 &  0.011\\ 
      &      & 0.15 & 0.20& 0.094 &  0.016& 0.117 &  0.010& 0.117 &  0.008& 0.127 &  0.009\\ 
      &      & 0.20 & 0.25& 0.045 &  0.010& 0.065 &  0.006& 0.077 &  0.005& 0.087 &  0.007\\ 
      &      & 0.25 & 0.30& 0.022 &  0.007& 0.057 &  0.006& 0.061 &  0.005& 0.061 &  0.005\\ 
      &      & 0.30 & 0.35& 0.010 &  0.004& 0.036 &  0.004& 0.048 &  0.004& 0.043 &  0.004\\ 
      &      & 0.35 & 0.40& 0.006 &  0.003& 0.028 &  0.003& 0.038 &  0.003& 0.032 &  0.003\\ 
      &      & 0.40 & 0.45& 0.008 &  0.004& 0.018 &  0.003& 0.026 &  0.003& 0.024 &  0.002\\ 
      &      & 0.45 & 0.50& 0.007 &  0.004& 0.012 &  0.002& 0.017 &  0.003& 0.017 &  0.002\\ 
%      &      & 0.50 & 0.60& 0.003 &  0.002& 0.008 &  0.001& 0.009 &  0.002& 0.011 &  0.002\\ 
%      &      & 0.60 & 0.70& 0.001 &  0.002& 0.005 &  0.001& 0.004 &  0.001& 0.007 &  0.001\\ 
%      &      & 0.70 & 0.80& 0.000 &  0.001& 0.003 &  0.001& 0.002 &  0.001& 0.005 &  0.001\\ 
\hline
 1.75 & 1.95 & 0.10 & 0.15& 0.077 &  0.016& 0.074 &  0.010& 0.087 &  0.011& 0.069 &  0.008\\ 
      &      & 0.15 & 0.20& 0.082 &  0.015& 0.095 &  0.008& 0.101 &  0.007& 0.103 &  0.008\\ 
      &      & 0.20 & 0.25& 0.025 &  0.007& 0.067 &  0.007& 0.069 &  0.006& 0.070 &  0.005\\ 
      &      & 0.25 & 0.30& 0.021 &  0.007& 0.040 &  0.005& 0.046 &  0.005& 0.051 &  0.004\\ 
      &      & 0.30 & 0.35& 0.013 &  0.005& 0.024 &  0.003& 0.026 &  0.003& 0.032 &  0.003\\ 
      &      & 0.35 & 0.40& 0.011 &  0.006& 0.016 &  0.002& 0.020 &  0.002& 0.022 &  0.002\\ 
      &      & 0.40 & 0.45& 0.003 &  0.003& 0.011 &  0.002& 0.018 &  0.002& 0.017 &  0.002\\ 
      &      & 0.45 & 0.50& 0.001 &  0.001& 0.007 &  0.002& 0.013 &  0.002& 0.011 &  0.002\\ 
%      &      & 0.50 & 0.60& 0.000 &  0.001& 0.003 &  0.001& 0.007 &  0.001& 0.006 &  0.001\\ 
%      &      & 0.60 & 0.70& 0.000 &  0.001& 0.002 &  0.001& 0.002 &  0.001& 0.003 &  0.001\\ 
%      &      & 0.70 & 0.80& 0.000 &  0.001& 0.001 &  0.001& 0.001 &  0.001& 0.002 &  0.001\\ 
\hline
 1.95 & 2.15 & 0.10 & 0.15& 0.040 &  0.010& 0.058 &  0.008& 0.066 &  0.007& 0.069 &  0.007\\ 
      &      & 0.15 & 0.20& 0.071 &  0.014& 0.081 &  0.009& 0.081 &  0.008& 0.080 &  0.007\\ 
      &      & 0.20 & 0.25& 0.048 &  0.012& 0.044 &  0.005& 0.064 &  0.005& 0.062 &  0.005\\ 
      &      & 0.25 & 0.30& 0.012 &  0.006& 0.025 &  0.004& 0.038 &  0.004& 0.038 &  0.003\\ 
      &      & 0.30 & 0.35& 0.003 &  0.002& 0.010 &  0.002& 0.025 &  0.003& 0.022 &  0.003\\ 
      &      & 0.35 & 0.40& 0.002 &  0.002& 0.007 &  0.001& 0.014 &  0.002& 0.015 &  0.002\\ 
      &      & 0.40 & 0.45& 0.002 &  0.002& 0.006 &  0.001& 0.008 &  0.002& 0.012 &  0.002\\ 
      &      & 0.45 & 0.50& 0.003 &  0.003& 0.004 &  0.001& 0.005 &  0.001& 0.007 &  0.001\\ 
%      &      & 0.50 & 0.60& 0.002 &  0.003& 0.003 &  0.001& 0.003 &  0.001& 0.004 &  0.001\\ 
%      &      & 0.60 & 0.70& 0.001 &  0.002& 0.001 &  0.001& 0.002 &  0.001& 0.001 &  0.000\\ 
%      &      & 0.70 & 0.80& 0.000 &  0.001& 0.001 &  0.001& 0.001 &  0.000& 0.001 &  0.000\\ 
\hline
\end{tabular}
\end{center}
\end{table}
%

%% file: Cu5_result-table-scl.tex
% copper
\begin{table}[hp!] 
\begin{center}
  \caption{\label{tab:xsec-p:cu}
    HARP results for the double-differential $\pi^+$ production
    cross-section in the laboratory system,
    $d^2\sigma^{\pi^+}/(dpd\theta)$ for copper. Each row refers to a
    different $(p_{\hbox{\small min}} \le p<p_{\hbox{\small max}},
    \theta_{\hbox{\small min}} \le \theta<\theta_{\hbox{\small max}})$ bin,
    where $p$ and $\theta$ are the pion momentum and polar angle, respectively.
    The central value as well as the square-root of the diagonal elements
    of the covariance matrix are given.}
\vspace{2mm}
%\begin{tabular}{rrrr|r@{$\pm$}lr{$\pm$}lr{$\pm$}lr{$\pm$}l} 
\begin{tabular}{rrrr|r@{$\pm$}lr@{$\pm$}lr@{$\pm$}lr@{$\pm$}l} 
\hline
$\theta_{\hbox{\small min}}$ &
$\theta_{\hbox{\small max}}$ &
$p_{\hbox{\small min}}$ &
$p_{\hbox{\small max}}$ &
\multicolumn{8}{c}{$d^2\sigma^{\pi^+}/(dpd\theta)$} 
\\
(rad) & (rad) & (\GeVc) & (\GeVc) &
\multicolumn{8}{c}{(barn/(\GeVc rad))}
\\
  &  &  & 
&\multicolumn{2}{c}{$ \bf{3 \ \GeVc}$} 
&\multicolumn{2}{c}{$ \bf{5 \ \GeVc}$} 
&\multicolumn{2}{c}{$ \bf{8 \ \GeVc}$} 
&\multicolumn{2}{c}{$ \bf{12 \ \GeVc}$} 
\\ 
\hline
 0.35 & 0.55 & 0.10 & 0.15& 0.31 &  0.14& 0.41 &  0.17& 0.46 &  0.20& 0.69 &  0.29\\ 
      &      & 0.15 & 0.20& 0.17 &  0.05& 0.41 &  0.09& 0.63 &  0.11& 0.84 &  0.14\\ 
      &      & 0.20 & 0.25& 0.22 &  0.07& 0.61 &  0.06& 0.82 &  0.07& 0.83 &  0.07\\ 
      &      & 0.25 & 0.30& 0.52 &  0.08& 0.75 &  0.07& 0.96 &  0.07& 1.13 &  0.10\\ 
      &      & 0.30 & 0.35& 0.43 &  0.06& 0.79 &  0.06& 1.08 &  0.09& 1.22 &  0.10\\ 
      &      & 0.35 & 0.40& 0.42 &  0.05& 0.75 &  0.05& 1.13 &  0.06& 1.41 &  0.09\\ 
      &      & 0.40 & 0.45& 0.33 &  0.04& 0.77 &  0.05& 1.09 &  0.06& 1.19 &  0.08\\ 
      &      & 0.45 & 0.50& 0.41 &  0.06& 0.70 &  0.05& 1.07 &  0.06& 1.17 &  0.09\\ 
      &      & 0.50 & 0.60& 0.44 &  0.05& 0.69 &  0.04& 1.08 &  0.06& 1.16 &  0.08\\ 
      &      & 0.60 & 0.70& 0.25 &  0.05& 0.56 &  0.06& 0.92 &  0.09& 1.18 &  0.11\\ 
      &      & 0.70 & 0.80& 0.13 &  0.03& 0.37 &  0.07& 0.72 &  0.10& 0.92 &  0.14\\ 
\hline
 0.55 & 0.75 & 0.10 & 0.15& 0.34 &  0.11& 0.37 &  0.11& 0.48 &  0.14& 0.53 &  0.15\\ 
      &      & 0.15 & 0.20& 0.48 &  0.07& 0.64 &  0.06& 0.82 &  0.08& 0.80 &  0.08\\ 
      &      & 0.20 & 0.25& 0.44 &  0.06& 0.76 &  0.06& 1.04 &  0.09& 1.18 &  0.11\\ 
      &      & 0.25 & 0.30& 0.50 &  0.06& 0.82 &  0.08& 1.01 &  0.06& 1.29 &  0.09\\ 
      &      & 0.30 & 0.35& 0.51 &  0.06& 0.89 &  0.06& 1.06 &  0.08& 1.12 &  0.07\\ 
      &      & 0.35 & 0.40& 0.47 &  0.05& 0.75 &  0.04& 1.06 &  0.06& 1.19 &  0.08\\ 
      &      & 0.40 & 0.45& 0.38 &  0.04& 0.64 &  0.04& 0.97 &  0.05& 1.12 &  0.06\\ 
      &      & 0.45 & 0.50& 0.33 &  0.04& 0.57 &  0.03& 0.91 &  0.04& 0.99 &  0.06\\ 
      &      & 0.50 & 0.60& 0.26 &  0.03& 0.47 &  0.04& 0.75 &  0.05& 0.85 &  0.05\\ 
      &      & 0.60 & 0.70& 0.15 &  0.03& 0.33 &  0.04& 0.49 &  0.06& 0.63 &  0.07\\ 
      &      & 0.70 & 0.80& 0.06 &  0.02& 0.20 &  0.04& 0.30 &  0.06& 0.43 &  0.08\\ 
\hline
 0.75 & 0.95 & 0.10 & 0.15& 0.41 &  0.10& 0.43 &  0.10& 0.52 &  0.12& 0.63 &  0.13\\ 
      &      & 0.15 & 0.20& 0.54 &  0.06& 0.73 &  0.06& 0.89 &  0.06& 1.04 &  0.10\\ 
      &      & 0.20 & 0.25& 0.56 &  0.07& 0.81 &  0.06& 1.06 &  0.09& 1.23 &  0.08\\ 
      &      & 0.25 & 0.30& 0.53 &  0.06& 0.73 &  0.05& 1.03 &  0.06& 1.06 &  0.07\\ 
      &      & 0.30 & 0.35& 0.38 &  0.05& 0.66 &  0.04& 0.92 &  0.06& 1.04 &  0.06\\ 
      &      & 0.35 & 0.40& 0.27 &  0.03& 0.54 &  0.03& 0.81 &  0.04& 0.92 &  0.05\\ 
      &      & 0.40 & 0.45& 0.24 &  0.03& 0.45 &  0.03& 0.63 &  0.03& 0.82 &  0.04\\ 
      &      & 0.45 & 0.50& 0.21 &  0.03& 0.37 &  0.02& 0.55 &  0.03& 0.67 &  0.04\\ 
      &      & 0.50 & 0.60& 0.10 &  0.02& 0.27 &  0.02& 0.43 &  0.03& 0.51 &  0.04\\ 
      &      & 0.60 & 0.70& 0.05 &  0.01& 0.16 &  0.03& 0.26 &  0.04& 0.32 &  0.05\\ 
%      &      & 0.70 & 0.80& 0.021 &  0.009& 0.104 &  0.023& 0.147 &  0.032& 0.21 &  0.04\\ 
\hline
 0.95 & 1.15 & 0.10 & 0.15& 0.35 &  0.07& 0.62 &  0.11& 0.63 &  0.11& 0.77 &  0.15\\ 
      &      & 0.15 & 0.20& 0.43 &  0.06& 0.81 &  0.05& 1.02 &  0.07& 1.06 &  0.08\\ 
      &      & 0.20 & 0.25& 0.41 &  0.05& 0.59 &  0.05& 0.97 &  0.05& 1.03 &  0.07\\ 
      &      & 0.25 & 0.30& 0.37 &  0.05& 0.63 &  0.05& 0.81 &  0.05& 0.82 &  0.05\\ 
      &      & 0.30 & 0.35& 0.28 &  0.04& 0.51 &  0.04& 0.64 &  0.04& 0.72 &  0.04\\ 
      &      & 0.35 & 0.40& 0.17 &  0.02& 0.35 &  0.02& 0.52 &  0.03& 0.59 &  0.03\\ 
      &      & 0.40 & 0.45& 0.16 &  0.03& 0.29 &  0.02& 0.39 &  0.03& 0.47 &  0.03\\ 
      &      & 0.45 & 0.50& 0.11 &  0.02& 0.23 &  0.02& 0.29 &  0.02& 0.39 &  0.03\\ 
      &      & 0.50 & 0.60& 0.05 &  0.01& 0.15 &  0.02& 0.19 &  0.02& 0.25 &  0.03\\ 
%      &      & 0.60 & 0.70& 0.015 &  0.006& 0.063 &  0.015& 0.110 &  0.019& 0.128 &  0.023\\ 
%      &      & 0.70 & 0.80& 0.009 &  0.004& 0.044 &  0.011& 0.066 &  0.014& 0.084 &  0.017\\ 
\hline
\end{tabular}
\end{center}
\end{table}

\begin{table}[hp!] 
\begin{center}
\begin{tabular}{rrrr|r@{$\pm$}lr@{$\pm$}lr@{$\pm$}lr@{$\pm$}l} 
\hline
$\theta_{\hbox{\small min}}$ &
$\theta_{\hbox{\small max}}$ &
$p_{\hbox{\small min}}$ &
$p_{\hbox{\small max}}$ &
\multicolumn{8}{c}{$d^2\sigma^{\pi^+}/(dpd\theta)$} 
\\
(rad) & (rad) & (\GeVc) & (\GeVc) &
\multicolumn{8}{c}{(barn/(\GeVc rad))}
\\
  &  &  & 
&\multicolumn{2}{c}{$ \bf{3 \ \GeVc}$} 
&\multicolumn{2}{c}{$ \bf{5 \ \GeVc}$} 
&\multicolumn{2}{c}{$ \bf{8 \ \GeVc}$} 
&\multicolumn{2}{c}{$ \bf{12 \ \GeVc}$} 
\\ 
\hline
 1.15 & 1.35 & 0.10 & 0.15& 0.43 &  0.09& 0.60 &  0.13& 0.66 &  0.14& 0.67 &  0.14\\ 
      &      & 0.15 & 0.20& 0.51 &  0.06& 0.80 &  0.05& 0.96 &  0.07& 0.96 &  0.10\\ 
      &      & 0.20 & 0.25& 0.42 &  0.05& 0.56 &  0.04& 0.87 &  0.05& 1.04 &  0.06\\ 
      &      & 0.25 & 0.30& 0.37 &  0.05& 0.44 &  0.03& 0.60 &  0.04& 0.74 &  0.04\\ 
      &      & 0.30 & 0.35& 0.18 &  0.03& 0.31 &  0.03& 0.42 &  0.03& 0.51 &  0.03\\ 
      &      & 0.35 & 0.40& 0.11 &  0.02& 0.23 &  0.02& 0.33 &  0.02& 0.37 &  0.03\\ 
      &      & 0.40 & 0.45& 0.08 &  0.02& 0.16 &  0.02& 0.25 &  0.02& 0.27 &  0.02\\ 
      &      & 0.45 & 0.50& 0.05 &  0.01& 0.11 &  0.01& 0.18 &  0.02& 0.19 &  0.02\\ 
%      &      & 0.50 & 0.60& 0.028 &  0.007& 0.061 &  0.010& 0.099 &  0.014& 0.119 &  0.016\\ 
%      &      & 0.60 & 0.70& 0.014 &  0.005& 0.025 &  0.006& 0.049 &  0.009& 0.066 &  0.012\\ 
%      &      & 0.70 & 0.80& 0.009 &  0.005& 0.015 &  0.004& 0.031 &  0.007& 0.045 &  0.011\\ 
\hline
 1.35 & 1.55 & 0.10 & 0.15& 0.43 &  0.12& 0.64 &  0.15& 0.69 &  0.17& 0.75 &  0.16\\ 
      &      & 0.15 & 0.20& 0.46 &  0.06& 0.73 &  0.06& 0.89 &  0.08& 0.88 &  0.09\\ 
      &      & 0.20 & 0.25& 0.30 &  0.04& 0.48 &  0.04& 0.70 &  0.04& 0.79 &  0.05\\ 
      &      & 0.25 & 0.30& 0.18 &  0.03& 0.34 &  0.03& 0.45 &  0.04& 0.50 &  0.04\\ 
      &      & 0.30 & 0.35& 0.13 &  0.02& 0.24 &  0.02& 0.30 &  0.02& 0.36 &  0.03\\ 
      &      & 0.35 & 0.40& 0.09 &  0.02& 0.15 &  0.01& 0.22 &  0.02& 0.28 &  0.02\\ 
      &      & 0.40 & 0.45& 0.06 &  0.01& 0.09 &  0.01& 0.14 &  0.01& 0.20 &  0.02\\ 
      &      & 0.45 & 0.50& 0.03 &  0.01& 0.06 &  0.01& 0.10 &  0.01& 0.12 &  0.02\\ 
%      &      & 0.50 & 0.60& 0.015 &  0.006& 0.026 &  0.006& 0.057 &  0.009& 0.060 &  0.011\\ 
%      &      & 0.60 & 0.70& 0.007 &  0.004& 0.011 &  0.003& 0.026 &  0.005& 0.026 &  0.006\\ 
%      &      & 0.70 & 0.80& 0.006 &  0.004& 0.009 &  0.003& 0.019 &  0.005& 0.023 &  0.006\\ 
\hline
 1.55 & 1.75 & 0.10 & 0.15& 0.52 &  0.13& 0.63 &  0.16& 0.79 &  0.18& 0.75 &  0.18\\ 
      &      & 0.15 & 0.20& 0.39 &  0.05& 0.77 &  0.06& 0.79 &  0.06& 0.80 &  0.07\\ 
      &      & 0.20 & 0.25& 0.30 &  0.04& 0.40 &  0.04& 0.60 &  0.04& 0.62 &  0.05\\ 
      &      & 0.25 & 0.30& 0.25 &  0.04& 0.23 &  0.02& 0.32 &  0.03& 0.34 &  0.03\\ 
      &      & 0.30 & 0.35& 0.11 &  0.03& 0.16 &  0.02& 0.23 &  0.02& 0.23 &  0.02\\ 
      &      & 0.35 & 0.40& 0.04 &  0.01& 0.10 &  0.01& 0.17 &  0.02& 0.18 &  0.02\\ 
      &      & 0.40 & 0.45& 0.03 &  0.01& 0.07 &  0.01& 0.10 &  0.01& 0.12 &  0.02\\ 
      &      & 0.45 & 0.50& 0.02 &  0.01& 0.05 &  0.01& 0.06 &  0.01& 0.07 &  0.01\\ 
%      &      & 0.50 & 0.60& 0.011 &  0.005& 0.018 &  0.005& 0.026 &  0.005& 0.029 &  0.006\\ 
%      &      & 0.60 & 0.70& 0.005 &  0.004& 0.008 &  0.003& 0.011 &  0.003& 0.014 &  0.004\\ 
%      &      & 0.70 & 0.80& 0.005 &  0.004& 0.006 &  0.002& 0.010 &  0.003& 0.010 &  0.003\\ 
\hline
 1.75 & 1.95 & 0.10 & 0.15& 0.59 &  0.11& 0.54 &  0.10& 0.66 &  0.12& 0.66 &  0.12\\ 
      &      & 0.15 & 0.20& 0.46 &  0.05& 0.61 &  0.04& 0.67 &  0.04& 0.68 &  0.05\\ 
      &      & 0.20 & 0.25& 0.36 &  0.05& 0.34 &  0.03& 0.41 &  0.03& 0.43 &  0.04\\ 
      &      & 0.25 & 0.30& 0.08 &  0.03& 0.16 &  0.02& 0.21 &  0.02& 0.25 &  0.02\\ 
      &      & 0.30 & 0.35& 0.03 &  0.01& 0.11 &  0.01& 0.12 &  0.01& 0.14 &  0.02\\ 
      &      & 0.35 & 0.40& 0.02 &  0.01& 0.08 &  0.01& 0.09 &  0.01& 0.09 &  0.01\\ 
      &      & 0.40 & 0.45& 0.02 &  0.01& 0.05 &  0.01& 0.05 &  0.01& 0.05 &  0.01\\ 
      &      & 0.45 & 0.50& 0.01 &  0.01& 0.02 &  0.01& 0.03 &  0.01& 0.03 &  0.01\\ 
%      &      & 0.50 & 0.60& 0.004 &  0.005& 0.007 &  0.003& 0.012 &  0.003& 0.016 &  0.004\\ 
%      &      & 0.60 & 0.70& 0.001 &  0.002& 0.003 &  0.001& 0.005 &  0.002& 0.007 &  0.003\\ 
%      &      & 0.70 & 0.80& 0.001 &  0.003& 0.002 &  0.001& 0.004 &  0.001& 0.005 &  0.002\\ 
\hline
 1.95 & 2.15 & 0.10 & 0.15& 0.31 &  0.07& 0.45 &  0.07& 0.49 &  0.07& 0.47 &  0.08\\ 
      &      & 0.15 & 0.20& 0.41 &  0.05& 0.44 &  0.04& 0.52 &  0.03& 0.50 &  0.04\\ 
      &      & 0.20 & 0.25& 0.16 &  0.04& 0.22 &  0.02& 0.28 &  0.02& 0.28 &  0.02\\ 
      &      & 0.25 & 0.30& 0.06 &  0.02& 0.10 &  0.02& 0.14 &  0.02& 0.20 &  0.02\\ 
      &      & 0.30 & 0.35& 0.03 &  0.01& 0.06 &  0.01& 0.09 &  0.01& 0.11 &  0.02\\ 
      &      & 0.35 & 0.40& 0.02 &  0.01& 0.05 &  0.01& 0.04 &  0.01& 0.06 &  0.01\\ 
      &      & 0.40 & 0.45& 0.01 &  0.01& 0.02 &  0.01& 0.03 &  0.01& 0.04 &  0.01\\ 
      &      & 0.45 & 0.50& \multicolumn{2}{c}{}&  \multicolumn{2}{c}{}& 0.02 &  0.01& 0.02 &  0.01\\ 
%      &      & 0.50 & 0.60& 0.001 &  0.002& 0.003 &  0.002& 0.007 &  0.002& 0.006 &  0.003\\ 
%      &      & 0.60 & 0.70& 0.000 &  0.001& 0.001 &  0.001& 0.002 &  0.001& 0.002 &  0.001\\ 
%      &      & 0.70 & 0.80& 0.000 &  0.001& 0.000 &  0.001& 0.001 &  0.001& 0.002 &  0.001\\ 
\hline

\end{tabular}
\end{center}
\end{table}

\begin{table}[hp!] 
\begin{center}
  \caption{\label{tab:xsec-n:cu}
    HARP results for the double-differential $\pi^-$ production
    cross-section in the laboratory system,
    $d^2\sigma^{\pi^-}/(dpd\theta)$ for copper. Each row refers to a
    different $(p_{\hbox{\small min}} \le p<p_{\hbox{\small max}},
    \theta_{\hbox{\small min}} \le \theta<\theta_{\hbox{\small max}})$ bin,
    where $p$ and $\theta$ are the pion momentum and polar angle, respectively.
    The central value as well as the square-root of the diagonal elements
    of the covariance matrix are given.}
\vspace{2mm}
\begin{tabular}{rrrr|r@{$\pm$}lr@{$\pm$}lr@{$\pm$}lr@{$\pm$}l} 
\hline
$\theta_{\hbox{\small min}}$ &
$\theta_{\hbox{\small max}}$ &
$p_{\hbox{\small min}}$ &
$p_{\hbox{\small max}}$ &
\multicolumn{8}{c}{$d^2\sigma^{\pi^-}/(dpd\theta)$} 
\\
(rad) & (rad) & (\GeVc) & (\GeVc) &
\multicolumn{8}{c}{(barn/(\GeVc rad))}
\\
  &  &  & 
&\multicolumn{2}{c}{$ \bf{3 \ \GeVc}$} 
&\multicolumn{2}{c}{$ \bf{5 \ \GeVc}$} 
&\multicolumn{2}{c}{$ \bf{8 \ \GeVc}$} 
&\multicolumn{2}{c}{$ \bf{12 \ \GeVc}$} 
\\ 
\hline
 0.35 & 0.55 & 0.10 & 0.15& 0.19 &  0.11& 0.45 &  0.19& 0.54 &  0.23& 0.55 &  0.27\\ 
      &      & 0.15 & 0.20& 0.12 &  0.07& 0.44 &  0.09& 0.61 &  0.11& 0.81 &  0.15\\ 
      &      & 0.20 & 0.25& 0.19 &  0.05& 0.48 &  0.06& 0.76 &  0.08& 0.96 &  0.09\\ 
      &      & 0.25 & 0.30& 0.20 &  0.06& 0.49 &  0.05& 0.80 &  0.06& 0.94 &  0.07\\ 
      &      & 0.30 & 0.35& 0.25 &  0.05& 0.45 &  0.04& 0.78 &  0.05& 0.92 &  0.07\\ 
      &      & 0.35 & 0.40& 0.16 &  0.03& 0.43 &  0.04& 0.76 &  0.04& 0.89 &  0.06\\ 
      &      & 0.40 & 0.45& 0.13 &  0.02& 0.40 &  0.03& 0.72 &  0.04& 0.79 &  0.04\\ 
      &      & 0.45 & 0.50& 0.21 &  0.05& 0.41 &  0.03& 0.70 &  0.04& 0.79 &  0.05\\ 
      &      & 0.50 & 0.60& 0.17 &  0.03& 0.41 &  0.03& 0.70 &  0.04& 0.79 &  0.05\\ 
      &      & 0.60 & 0.70& 0.11 &  0.02& 0.32 &  0.03& 0.64 &  0.05& 0.69 &  0.06\\ 
      &      & 0.70 & 0.80& 0.10 &  0.02& 0.25 &  0.03& 0.53 &  0.07& 0.63 &  0.08\\ 
\hline
 0.55 & 0.75 & 0.10 & 0.15& 0.26 &  0.09& 0.43 &  0.13& 0.49 &  0.15& 0.56 &  0.19\\ 
      &      & 0.15 & 0.20& 0.21 &  0.06& 0.53 &  0.08& 0.70 &  0.08& 0.92 &  0.09\\ 
      &      & 0.20 & 0.25& 0.43 &  0.07& 0.59 &  0.06& 0.79 &  0.06& 1.02 &  0.09\\ 
      &      & 0.25 & 0.30& 0.21 &  0.04& 0.62 &  0.05& 0.84 &  0.07& 0.98 &  0.09\\ 
      &      & 0.30 & 0.35& 0.22 &  0.03& 0.52 &  0.04& 0.78 &  0.04& 0.86 &  0.04\\ 
      &      & 0.35 & 0.40& 0.25 &  0.04& 0.44 &  0.03& 0.67 &  0.03& 0.81 &  0.05\\ 
      &      & 0.40 & 0.45& 0.18 &  0.03& 0.43 &  0.03& 0.65 &  0.04& 0.81 &  0.04\\ 
      &      & 0.45 & 0.50& 0.13 &  0.02& 0.39 &  0.03& 0.60 &  0.03& 0.79 &  0.04\\ 
      &      & 0.50 & 0.60& 0.14 &  0.02& 0.28 &  0.02& 0.54 &  0.03& 0.65 &  0.05\\ 
      &      & 0.60 & 0.70& 0.09 &  0.02& 0.22 &  0.02& 0.44 &  0.04& 0.52 &  0.04\\ 
      &      & 0.70 & 0.80& 0.06 &  0.02& 0.18 &  0.02& 0.34 &  0.05& 0.46 &  0.05\\ 
\hline
 0.75 & 0.95 & 0.10 & 0.15& 0.24 &  0.07& 0.47 &  0.10& 0.54 &  0.13& 0.55 &  0.15\\ 
      &      & 0.15 & 0.20& 0.21 &  0.04& 0.63 &  0.06& 0.83 &  0.07& 0.88 &  0.07\\ 
      &      & 0.20 & 0.25& 0.21 &  0.04& 0.52 &  0.04& 0.79 &  0.05& 0.93 &  0.07\\ 
      &      & 0.25 & 0.30& 0.21 &  0.04& 0.49 &  0.03& 0.70 &  0.04& 0.85 &  0.06\\ 
      &      & 0.30 & 0.35& 0.23 &  0.04& 0.40 &  0.03& 0.65 &  0.04& 0.83 &  0.06\\ 
      &      & 0.35 & 0.40& 0.23 &  0.03& 0.34 &  0.02& 0.60 &  0.03& 0.67 &  0.04\\ 
      &      & 0.40 & 0.45& 0.23 &  0.03& 0.30 &  0.02& 0.48 &  0.03& 0.54 &  0.03\\ 
      &      & 0.45 & 0.50& 0.15 &  0.03& 0.28 &  0.02& 0.41 &  0.02& 0.48 &  0.03\\ 
      &      & 0.50 & 0.60& 0.09 &  0.02& 0.25 &  0.02& 0.32 &  0.02& 0.40 &  0.03\\ 
      &      & 0.60 & 0.70& 0.06 &  0.01& 0.17 &  0.02& 0.26 &  0.02& 0.31 &  0.03\\ 
%      &      & 0.70 & 0.80& 0.038 &  0.014& 0.112 &  0.021& 0.188 &  0.032& 0.258 &  0.035\\ 
\hline
 0.95 & 1.15 & 0.10 & 0.15& 0.30 &  0.07& 0.53 &  0.10& 0.65 &  0.11& 0.69 &  0.14\\ 
      &      & 0.15 & 0.20& 0.35 &  0.05& 0.65 &  0.06& 0.80 &  0.06& 0.86 &  0.07\\ 
      &      & 0.20 & 0.25& 0.31 &  0.04& 0.49 &  0.04& 0.74 &  0.04& 0.74 &  0.06\\ 
      &      & 0.25 & 0.30& 0.21 &  0.03& 0.44 &  0.03& 0.61 &  0.03& 0.73 &  0.05\\ 
      &      & 0.30 & 0.35& 0.13 &  0.02& 0.36 &  0.03& 0.50 &  0.03& 0.57 &  0.04\\ 
      &      & 0.35 & 0.40& 0.11 &  0.02& 0.25 &  0.02& 0.44 &  0.02& 0.48 &  0.03\\ 
      &      & 0.40 & 0.45& 0.10 &  0.02& 0.20 &  0.01& 0.35 &  0.02& 0.43 &  0.03\\ 
      &      & 0.45 & 0.50& 0.10 &  0.02& 0.17 &  0.01& 0.27 &  0.02& 0.37 &  0.02\\ 
      &      & 0.50 & 0.60& 0.08 &  0.02& 0.13 &  0.01& 0.20 &  0.01& 0.25 &  0.02\\ 
%      &      & 0.60 & 0.70& 0.049 &  0.014& 0.086 &  0.013& 0.145 &  0.015& 0.162 &  0.025\\ 
%      &      & 0.70 & 0.80& 0.018 &  0.010& 0.061 &  0.012& 0.106 &  0.016& 0.127 &  0.020\\ 
\hline
\end{tabular}
\end{center}
\end{table}

\begin{table}[hp!] 
\begin{center}
\begin{tabular}{rrrr|r@{$\pm$}lr@{$\pm$}lr@{$\pm$}lr@{$\pm$}l} 
\hline
$\theta_{\hbox{\small min}}$ &
$\theta_{\hbox{\small max}}$ &
$p_{\hbox{\small min}}$ &
$p_{\hbox{\small max}}$ &
\multicolumn{8}{c}{$d^2\sigma^{\pi^-}/(dpd\theta)$} 
\\
(rad) & (rad) & (\GeVc) & (\GeVc) &
\multicolumn{8}{c}{(barn/(\GeVc rad))}
\\
  &  &  & 
&\multicolumn{2}{c}{$ \bf{3 \ \GeVc}$} 
&\multicolumn{2}{c}{$ \bf{5 \ \GeVc}$} 
&\multicolumn{2}{c}{$ \bf{8 \ \GeVc}$} 
&\multicolumn{2}{c}{$ \bf{12 \ \GeVc}$} 
\\ 
\hline
 1.15 & 1.35 & 0.10 & 0.15& 0.29 &  0.07& 0.60 &  0.11& 0.68 &  0.13& 0.65 &  0.15\\ 
      &      & 0.15 & 0.20& 0.36 &  0.05& 0.57 &  0.05& 0.70 &  0.06& 0.91 &  0.09\\ 
      &      & 0.20 & 0.25& 0.27 &  0.04& 0.47 &  0.04& 0.66 &  0.04& 0.74 &  0.05\\ 
      &      & 0.25 & 0.30& 0.25 &  0.04& 0.42 &  0.03& 0.51 &  0.03& 0.59 &  0.04\\ 
      &      & 0.30 & 0.35& 0.20 &  0.03& 0.29 &  0.03& 0.37 &  0.03& 0.46 &  0.03\\ 
      &      & 0.35 & 0.40& 0.11 &  0.02& 0.20 &  0.02& 0.30 &  0.02& 0.38 &  0.03\\ 
      &      & 0.40 & 0.45& 0.05 &  0.01& 0.16 &  0.01& 0.24 &  0.02& 0.28 &  0.02\\ 
      &      & 0.45 & 0.50& 0.03 &  0.01& 0.12 &  0.01& 0.18 &  0.01& 0.20 &  0.02\\ 
%      &      & 0.50 & 0.60& 0.018 &  0.007& 0.068 &  0.010& 0.123 &  0.011& 0.142 &  0.014\\ 
%      &      & 0.60 & 0.70& 0.010 &  0.005& 0.038 &  0.007& 0.082 &  0.009& 0.090 &  0.012\\ 
%      &      & 0.70 & 0.80& 0.007 &  0.004& 0.031 &  0.006& 0.062 &  0.011& 0.079 &  0.012\\ 
\hline
 1.35 & 1.55 & 0.10 & 0.15& 0.33 &  0.10& 0.56 &  0.12& 0.66 &  0.15& 0.73 &  0.18\\ 
      &      & 0.15 & 0.20& 0.36 &  0.05& 0.49 &  0.05& 0.68 &  0.06& 0.85 &  0.08\\ 
      &      & 0.20 & 0.25& 0.25 &  0.04& 0.35 &  0.03& 0.54 &  0.04& 0.68 &  0.05\\ 
      &      & 0.25 & 0.30& 0.21 &  0.03& 0.28 &  0.03& 0.40 &  0.03& 0.47 &  0.04\\ 
      &      & 0.30 & 0.35& 0.17 &  0.03& 0.18 &  0.02& 0.30 &  0.02& 0.32 &  0.03\\ 
      &      & 0.35 & 0.40& 0.08 &  0.02& 0.14 &  0.01& 0.21 &  0.02& 0.21 &  0.02\\ 
      &      & 0.40 & 0.45& 0.04 &  0.01& 0.09 &  0.01& 0.16 &  0.01& 0.16 &  0.01\\ 
      &      & 0.45 & 0.50& 0.02 &  0.01& 0.06 &  0.01& 0.13 &  0.01& 0.13 &  0.01\\ 
%      &      & 0.50 & 0.60& 0.010 &  0.005& 0.037 &  0.006& 0.077 &  0.009& 0.087 &  0.010\\ 
%      &      & 0.60 & 0.70& 0.007 &  0.004& 0.024 &  0.004& 0.042 &  0.006& 0.053 &  0.008\\ 
%      &      & 0.70 & 0.80& 0.006 &  0.005& 0.017 &  0.004& 0.032 &  0.006& 0.041 &  0.007\\ 
\hline
 1.55 & 1.75 & 0.10 & 0.15& 0.25 &  0.07& 0.57 &  0.12& 0.66 &  0.16& 0.78 &  0.20\\ 
      &      & 0.15 & 0.20& 0.32 &  0.05& 0.43 &  0.05& 0.61 &  0.05& 0.70 &  0.06\\ 
      &      & 0.20 & 0.25& 0.23 &  0.04& 0.31 &  0.03& 0.41 &  0.03& 0.47 &  0.04\\ 
      &      & 0.25 & 0.30& 0.11 &  0.03& 0.20 &  0.02& 0.30 &  0.02& 0.35 &  0.03\\ 
      &      & 0.30 & 0.35& 0.04 &  0.01& 0.13 &  0.01& 0.21 &  0.02& 0.24 &  0.02\\ 
      &      & 0.35 & 0.40& 0.03 &  0.01& 0.10 &  0.01& 0.15 &  0.01& 0.16 &  0.01\\ 
      &      & 0.40 & 0.45& 0.03 &  0.01& 0.07 &  0.01& 0.11 &  0.01& 0.10 &  0.01\\ 
      &      & 0.45 & 0.50& 0.01 &  0.01& 0.05 &  0.01& 0.08 &  0.01& 0.07 &  0.01\\ 
%      &      & 0.50 & 0.60& 0.003 &  0.003& 0.027 &  0.006& 0.045 &  0.007& 0.039 &  0.006\\ 
%      &      & 0.60 & 0.70& 0.001 &  0.002& 0.012 &  0.003& 0.022 &  0.004& 0.022 &  0.004\\ 
%      &      & 0.70 & 0.80& 0.001 &  0.003& 0.009 &  0.003& 0.016 &  0.004& 0.020 &  0.004\\ 
\hline
 1.75 & 1.95 & 0.10 & 0.15& 0.19 &  0.05& 0.52 &  0.09& 0.55 &  0.10& 0.68 &  0.12\\ 
      &      & 0.15 & 0.20& 0.26 &  0.04& 0.43 &  0.03& 0.51 &  0.03& 0.58 &  0.04\\ 
      &      & 0.20 & 0.25& 0.19 &  0.03& 0.26 &  0.02& 0.31 &  0.02& 0.37 &  0.03\\ 
      &      & 0.25 & 0.30& 0.08 &  0.02& 0.16 &  0.02& 0.20 &  0.02& 0.24 &  0.02\\ 
      &      & 0.30 & 0.35& 0.02 &  0.01& 0.10 &  0.01& 0.14 &  0.01& 0.15 &  0.02\\ 
      &      & 0.35 & 0.40& 0.03 &  0.01& 0.06 &  0.01& 0.10 &  0.01& 0.10 &  0.01\\ 
      &      & 0.40 & 0.45& 0.01 &  0.01& 0.04 &  0.01& 0.07 &  0.01& 0.07 &  0.01\\ 
      &      & 0.45 & 0.50&  \multicolumn{2}{c}{}  & 0.02 &  0.01& 0.05 &  0.01& 0.06 &  0.01\\ 
%      &      & 0.50 & 0.60& 0.002 &  0.004& 0.012 &  0.003& 0.027 &  0.004& 0.036 &  0.006\\ 
%      &      & 0.60 & 0.70& 0.001 &  0.003& 0.007 &  0.002& 0.015 &  0.002& 0.022 &  0.005\\ 
%      &      & 0.70 & 0.80& 0.001 &  0.003& 0.004 &  0.002& 0.010 &  0.003& 0.015 &  0.003\\ 
\hline
 1.95 & 2.15 & 0.10 & 0.15& 0.25 &  0.06& 0.40 &  0.06& 0.47 &  0.07& 0.52 &  0.08\\ 
      &      & 0.15 & 0.20& 0.13 &  0.03& 0.28 &  0.03& 0.43 &  0.03& 0.47 &  0.04\\ 
      &      & 0.20 & 0.25& 0.12 &  0.03& 0.14 &  0.01& 0.28 &  0.02& 0.29 &  0.03\\ 
      &      & 0.25 & 0.30& 0.04 &  0.02& 0.10 &  0.01& 0.16 &  0.02& 0.12 &  0.02\\ 
      &      & 0.30 & 0.35& 0.01 &  0.01& 0.07 &  0.01& 0.08 &  0.01& 0.11 &  0.01\\ 
      &      & 0.35 & 0.40& 0.02 &  0.01& 0.05 &  0.01& 0.04 &  0.01& 0.07 &  0.01\\ 
      &      & 0.40 & 0.45& 0.03 &  0.02& 0.03 &  0.01& 0.03 &  0.01& 0.06 &  0.01\\ 
      &      & 0.45 & 0.50& \multicolumn{2}{c}{}& 0.02 &  0.01& 0.02 &  0.01& 0.05 &  0.01\\ 
%      &      & 0.50 & 0.60& 0.001 &  0.004& 0.010 &  0.003& 0.015 &  0.003& 0.022 &  0.005\\ 
%      &      & 0.60 & 0.70& 0.000 &  0.002& 0.006 &  0.002& 0.009 &  0.002& 0.009 &  0.003\\ 
%      &      & 0.70 & 0.80& 0.000 &  0.002& 0.004 &  0.002& 0.006 &  0.002& 0.006 &  0.003\\ 
\hline
\end{tabular}
\end{center}
\end{table}
%

%% file: Sn5_result-table-scl.tex
% tin 
\begin{table}[hp!] 
\begin{center}
  \caption{\label{tab:xsec-p:sn}
    HARP results for the double-differential $\pi^+$ production
    cross-section in the laboratory system,
    $d^2\sigma^{\pi^+}/(dpd\theta)$ for tin. Each row refers to a
    different $(p_{\hbox{\small min}} \le p<p_{\hbox{\small max}},
    \theta_{\hbox{\small min}} \le \theta<\theta_{\hbox{\small max}})$ bin,
    where $p$ and $\theta$ are the pion momentum and polar angle, respectively.
    The central value as well as the square-root of the diagonal elements
    of the covariance matrix are given.}
\vspace{2mm}
%\begin{tabular}{rrrr|r@{$\pm$}lr{$\pm$}lr{$\pm$}lr{$\pm$}l} 
\begin{tabular}{rrrr|r@{$\pm$}lr@{$\pm$}lr@{$\pm$}lr@{$\pm$}l} 
\hline
$\theta_{\hbox{\small min}}$ &
$\theta_{\hbox{\small max}}$ &
$p_{\hbox{\small min}}$ &
$p_{\hbox{\small max}}$ &
\multicolumn{8}{c}{$d^2\sigma^{\pi^+}/(dpd\theta)$} 
\\
(rad) & (rad) & (\GeVc) & (\GeVc) &
\multicolumn{8}{c}{(barn/(\GeVc rad))}
\\
  &  &  & 
&\multicolumn{2}{c}{$ \bf{3 \ \GeVc}$} 
&\multicolumn{2}{c}{$ \bf{5 \ \GeVc}$} 
&\multicolumn{2}{c}{$ \bf{8 \ \GeVc}$} 
&\multicolumn{2}{c}{$ \bf{12 \ \GeVc}$} 
\\ 
\hline
 0.35 & 0.55 & 0.10 & 0.15& 0.07 &  0.08& 0.53 &  0.22& 1.06 &  0.40& 1.55 &  0.49\\ 
      &      & 0.15 & 0.20& 0.29 &  0.12& 0.63 &  0.14& 1.15 &  0.17& 1.41 &  0.21\\ 
      &      & 0.20 & 0.25& 0.46 &  0.12& 0.80 &  0.09& 1.29 &  0.12& 1.57 &  0.17\\ 
      &      & 0.25 & 0.30& 0.65 &  0.12& 1.02 &  0.10& 1.44 &  0.12& 1.90 &  0.14\\ 
      &      & 0.30 & 0.35& 0.53 &  0.09& 0.92 &  0.07& 1.54 &  0.11& 1.99 &  0.14\\ 
      &      & 0.35 & 0.40& 0.50 &  0.09& 1.01 &  0.09& 1.66 &  0.11& 1.86 &  0.10\\ 
      &      & 0.40 & 0.45& 0.43 &  0.07& 1.07 &  0.07& 1.53 &  0.09& 2.07 &  0.20\\ 
      &      & 0.45 & 0.50& 0.42 &  0.06& 1.02 &  0.06& 1.43 &  0.07& 1.92 &  0.11\\ 
      &      & 0.50 & 0.60& 0.47 &  0.06& 0.87 &  0.06& 1.40 &  0.08& 1.92 &  0.12\\ 
      &      & 0.60 & 0.70& 0.36 &  0.06& 0.74 &  0.08& 1.22 &  0.13& 1.77 &  0.18\\ 
      &      & 0.70 & 0.80& 0.26 &  0.06& 0.42 &  0.07& 0.89 &  0.14& 1.32 &  0.19\\ 
\hline
 0.55 & 0.75 & 0.10 & 0.15& 0.23 &  0.13& 0.65 &  0.17& 0.87 &  0.21& 1.19 &  0.27\\ 
      &      & 0.15 & 0.20& 0.50 &  0.10& 1.05 &  0.10& 1.43 &  0.10& 1.59 &  0.15\\ 
      &      & 0.20 & 0.25& 0.60 &  0.10& 1.15 &  0.09& 1.62 &  0.15& 2.01 &  0.18\\ 
      &      & 0.25 & 0.30& 0.68 &  0.08& 1.05 &  0.08& 1.58 &  0.09& 1.99 &  0.16\\ 
      &      & 0.30 & 0.35& 0.50 &  0.06& 1.09 &  0.08& 1.48 &  0.08& 1.87 &  0.12\\ 
      &      & 0.35 & 0.40& 0.49 &  0.07& 0.96 &  0.07& 1.47 &  0.08& 1.84 &  0.11\\ 
      &      & 0.40 & 0.45& 0.46 &  0.06& 0.86 &  0.05& 1.29 &  0.06& 2.01 &  0.11\\ 
      &      & 0.45 & 0.50& 0.39 &  0.06& 0.75 &  0.05& 1.21 &  0.06& 1.70 &  0.11\\ 
      &      & 0.50 & 0.60& 0.27 &  0.04& 0.57 &  0.05& 0.97 &  0.06& 1.36 &  0.10\\ 
      &      & 0.60 & 0.70& 0.18 &  0.03& 0.37 &  0.04& 0.70 &  0.09& 1.00 &  0.11\\ 
      &      & 0.70 & 0.80& 0.12 &  0.03& 0.24 &  0.05& 0.43 &  0.08& 0.58 &  0.11\\ 
\hline
 0.75 & 0.95 & 0.10 & 0.15& 0.50 &  0.11& 0.86 &  0.14& 1.03 &  0.16& 1.14 &  0.19\\ 
      &      & 0.15 & 0.20& 0.67 &  0.10& 1.12 &  0.09& 1.60 &  0.10& 1.88 &  0.14\\ 
      &      & 0.20 & 0.25& 0.65 &  0.08& 1.24 &  0.09& 1.55 &  0.10& 1.86 &  0.13\\ 
      &      & 0.25 & 0.30& 0.60 &  0.07& 1.02 &  0.06& 1.34 &  0.09& 1.77 &  0.13\\ 
      &      & 0.30 & 0.35& 0.52 &  0.06& 0.89 &  0.06& 1.33 &  0.07& 1.65 &  0.09\\ 
      &      & 0.35 & 0.40& 0.51 &  0.06& 0.74 &  0.05& 1.18 &  0.07& 1.29 &  0.06\\ 
      &      & 0.40 & 0.45& 0.34 &  0.05& 0.57 &  0.04& 0.95 &  0.05& 1.16 &  0.07\\ 
      &      & 0.45 & 0.50& 0.27 &  0.03& 0.47 &  0.03& 0.77 &  0.05& 1.01 &  0.06\\ 
      &      & 0.50 & 0.60& 0.20 &  0.03& 0.33 &  0.03& 0.55 &  0.05& 0.79 &  0.06\\ 
      &      & 0.60 & 0.70& 0.09 &  0.02& 0.19 &  0.03& 0.31 &  0.05& 0.49 &  0.07\\ 
%      &      & 0.70 & 0.80& 0.058 &  0.018& 0.115 &  0.028& 0.22 &  0.05& 0.29 &  0.06\\ 
\hline
 0.95 & 1.15 & 0.10 & 0.15& 0.46 &  0.09& 0.85 &  0.11& 1.04 &  0.14& 1.21 &  0.17\\ 
      &      & 0.15 & 0.20& 0.63 &  0.09& 1.01 &  0.09& 1.45 &  0.09& 1.83 &  0.12\\ 
      &      & 0.20 & 0.25& 0.56 &  0.07& 1.11 &  0.07& 1.39 &  0.10& 1.58 &  0.09\\ 
      &      & 0.25 & 0.30& 0.38 &  0.05& 0.81 &  0.06& 1.16 &  0.06& 1.50 &  0.11\\ 
      &      & 0.30 & 0.35& 0.24 &  0.04& 0.55 &  0.05& 0.90 &  0.07& 1.18 &  0.08\\ 
      &      & 0.35 & 0.40& 0.21 &  0.04& 0.47 &  0.04& 0.76 &  0.05& 0.94 &  0.06\\ 
      &      & 0.40 & 0.45& 0.16 &  0.03& 0.43 &  0.03& 0.67 &  0.04& 0.81 &  0.05\\ 
      &      & 0.45 & 0.50& 0.12 &  0.03& 0.32 &  0.03& 0.48 &  0.04& 0.64 &  0.05\\ 
      &      & 0.50 & 0.60& 0.06 &  0.02& 0.20 &  0.03& 0.28 &  0.03& 0.43 &  0.05\\ 
%      &      & 0.60 & 0.70& 0.026 &  0.010& 0.093 &  0.020& 0.149 &  0.026& 0.23 &  0.05\\ 
%      &      & 0.70 & 0.80& 0.015 &  0.006& 0.056 &  0.017& 0.104 &  0.026& 0.123 &  0.030\\ 
\hline
\end{tabular}
\end{center}
\end{table}

\begin{table}[hp!] 
\begin{center}
\begin{tabular}{rrrr|r@{$\pm$}lr@{$\pm$}lr@{$\pm$}lr@{$\pm$}l} 
\hline
$\theta_{\hbox{\small min}}$ &
$\theta_{\hbox{\small max}}$ &
$p_{\hbox{\small min}}$ &
$p_{\hbox{\small max}}$ &
\multicolumn{8}{c}{$d^2\sigma^{\pi^+}/(dpd\theta)$} 
\\
(rad) & (rad) & (\GeVc) & (\GeVc) &
\multicolumn{8}{c}{(barn/(\GeVc rad))}
\\
  &  &  & 
&\multicolumn{2}{c}{$ \bf{3 \ \GeVc}$} 
&\multicolumn{2}{c}{$ \bf{5 \ \GeVc}$} 
&\multicolumn{2}{c}{$ \bf{8 \ \GeVc}$} 
&\multicolumn{2}{c}{$ \bf{12 \ \GeVc}$} 
\\ 
\hline
 1.15 & 1.35 & 0.10 & 0.15& 0.58 &  0.12& 0.78 &  0.11& 1.00 &  0.13& 1.17 &  0.18\\ 
      &      & 0.15 & 0.20& 0.74 &  0.09& 1.05 &  0.08& 1.34 &  0.10& 1.53 &  0.13\\ 
      &      & 0.20 & 0.25& 0.51 &  0.08& 0.81 &  0.06& 1.15 &  0.07& 1.50 &  0.09\\ 
      &      & 0.25 & 0.30& 0.43 &  0.07& 0.63 &  0.05& 0.87 &  0.05& 1.03 &  0.07\\ 
      &      & 0.30 & 0.35& 0.26 &  0.04& 0.47 &  0.04& 0.63 &  0.04& 0.81 &  0.06\\ 
      &      & 0.35 & 0.40& 0.16 &  0.03& 0.33 &  0.03& 0.46 &  0.03& 0.60 &  0.04\\ 
      &      & 0.40 & 0.45& 0.11 &  0.02& 0.24 &  0.02& 0.35 &  0.02& 0.47 &  0.03\\ 
      &      & 0.45 & 0.50& 0.09 &  0.02& 0.19 &  0.02& 0.26 &  0.03& 0.35 &  0.03\\ 
%      &      & 0.50 & 0.60& 0.059 &  0.015& 0.087 &  0.015& 0.147 &  0.021& 0.205 &  0.026\\ 
%      &      & 0.60 & 0.70& 0.023 &  0.008& 0.043 &  0.010& 0.066 &  0.013& 0.097 &  0.020\\ 
%      &      & 0.70 & 0.80& 0.015 &  0.007& 0.031 &  0.009& 0.044 &  0.013& 0.065 &  0.017\\ 
\hline
 1.35 & 1.55 & 0.10 & 0.15& 0.61 &  0.14& 0.79 &  0.14& 1.18 &  0.20& 1.38 &  0.24\\ 
      &      & 0.15 & 0.20& 0.76 &  0.08& 1.12 &  0.10& 1.41 &  0.09& 1.66 &  0.15\\ 
      &      & 0.20 & 0.25& 0.47 &  0.06& 0.86 &  0.06& 1.06 &  0.08& 1.35 &  0.09\\ 
      &      & 0.25 & 0.30& 0.30 &  0.05& 0.51 &  0.05& 0.73 &  0.05& 0.94 &  0.08\\ 
      &      & 0.30 & 0.35& 0.15 &  0.03& 0.37 &  0.03& 0.55 &  0.04& 0.58 &  0.05\\ 
      &      & 0.35 & 0.40& 0.09 &  0.02& 0.25 &  0.03& 0.36 &  0.03& 0.47 &  0.04\\ 
      &      & 0.40 & 0.45& 0.06 &  0.01& 0.14 &  0.02& 0.22 &  0.02& 0.39 &  0.04\\ 
      &      & 0.45 & 0.50& 0.03 &  0.01& 0.09 &  0.01& 0.15 &  0.02& 0.23 &  0.03\\ 
%      &      & 0.50 & 0.60& 0.016 &  0.007& 0.040 &  0.008& 0.071 &  0.012& 0.106 &  0.022\\ 
%      &      & 0.60 & 0.70& 0.011 &  0.006& 0.019 &  0.005& 0.032 &  0.008& 0.051 &  0.012\\ 
%      &      & 0.70 & 0.80& 0.008 &  0.005& 0.019 &  0.006& 0.025 &  0.007& 0.031 &  0.010\\ 
\hline
 1.55 & 1.75 & 0.10 & 0.15& 0.73 &  0.13& 0.86 &  0.15& 1.13 &  0.19& 1.20 &  0.22\\ 
      &      & 0.15 & 0.20& 0.68 &  0.08& 1.02 &  0.09& 1.34 &  0.09& 1.52 &  0.12\\ 
      &      & 0.20 & 0.25& 0.43 &  0.06& 0.70 &  0.06& 0.91 &  0.06& 1.03 &  0.08\\ 
      &      & 0.25 & 0.30& 0.18 &  0.04& 0.42 &  0.05& 0.56 &  0.05& 0.68 &  0.06\\ 
      &      & 0.30 & 0.35& 0.09 &  0.02& 0.24 &  0.02& 0.39 &  0.03& 0.43 &  0.04\\ 
      &      & 0.35 & 0.40& 0.06 &  0.02& 0.20 &  0.03& 0.24 &  0.02& 0.28 &  0.03\\ 
      &      & 0.40 & 0.45& 0.04 &  0.02& 0.10 &  0.02& 0.15 &  0.02& 0.18 &  0.02\\ 
      &      & 0.45 & 0.50& 0.04 &  0.01& 0.06 &  0.01& 0.09 &  0.02& 0.10 &  0.02\\ 
%      &      & 0.50 & 0.60& 0.017 &  0.011& 0.026 &  0.007& 0.037 &  0.007& 0.052 &  0.010\\ 
%      &      & 0.60 & 0.70& 0.005 &  0.003& 0.014 &  0.004& 0.022 &  0.004& 0.026 &  0.007\\ 
%      &      & 0.70 & 0.80& 0.004 &  0.004& 0.012 &  0.004& 0.017 &  0.004& 0.023 &  0.008\\ 
\hline
 1.75 & 1.95 & 0.10 & 0.15& 0.69 &  0.11& 0.86 &  0.11& 0.91 &  0.12& 0.91 &  0.12\\ 
      &      & 0.15 & 0.20& 0.49 &  0.06& 0.74 &  0.06& 1.02 &  0.05& 1.02 &  0.07\\ 
      &      & 0.20 & 0.25& 0.35 &  0.05& 0.46 &  0.04& 0.63 &  0.05& 0.74 &  0.05\\ 
      &      & 0.25 & 0.30& 0.12 &  0.04& 0.22 &  0.03& 0.34 &  0.03& 0.42 &  0.05\\ 
      &      & 0.30 & 0.35& 0.03 &  0.02& 0.13 &  0.02& 0.22 &  0.02& 0.22 &  0.02\\ 
      &      & 0.35 & 0.40& 0.01 &  0.01& 0.08 &  0.01& 0.14 &  0.02& 0.13 &  0.02\\ 
      &      & 0.40 & 0.45& \multicolumn{2}{c}{}& 0.05 &  0.01& 0.07 &  0.01& 0.08 &  0.01\\ 
      &      & 0.45 & 0.50& \multicolumn{2}{c}{}& 0.02 &  0.01& 0.04 &  0.01& 0.05 &  0.01\\ 
%      &      & 0.50 & 0.60& 0.001 &  0.002& 0.008 &  0.003& 0.015 &  0.004& 0.021 &  0.005\\ 
%      &      & 0.60 & 0.70& 0.001 &  0.002& 0.002 &  0.001& 0.007 &  0.002& 0.013 &  0.004\\ 
%      &      & 0.70 & 0.80& 0.001 &  0.004& 0.002 &  0.002& 0.004 &  0.001& 0.011 &  0.003\\ 
\hline
 1.95 & 2.15 & 0.10 & 0.15& 0.47 &  0.09& 0.64 &  0.08& 0.69 &  0.07& 0.69 &  0.08\\ 
      &      & 0.15 & 0.20& 0.44 &  0.07& 0.64 &  0.05& 0.66 &  0.04& 0.80 &  0.06\\ 
      &      & 0.20 & 0.25& 0.27 &  0.05& 0.38 &  0.04& 0.40 &  0.03& 0.50 &  0.05\\ 
      &      & 0.25 & 0.30& 0.08 &  0.03& 0.18 &  0.03& 0.22 &  0.02& 0.25 &  0.03\\ 
      &      & 0.30 & 0.35& 0.02 &  0.01& 0.10 &  0.01& 0.12 &  0.01& 0.15 &  0.02\\ 
      &      & 0.35 & 0.40& 0.01 &  0.01& 0.05 &  0.01& 0.09 &  0.01& 0.08 &  0.02\\ 
      &      & 0.40 & 0.45& 0.01 &  0.01& 0.03 &  0.01& 0.05 &  0.01& 0.05 &  0.01\\ 
      &      & 0.45 & 0.50& \multicolumn{2}{c}{}& 0.02 &  0.01& 0.02 &  0.01& 0.03 &  0.01\\ 
%      &      & 0.50 & 0.60& 0.001 &  0.004& 0.008 &  0.003& 0.010 &  0.003& 0.010 &  0.004\\ 
%      &      & 0.60 & 0.70& 0.000 &  0.002& 0.004 &  0.003& 0.004 &  0.001& 0.003 &  0.002\\ 
%      &      & 0.70 & 0.80& 0.000 &  0.003& 0.002 &  0.002& 0.002 &  0.001& 0.002 &  0.002\\ 
\hline
\end{tabular}
\end{center}
\end{table}

\begin{table}[hp!] 
\begin{center}
  \caption{\label{tab:xsec-n:sn}
    HARP results for the double-differential $\pi^-$ production
    cross-section in the laboratory system,
    $d^2\sigma^{\pi^-}/(dpd\theta)$ for tin. Each row refers to a
    different $(p_{\hbox{\small min}} \le p<p_{\hbox{\small max}},
    \theta_{\hbox{\small min}} \le \theta<\theta_{\hbox{\small max}})$ bin,
    where $p$ and $\theta$ are the pion momentum and polar angle, respectively.
    The central value as well as the square-root of the diagonal elements
    of the covariance matrix are given.}
\vspace{2mm}
\begin{tabular}{rrrr|r@{$\pm$}lr@{$\pm$}lr@{$\pm$}lr@{$\pm$}l} 
\hline
$\theta_{\hbox{\small min}}$ &
$\theta_{\hbox{\small max}}$ &
$p_{\hbox{\small min}}$ &
$p_{\hbox{\small max}}$ &
\multicolumn{8}{c}{$d^2\sigma^{\pi^-}/(dpd\theta)$} 
\\
(rad) & (rad) & (\GeVc) & (\GeVc) &
\multicolumn{8}{c}{(barn/(\GeVc rad))}
\\
  &  &  & 
&\multicolumn{2}{c}{$ \bf{3 \ \GeVc}$} 
&\multicolumn{2}{c}{$ \bf{5 \ \GeVc}$} 
&\multicolumn{2}{c}{$ \bf{8 \ \GeVc}$} 
&\multicolumn{2}{c}{$ \bf{12 \ \GeVc}$} 
\\ 
\hline
 0.35 & 0.55 & 0.10 & 0.15& 0.28 &  0.20& 0.40 &  0.20& 0.91 &  0.36& 1.16 &  0.48\\ 
      &      & 0.15 & 0.20& 0.30 &  0.13& 0.76 &  0.14& 1.21 &  0.20& 1.56 &  0.25\\ 
      &      & 0.20 & 0.25& 0.28 &  0.09& 0.70 &  0.11& 1.31 &  0.12& 1.77 &  0.20\\ 
      &      & 0.25 & 0.30& 0.46 &  0.10& 0.76 &  0.08& 1.38 &  0.10& 1.85 &  0.14\\ 
      &      & 0.30 & 0.35& 0.32 &  0.06& 0.72 &  0.06& 1.26 &  0.07& 1.78 &  0.10\\ 
      &      & 0.35 & 0.40& 0.20 &  0.04& 0.64 &  0.06& 1.22 &  0.07& 1.56 &  0.08\\ 
      &      & 0.40 & 0.45& 0.29 &  0.07& 0.62 &  0.04& 1.08 &  0.06& 1.37 &  0.07\\ 
      &      & 0.45 & 0.50& 0.34 &  0.06& 0.55 &  0.04& 1.05 &  0.05& 1.32 &  0.07\\ 
      &      & 0.50 & 0.60& 0.17 &  0.04& 0.52 &  0.04& 0.90 &  0.05& 1.26 &  0.08\\ 
      &      & 0.60 & 0.70& 0.17 &  0.03& 0.43 &  0.05& 0.84 &  0.06& 1.26 &  0.10\\ 
      &      & 0.70 & 0.80& 0.16 &  0.04& 0.31 &  0.04& 0.66 &  0.08& 0.98 &  0.14\\ 
\hline
 0.55 & 0.75 & 0.10 & 0.15& 0.28 &  0.13& 0.50 &  0.15& 0.96 &  0.22& 1.15 &  0.30\\ 
      &      & 0.15 & 0.20& 0.37 &  0.08& 1.00 &  0.11& 1.41 &  0.11& 1.58 &  0.21\\ 
      &      & 0.20 & 0.25& 0.37 &  0.08& 0.83 &  0.08& 1.26 &  0.08& 1.80 &  0.14\\ 
      &      & 0.25 & 0.30& 0.43 &  0.08& 0.80 &  0.06& 1.27 &  0.08& 1.52 &  0.10\\ 
      &      & 0.30 & 0.35& 0.30 &  0.06& 0.63 &  0.04& 1.26 &  0.08& 1.60 &  0.12\\ 
      &      & 0.35 & 0.40& 0.32 &  0.06& 0.61 &  0.04& 1.08 &  0.05& 1.53 &  0.09\\ 
      &      & 0.40 & 0.45& 0.23 &  0.05& 0.60 &  0.04& 0.97 &  0.04& 1.27 &  0.08\\ 
      &      & 0.45 & 0.50& 0.18 &  0.03& 0.54 &  0.04& 0.89 &  0.04& 1.14 &  0.06\\ 
      &      & 0.50 & 0.60& 0.18 &  0.03& 0.49 &  0.03& 0.79 &  0.04& 1.04 &  0.06\\ 
      &      & 0.60 & 0.70& 0.15 &  0.03& 0.39 &  0.05& 0.59 &  0.05& 0.80 &  0.08\\ 
      &      & 0.70 & 0.80& 0.13 &  0.04& 0.25 &  0.04& 0.46 &  0.06& 0.61 &  0.08\\ 
\hline
 0.75 & 0.95 & 0.10 & 0.15& 0.31 &  0.09& 0.76 &  0.12& 0.98 &  0.14& 1.26 &  0.19\\ 
      &      & 0.15 & 0.20& 0.49 &  0.08& 0.91 &  0.07& 1.42 &  0.10& 1.76 &  0.14\\ 
      &      & 0.20 & 0.25& 0.20 &  0.04& 0.73 &  0.06& 1.30 &  0.08& 1.66 &  0.13\\ 
      &      & 0.25 & 0.30& 0.33 &  0.06& 0.75 &  0.06& 1.16 &  0.06& 1.41 &  0.09\\ 
      &      & 0.30 & 0.35& 0.32 &  0.05& 0.69 &  0.05& 0.96 &  0.05& 1.24 &  0.07\\ 
      &      & 0.35 & 0.40& 0.26 &  0.04& 0.53 &  0.04& 0.89 &  0.05& 1.08 &  0.06\\ 
      &      & 0.40 & 0.45& 0.16 &  0.03& 0.46 &  0.03& 0.77 &  0.04& 0.89 &  0.05\\ 
      &      & 0.45 & 0.50& 0.13 &  0.02& 0.38 &  0.03& 0.60 &  0.03& 0.88 &  0.06\\ 
      &      & 0.50 & 0.60& 0.13 &  0.03& 0.31 &  0.02& 0.50 &  0.03& 0.76 &  0.05\\ 
      &      & 0.60 & 0.70& 0.09 &  0.02& 0.23 &  0.03& 0.37 &  0.04& 0.52 &  0.07\\ 
%      &      & 0.70 & 0.80& 0.064 &  0.021& 0.161 &  0.025& 0.30 &  0.04& 0.38 &  0.06\\ 
\hline
 0.95 & 1.15 & 0.10 & 0.15& 0.36 &  0.08& 0.64 &  0.10& 0.86 &  0.11& 1.28 &  0.15\\ 
      &      & 0.15 & 0.20& 0.56 &  0.10& 0.87 &  0.06& 1.25 &  0.08& 1.65 &  0.12\\ 
      &      & 0.20 & 0.25& 0.46 &  0.06& 0.78 &  0.06& 1.07 &  0.06& 1.39 &  0.08\\ 
      &      & 0.25 & 0.30& 0.36 &  0.05& 0.69 &  0.05& 0.99 &  0.06& 1.20 &  0.07\\ 
      &      & 0.30 & 0.35& 0.22 &  0.04& 0.55 &  0.04& 0.82 &  0.04& 0.95 &  0.06\\ 
      &      & 0.35 & 0.40& 0.14 &  0.03& 0.38 &  0.03& 0.68 &  0.04& 0.77 &  0.05\\ 
      &      & 0.40 & 0.45& 0.10 &  0.02& 0.28 &  0.02& 0.52 &  0.03& 0.65 &  0.04\\ 
      &      & 0.45 & 0.50& 0.09 &  0.02& 0.27 &  0.02& 0.42 &  0.03& 0.54 &  0.03\\ 
      &      & 0.50 & 0.60& 0.08 &  0.02& 0.21 &  0.02& 0.30 &  0.02& 0.41 &  0.03\\ 
%      &      & 0.60 & 0.70& 0.047 &  0.014& 0.125 &  0.020& 0.221 &  0.024& 0.29 &  0.04\\ 
%      &      & 0.70 & 0.80& 0.040 &  0.014& 0.102 &  0.017& 0.146 &  0.025& 0.214 &  0.033\\ 
\hline
\end{tabular}
\end{center}
\end{table}

\begin{table}[hp!] 
\begin{center}
\begin{tabular}{rrrr|r@{$\pm$}lr@{$\pm$}lr@{$\pm$}lr@{$\pm$}l} 
\hline
$\theta_{\hbox{\small min}}$ &
$\theta_{\hbox{\small max}}$ &
$p_{\hbox{\small min}}$ &
$p_{\hbox{\small max}}$ &
\multicolumn{8}{c}{$d^2\sigma^{\pi^-}/(dpd\theta)$} 
\\
(rad) & (rad) & (\GeVc) & (\GeVc) &
\multicolumn{8}{c}{(barn/(\GeVc rad))}
\\
  &  &  & 
&\multicolumn{2}{c}{$ \bf{3 \ \GeVc}$} 
&\multicolumn{2}{c}{$ \bf{5 \ \GeVc}$} 
&\multicolumn{2}{c}{$ \bf{8 \ \GeVc}$} 
&\multicolumn{2}{c}{$ \bf{12 \ \GeVc}$} 
\\ 
\hline
 1.15 & 1.35 & 0.10 & 0.15& 0.43 &  0.09& 0.58 &  0.09& 0.91 &  0.13& 1.15 &  0.19\\ 
      &      & 0.15 & 0.20& 0.46 &  0.06& 0.90 &  0.07& 1.21 &  0.07& 1.49 &  0.11\\ 
      &      & 0.20 & 0.25& 0.37 &  0.06& 0.76 &  0.05& 0.98 &  0.05& 1.23 &  0.08\\ 
      &      & 0.25 & 0.30& 0.32 &  0.06& 0.54 &  0.04& 0.86 &  0.05& 1.03 &  0.07\\ 
      &      & 0.30 & 0.35& 0.17 &  0.03& 0.36 &  0.03& 0.64 &  0.05& 0.80 &  0.07\\ 
      &      & 0.35 & 0.40& 0.18 &  0.04& 0.30 &  0.02& 0.44 &  0.03& 0.60 &  0.04\\ 
      &      & 0.40 & 0.45& 0.10 &  0.03& 0.25 &  0.02& 0.34 &  0.02& 0.48 &  0.03\\ 
      &      & 0.45 & 0.50& 0.06 &  0.02& 0.18 &  0.02& 0.26 &  0.02& 0.36 &  0.03\\ 
%      &      & 0.50 & 0.60& 0.034 &  0.011& 0.104 &  0.017& 0.181 &  0.015& 0.269 &  0.025\\ 
%      &      & 0.60 & 0.70& 0.021 &  0.009& 0.065 &  0.010& 0.119 &  0.016& 0.172 &  0.019\\ 
%      &      & 0.70 & 0.80& 0.018 &  0.009& 0.047 &  0.008& 0.094 &  0.015& 0.128 &  0.028\\ 
\hline
 1.35 & 1.55 & 0.10 & 0.15& 0.34 &  0.08& 0.72 &  0.12& 1.05 &  0.17& 1.47 &  0.27\\ 
      &      & 0.15 & 0.20& 0.52 &  0.07& 0.99 &  0.08& 1.27 &  0.09& 1.63 &  0.12\\ 
      &      & 0.20 & 0.25& 0.40 &  0.06& 0.72 &  0.06& 0.84 &  0.06& 1.17 &  0.09\\ 
      &      & 0.25 & 0.30& 0.28 &  0.05& 0.47 &  0.05& 0.69 &  0.05& 0.89 &  0.08\\ 
      &      & 0.30 & 0.35& 0.13 &  0.03& 0.28 &  0.03& 0.52 &  0.05& 0.59 &  0.05\\ 
      &      & 0.35 & 0.40& 0.09 &  0.02& 0.22 &  0.02& 0.33 &  0.02& 0.40 &  0.03\\ 
      &      & 0.40 & 0.45& 0.07 &  0.02& 0.16 &  0.02& 0.26 &  0.02& 0.30 &  0.03\\ 
      &      & 0.45 & 0.50& 0.05 &  0.02& 0.12 &  0.01& 0.19 &  0.02& 0.22 &  0.02\\ 
%      &      & 0.50 & 0.60& 0.027 &  0.010& 0.071 &  0.011& 0.119 &  0.013& 0.141 &  0.016\\ 
%      &      & 0.60 & 0.70& 0.014 &  0.007& 0.034 &  0.008& 0.071 &  0.009& 0.090 &  0.014\\ 
%      &      & 0.70 & 0.80& 0.011 &  0.007& 0.029 &  0.007& 0.055 &  0.010& 0.073 &  0.013\\ 
\hline
 1.55 & 1.75 & 0.10 & 0.15& 0.38 &  0.08& 0.63 &  0.11& 0.88 &  0.14& 1.33 &  0.24\\ 
      &      & 0.15 & 0.20& 0.43 &  0.06& 0.85 &  0.07& 1.23 &  0.08& 1.37 &  0.11\\ 
      &      & 0.20 & 0.25& 0.21 &  0.04& 0.57 &  0.05& 0.70 &  0.05& 0.85 &  0.07\\ 
      &      & 0.25 & 0.30& 0.14 &  0.03& 0.34 &  0.03& 0.48 &  0.04& 0.59 &  0.05\\ 
      &      & 0.30 & 0.35& 0.10 &  0.03& 0.25 &  0.03& 0.36 &  0.03& 0.47 &  0.04\\ 
      &      & 0.35 & 0.40& 0.05 &  0.02& 0.16 &  0.02& 0.23 &  0.02& 0.34 &  0.03\\ 
      &      & 0.40 & 0.45& 0.03 &  0.01& 0.10 &  0.01& 0.16 &  0.01& 0.25 &  0.03\\ 
      &      & 0.45 & 0.50& 0.02 &  0.01& 0.08 &  0.01& 0.11 &  0.01& 0.18 &  0.02\\ 
%      &      & 0.50 & 0.60& 0.011 &  0.006& 0.050 &  0.008& 0.057 &  0.007& 0.103 &  0.014\\ 
%      &      & 0.60 & 0.70& 0.005 &  0.004& 0.021 &  0.006& 0.036 &  0.005& 0.059 &  0.009\\ 
%      &      & 0.70 & 0.80& 0.004 &  0.004& 0.016 &  0.005& 0.030 &  0.006& 0.046 &  0.010\\ 
\hline
 1.75 & 1.95 & 0.10 & 0.15& 0.38 &  0.07& 0.58 &  0.07& 0.79 &  0.09& 0.92 &  0.13\\ 
      &      & 0.15 & 0.20& 0.38 &  0.06& 0.60 &  0.05& 0.83 &  0.05& 1.03 &  0.07\\ 
      &      & 0.20 & 0.25& 0.17 &  0.04& 0.36 &  0.03& 0.54 &  0.03& 0.56 &  0.05\\ 
      &      & 0.25 & 0.30& 0.14 &  0.04& 0.21 &  0.02& 0.37 &  0.03& 0.39 &  0.04\\ 
      &      & 0.30 & 0.35& 0.11 &  0.03& 0.14 &  0.02& 0.25 &  0.02& 0.25 &  0.03\\ 
      &      & 0.35 & 0.40& 0.06 &  0.02& 0.09 &  0.01& 0.18 &  0.02& 0.17 &  0.02\\ 
      &      & 0.40 & 0.45& 0.03 &  0.02& 0.08 &  0.01& 0.12 &  0.01& 0.15 &  0.02\\ 
      &      & 0.45 & 0.50& 0.02 &  0.01& 0.06 &  0.01& 0.08 &  0.01& 0.12 &  0.01\\ 
%      &      & 0.50 & 0.60& 0.008 &  0.006& 0.030 &  0.007& 0.040 &  0.006& 0.067 &  0.012\\ 
%      &      & 0.60 & 0.70& 0.005 &  0.006& 0.013 &  0.004& 0.021 &  0.004& 0.036 &  0.009\\ 
%      &      & 0.70 & 0.80& 0.004 &  0.005& 0.008 &  0.003& 0.016 &  0.004& 0.020 &  0.006\\ 
\hline
 1.95 & 2.15 & 0.10 & 0.15& 0.32 &  0.06& 0.58 &  0.07& 0.69 &  0.06& 0.78 &  0.08\\ 
      &      & 0.15 & 0.20& 0.22 &  0.05& 0.53 &  0.05& 0.66 &  0.04& 0.77 &  0.06\\ 
      &      & 0.20 & 0.25& 0.10 &  0.03& 0.26 &  0.03& 0.39 &  0.03& 0.46 &  0.04\\ 
      &      & 0.25 & 0.30& 0.12 &  0.04& 0.17 &  0.02& 0.26 &  0.03& 0.34 &  0.03\\ 
      &      & 0.30 & 0.35& 0.05 &  0.03& 0.10 &  0.02& 0.16 &  0.01& 0.19 &  0.03\\ 
      &      & 0.35 & 0.40& 0.01 &  0.01& 0.06 &  0.01& 0.11 &  0.01& 0.10 &  0.01\\ 
      &      & 0.40 & 0.45& 0.01 &  0.01& 0.04 &  0.01& 0.08 &  0.01& 0.08 &  0.01\\ 
      &      & 0.45 & 0.50& 0.01 &  0.01& 0.04 &  0.01& 0.05 &  0.01& 0.05 &  0.01\\ 
%      &      & 0.50 & 0.60& 0.008 &  0.009& 0.014 &  0.005& 0.023 &  0.005& 0.036 &  0.007\\ 
%      &      & 0.60 & 0.70& 0.005 &  0.008& 0.008 &  0.003& 0.012 &  0.003& 0.018 &  0.005\\ 
%      &      & 0.70 & 0.80& 0.003 &  0.012& 0.004 &  0.002& 0.011 &  0.003& 0.010 &  0.004\\ 
\hline
\end{tabular}
\end{center}
\end{table}
%